FEN BİLİMLERİ ENSTİTÜSÜ

BİLGİSAYAR MÜHENDİSLİĞİ ANABİLİM DALI

# KUANTUM FISHER BİLGİSİ OPTİMİZASYONU ÖNERİSİ VE DOLANIKLIK ÖLÇÜTLERİ İLE İLİŞKİSİ

DOKTORA TEZİ

VOLKAN EROL

tarafından

DOKTORA

derecesi şartını sağlamak için hazırlanmıştır.

Program: Bilgisayar Mühendisliği

KUANTUM FISHER BİLGİSİ OPTİMİZASYONU ÖNERİSİ

VE DOLANIKLIK ÖLÇÜTLERİ İLE İLİŞKİSİ

DOKTORA TEZİ

VOLKAN EROL

tarafından

OKAN ÜNİVERSİTESİ

Bilgisayar Mühendisliği Anabilim Dalına

Doktora

derecesi şartını sağlamak için sunulmuştur.

Onaylayan:

| | |
|---|---|
| Danışman | Eş-Danışman |
| Doç. Dr. Azmi Ali Altıntaş | Doç. Dr. Fatih Özaydın |
| Jüri Üyesi | Jüri Üyesi |
| Prof. Dr. Bekir Tevfik Akgün | Yrd. Doç. Dr. Birim Balcı Demirci |
| Jüri Üyesi | |
| Yrd Doç. Dr. Bahri Atay Özgövde | |

Mayıs 2015

Program: Bilgisayar Mühendisliği

# KISA ÖZET

Günümüzde Kuantum Bilgi Kuramı üzerine çalışmalar aktif bir şekilde devam etmektedir. Shor'un çarpanlara ayırma algoritması veya Grover'ın arama algoritması gibi bazı algoritmaların kuantum sistemlerde klasik sistemlere göre çok daha hızlı şekilde çalışabileceği gösterilmiştir. Son dönemde, pratikte Kuantum Bilgisayarlarının üretilmesi konusunda; Kuantum Tekrarlayıcı, Hafıza ve İşlemciler üretilerek ciddi oranda yol alınmıştır. İşlem hızı ve kapasitesi açısından Bilgisayar Bilimleri problemlerinde devrimsel bir dönemin kapıları yavaş yavaş aralanmaktadır. Kuantum Anahtar Dağıtımı Altyapıları çok uzunca bir süredir teknolojik olarak hayatımızdadır ve Bankacılık, Savunma vb. sektörler için ürünleşme noktasına kadar gelmiştir.

Kuantum Hesaplama açısından kullanılan en temel teorik altyapı dolanıklık olarak karşımıza çıkar. Dolanıklığın hesaplama açısından bize sağladığı fayda bahsedilen kuantum algoritmaların üretiminde önem arz etmektedir. Dolanıklığı ölçmenin çeşitli yöntemleri bulunmaktadır. Bunlardan en formel olanı Dolanıklık Ölçütleri ya da Dolanıklık Monotonları dediğimiz altyapıların kullanılmasıdır. Bu konularda açık bir konu olan Kuantum Sistem Durumlarının Sıralaması problemi özellikle çoklu dolanık sistemler için çözümlenmesi gereken önemli ve açık bir problemdir.

Fisher Bilgisi, Bilgi Kuramı açısından birçok günümüz probleminin çözümlenmesinde bize uygun bir altyapı sağlamaktadır. Keşifsel Veri Analizi dediğimiz yöntemle, Büyük Veri, Veri Madenciliği, Makina Öğrenimi gibi konularda çözümler elde edilebilmektedir. Kuantum Fisher Bilgisi, faz hassasiyeti gerektiren durumlarda işe yarayan bir değer olmakla beraber kuantum bilgisayarlarının icadı ile bahsettiğimiz altyapının kuantum eşleniğini oluşturacaktır. Kuantum Fisher Bilgisi tek başına bir dolanıklık ölçütü değildir.

Tez kapsamında Kuantum Fisher Bilgisi önerilen yeni bir optimizasyon yöntemi ile optimize edilmiş ve dolanıklık ölçütleri ile sistem durum sıralaması açısından ilişkileri incelenmiştir. Daha önce sistem durum sıralaması üzerine yapılan çalışmaların üzerine özgün ve oldukça ilginç sonuçlar bulunmuş ve bu sonuçlar tez kapsamında açıklanmıştır. Bulunan en ilginç sonuç Lokal Operasyon Klasik İletişim yöntemleri ile maksimize edilen Kuantum Fisher Bilgisi'nin iki kübit sistem durumları için özellikle Dolanıklığın Göreceli Entropisi ölçütü ile anlamlı bir sıralama ilişkisi içerisinde olmasıdır. Ayrıca diğer dolanıklık ölçütleri ile bulunmuş olan analiz çalışmaları da paylaşılmıştır. Çalışmamız kübit-kütrit sistem durumlarına da genişletilmiş ve elde edilen sıralama ilişkileri ve sınıflandırmaları paylaşılmıştır. Belli sistemler için Kuantum Fisher Bilgisi'nin belli uyum bozulması kanalları altındaki değişimleri de detaylı olarak incelenmiştir.

# ABSTRACT

Studies about Quantum Information Theory continue actively in many research institutions. Algorithms like Shor's factorization algorithm or Grover's search algorithm are shown that should work quite faster on quantum systems compared to classical systems. Very recently, pratical setups of large scale quantum computers are widely studied e.g. quantum repeaters, memories and processors. The doors of a revolunary quantum era in Computer Science is to be opened after some period of time. Technologies like Quantum Key Distribution were defined and developed since many years and they have been daily life products for some sectors like Banking and Military application.

In Quantum Computing, Entanglement is used for the base computational infrastructure. Entanglement provides us a computational advantage in realization of quantum algorithms. Some ways to quantifiying entanglement were defined. The best formal way to quantify it, is the methods that we call Entanglement Measures or Entanglement Monotones. In this research area, State Ordering Problem is defined and still an open problem especially for multiparticle entangled states.

Fisher Information provides a good background for the solution of some actual Information Theory problems. With the method called as Exploratory Data Analysis, problems related to Big Data, Data Mining and Machine Learning could be solved. Quantum Fisher Information (QFI) is a value that could be used in situations where phase sensitivity is important and this concept is expected to be the quantum version of the mentioned inftrastructure for Quantum Systems. QFI cannot be defined as an Entaglement Measure or Monotone.

In the scope of this thesis, a new optimization technique is proposed for optimizing QFI and thanks to this optimization method, the ordering relation between QFI and entanglement measures is studied. Based on the studies made in the area of quantum state ordering, new and interesting analysis results are found and reported. The main important and interesting result achieved is that for two qubit quantum states, QFI maximized under Local Operation and Classical Communication has an interesting ordering relation with entanglement measures especially with Relative Entropy of Entanglement. Our study is extended fort he qubit-qutrit systems and ordering relations and classification results are presented. For some quantum systems, the changes in QFI under decoherence channels are also considered.

Hayat arkadaşım, sevgili eşim, sevgilim Arzu’ya …

# TEŞEKKÜR

Öncelikle Tez kapsamındaki çalışmalardaki inanılmaz destek, yönlendirme ve yardımları için Danışmanlarım Doç. Dr. Azmi Ali Altıntaş ve Doç. Dr. Fatih Özaydın'a teşekkür etmek istiyorum. Kendilerinin motivasyonu ve çalışkanlığı sayesinde çok verimli ve güzel bir çalışma ortamı yakaladık. Umuyorum ki bundan sonraki dönemde de kendileriyle artan bir hızda çalışmaya devam edeceğiz.

Bilgisayar Müh. Bölüm Başkanı ve Doktora Program Koordinatörü Prof. Dr. B. Tevfik Akgün'e desteği ve yardımları için çok teşekkür ederim. Doktoraya başlama konusunda beni teşvik eden sevgili dostum Doç. Dr. Erhan Abdullah'a çok teşekkür ederim.

Doktora Tez İzleme Jürimde yer alan Yrd. Doç. Dr. Birim Balcı Demirci ve Yrd. Doç. Dr. Bahri Atay Özgövde'ye olumlu yorum ve yönlendirmeleri için çok teşekkür ederim. Galatasaray Üniversitesi'nden sevgili hocam Dr. Sultan Turhan'a desteği için teşekkür etmek istiyorum.

Eğitim hayatım boyunca benden desteğini esirgemeyen sevgili anneme ve babama çok teşekkür ederim.

Son olarak 3,5 yıllık Doktora maratonunda sürekli olarak yanımda duran, manevi desteğini esirgemeyen ve benim bütün çalışma tempoma sabırla katlanan sevgili eşim, hayat arkadaşım Arzu'ya ve kendisini bazen ihmal ettiğim biricik kızım Çağla İnci'ye teşekkür etmek istiyorum. Bundan sonra, söz veriyorum sizi ihmal etmeyeceğim.

Doktora kapsamındaki çalışmalar (şu an faal durumda olmayan) Provus A MasterCard Company Ar-Ge Merkezi, FMV Işık Üniversitesi BAP-14A101 projesi ve Netaş Telekomünikasyon A.Ş. tarafından kısmen fonlanmıştır.

# İÇİNDEKİLER

# TABLO LİSTESİ

# ŞEKİL LİSTESİ

# SİMGELER

$| \ \rangle$ : Ket

$\langle \ |$: Bra

$\otimes$ : Tensör Çarpımı (Kronecker Product)

AND: And Kapısı

OR: Or Kapısı

NOT: Not Kapısı

NAND: Nand Kapısı

NOR: Nor Kapısı

CNOT: Controlled Not Kapısı

U: Üniter-Unitary Matris

tr: Trace-iz İşlemi

$tr_B$ : B alt sistemi üzerinde Kısmi Trace İşlemi

$A_i$: Kraus Operatörleri

S: Kuantum Göreceli Entropi

$J_n$: Açısal Momentum Operatörleri

$\sigma_{x,y,z}$: Pauli Operatörleri

$U_{Rot}$: Euler Rotasyonu

# KISALTMALAR

KFB: Kuantum Fisher Bilgisi

LOCC: Local Operation Classical Communication-Lokal Operasyon Klasik Kanal

REE: Relative Entropy of Entanglement-Dolanıklığın Göreceli Entropisi

ADC: Amplitude Damping Channel-Genlik Azaltan Kanal

PDC: Phase Damping Channel-Faz Azaltan Kanal

DPC: Depolarizing Channel-Depolarize Kanal

RMQFI: Mean Quantum Fisher Information per Particle-Parçacık Başı Ortalama KFB

QED: Cavity Quantum Electrodynamics-Kavite Kuantum Elektrodinamikleri

NMR: Nüklear Manyetik Rezonans

DNA: Deoksiribo Nükleik Asit

Q: Cavity Quality Factor-Kavite Kalite Etmeni

QKD: Quantum Key Distribution-Kuantum Anahtar Dağıtımı

NICT: National Institute of Information and Communication Technologies – Tokyo

Z: Z Kapısı

H: Hadamard Kapısı

N: Negatiflik

C: Concurrence-Eş Zamanlılık

MKFB: Maksimize Kuantum Fisher Bilgisi

# I. GİRİŞ

Kuantum Bilgi Teorisi ve Kuantum Hesaplama konuları geleceğin bilgisayar teknolojisi olarak nitelendirilen ve çok yüksek hızlarda işlem yapacak olması öngörülen Kuantum Bilgisayarlarının teorik temelini oluşturan oldukça sıcak çalışma alanlarıdır.

Dolanıklık sayesinde, kuantum mekaniksel sistemler, birçok bilgi işleme görevini klasik mekaniksel sistemlere nazaran çok daha hızlı gerçekleştirebilmektedir. Örneğin Shor'un çarpanlara ayırma algoritması, Grover'ın arama algoritması, kuantum Fourier dönüşümü vs.[1]. Öte yandan kuantum durum (bilgi) ışınlanması gibi, klasik mekanikle mümkün olmayan birçok iş, kuantum mekaniksel sistemlerle gerçekleştirilebilmektedir. Dolayısıyla kuantum teknolojileri, bilgisayardan haberleşmeye ve şifrelemeye kadar birçok alanda çığır açma sürecindedir.

Kuantum dolanıklık, Kuantum Bilgisayarlarının veri işleyebilmesini sağlayan altyapıyı oluşturmaktadır. Dolanıklık iki taraflı ve çok taraflı halde gözlemlenebilmektedir. İki taraflı sistemlerin dolanıklık miktarını ölçmek üzere çeşitli ölçütler geliştirilmiştir. Örneğin, sistemin yoğunluk matrisinin kısmi transpozisyonunun özdeğerlerinin negatifliğine dayanan Negatiflik (Negativity), Logaritmik Negatiflik (Logarithmic Negativity) ve Ez Zamanlılık (Concurrence); sistemin, dolanık olmayan sistemler arasından kendisine en yakın olan sisteme uzaklığına dayanan Dolanıklığın Göreceli Entropisi vb. [2-4].

Öte yandan, çok taraflı dolanık sistemlerin dolanıklık miktarını ölçebilecek genel ve kabul edilmiş bir ölçüt, henüz bulunamamıştır. Ne var ki birçok bilgi işleme görevinde kullanılması gerektiğinden, çok taraflı kuantum dolanık sistemlerin üretimi ve işlenmesi, son yılların sıcak konularının başında gelmektedir [5-11].

Sistemlerin, faz hassasiyeti gerektiren işlerde sağlayabileceği hassasiyeti de ölçmeye yarayan Fisher Bilgisi'nin, kuantum sistemler için geliştirilmiş versiyonu olan Kuantum Fisher Bilgisi, son yıllarda yine, çokça çalışılan bir konu haline gelmiştir [12-30,64]. Klasik Bilgisayar Mühendisliği açısından önemli olan bazı problemlerin çözümünde; Örneğin büyük miktarda ve çok değişkenli verinin işlenmesi, veri madenciliği, vb. Fisher Bilgisi kullanılmaktadır [31]. Kuantum Bilgisayarları üretildiğinde benzer veya daha karmaşık problemlerin çözümünde Kuantum Fisher Bilgisi'nin kullanılacağı öngörülmektedir.

Çok taraflı dolanık kuantum sistemlerin bazıları, klasik sistemlerin sağlayabileceği en yüksek hassasiyet limitini aşabilmektedir. Bu, klasik limiti aşabilen sistemlere "kullanışlı" sistemler denmektedir [32,33]. Hangi sistemlerin, hangi çevresel gürültüler altında ne kadar kullanışlı olduğu, kuantum teknolojileri için temel bir alan haline gelmiştir [21]. Fisher bilgisi sayesinde çok taraflı dolanık sistemler için bir dolanıklık ölçütü geliştirilebileceği düşünülmekte ve bu konuda güncel olarak çalışılmaktadır [34].

Bu tezde, kuantum dolanıklık ve kuantum Fisher bilgisinin Bilgisayar Mühendisliği kavramlarına etkisinin anlaşılması amacıyla çalışılmıştır. Çok taraflı kuantum dolanıklık ve çokça çalışılmış olan iki taraflı kuantum dolanıklık ölçütleriyle kuantum Fisher Bilgisi'nin ilişkisi araştırılmıştır. Spesifik olarak, çeşitli sınıflardan çok sayıda iki taraflı dolanık sistemler ele alındığında, kimi dolanıklık ölçütlerinin diğerinden yüksek, kiminin de düşük çıktığı hal-i hazırda bulunmuştur ve bu çerçevede, sistemlerin sıralanması için çeşitli yöntemler geliştirilmiştir [35,36]. Daha ilginç olarak, gürültülü (karışık) sistemlerin, gürültüsüz (saf) sistemlerden, genellikle daha dolanık olduğu da bulunmuştur [37] .

Kuantum kriptografi, haberleşme, bilgisayarı gibi temel kuantum teknolojilerindeki birçok işte, GHZ, W gibi çok taraflı kuantum dolanık (multi-partite entangled) sistemlere ihtiyaç vardır [22,30]. Faz hassasiyeti gerektiren işlerde, klasik sistemlerden daha iyi faz hassasiyeti verebilmesi için, bir kuantum sistemin sahip olması gereken niteliklerin başında, çok taraflı dolanık bir sistem olması gerekmektedir, ancak bu nitelik tek başına yeterli değildir. Sistemin Fisher bilgisinin hesaplanması ve belli bir seviyeyi tutturduğu da kanıtlanmalıdır.

Bir deneyden ne kadar veri alabiliriz sorusuyla ortaya çıkan Fisher bilgisinin kuantum uzantısı, Helstrom (1976) [38] ve Holevo (1982) [39] tarafından geliştirilmiştir. Kuantum Fisher bilgisi (Petz vd. 2008) temel olarak, sistemlerin, kuantum teknolojilerinde sağlayabileceği faz hassasiyetinin hesaplanmasında kullanılmaktadır. Klasik sistemler, ancak, ("shot noise limit" denilen) belli bir seviyeye kadar hassasiyet sağlamaktayken, Hyllus vd. (2010) [34] kuantum sistemlerin hepsinin klasik sistemlerden daha yüksek faz hassasiyeti sağlayamadığını bulmuştur. Dolayısıyla hangi kuantum sistemlerin, klasikten daha iyi hassasiyet sağlayabileceği, önemli bir çalışma alanı haline gelmiştir ve bu kuantum sistemlere, "kullanılabilir" denmeye başlanmıştır.

Kuantum sistemlerin de standart bir limitinin olduğu ancak Heisenberg belirsizlik ilkesinin sınırlandırmasıyla, bazı özel durumlarda bu kuantum limitin de aşılabileceği, Lloyd vd.

(2004) tarafından Science dergisinde yayımlanmıştır [33] . EPR/Bell çifti yapısındaki yalnızca iki taraflı dolanık sistemlerin kullanılabilir olmadığı ve kullanılabilirliğin temel gerek şartının, çok taraflı dolanık kuantum sistem olduğu ortaya çıkmıştır ancak yeter şart için, sistemdeki parçacık başına düşen ortalama kuantum Fisher bilgisinin belli bir kriteri tutturması gerektiği, Pezze vd. (2009) [32] tarafından bulunmuştur. Bu bulguya göre, çok taraflı dolanık bir sistem için, taraf (yani sistemi oluşturan kuantum parçacığı) sayısının, sistemin verebileceği maksimum kuantum Fisher bilgisine olan oranı, "parçacık başına ortalama kuantum Fisher bilgisi" olarak bir parametre olarak tanımlanmıştır. Bu parametre, ne kadar küçükse, sistemin faz hassasiyet o kadar yüksek olmaktadır. En iyi sonuç verebilecek klasik sistemler için bu parametre, minimum 1 olmaktadır. Eldeki bir sistemin "kullanılabilir" yani klasikten daha iyi sonuç verebileceğini anlayabilmek üzere, o sistem için bu parametrenin 1'den küçük olduğunun anlaşılması gerek ve yeter koşuldur.

Kuantum teknolojilerinde bir çığır açan bu buluşla beraber, çok taraflı dolanık sistemlerin kuantum Fisher bilgisi üzerine araştırmalar ve kullanılabilir sistemlerin sınıflandırılması çalışmaları çığ gibi büyümüştür: Xiong vd. (2010) [40] , Dicke sistemlerinin üstkonumlu hallerinin, kendinden daha kullanılabilir olduğunu göstermiştir. GHZ ve W sistemlerinin saf ve tek halleri ile, çeşitli üstkonumlu hallerinin kuantum Fisher bilgisi üzerine birçok çalışma yapılmıştır. 3 kuantum parçacıktan oluşan böyle bir kuantum sistemdeki, üstkonumluluk arasındaki üstkonumluluk katsayıları ve göreceli fazlara göre kuantum Fisher bilgisinin nasıl değiştiği Wang vd (2012) [14] tarafından ve 4 kuantum parçacıktan oluşan benzeri bir sistemin kuantum Fisher bilgisi Ouyang vd.(2012) tarafından bulunmuştur.

Miranowicz ve arkadaşları, bir sistem için bu çeşitli dolanıklık ölçütlerinin farklı değerler verdiğinin keşfiyle birlikte, sistematik bir şekilde çok sayıda kuantum sistemi simüle ederek, dolanıklık ölçütlerinden verdiği değerleri karşılaştırmışlardır. Buldukları genel sonuca göre, örneğin kimi sistem çiftlerinde bir ölçüt değeri birinci sistemde büyük ikincide küçükken, başka bir ölçütten gelen değer bunun tam tersi davranış sergilemektedir [35,36]. Bu sonuç, kuantum dolanıklık ölçütlerinin, sistemlerin farklı özelliklerini yansıttığını önermektedir. Biz bu çalışmamızda, bu problemi Kuantum Fisher Bilgisi açısından ele alıp genişlettik.

Bu sistemlerin kuantum Fisher bilgisinin çalışılması ve dolanıklık ölçütleriyle olan, özellikle sıralama ilişkilerinin çalışılması, tez çalışması kapsamında yapılmıştır. [35,36]'daki sistemler üretilmiş, dolanıklık ölçütleri hesaplanmış ve bu sistemlerin, orijinal katkımız olarak, kuantum Fisher bilgilerinin hesaplanması ve dolanıklık ölçütleriyle birlikte analizi yapılmıştır.

Bu sayede, kuantum teknolojilerinin her bir özel bilgi işleme protokolünde gerekli seviyede dolanıklık içeriyor olarak ihtiyaç duyulan sistemlerin, aynı zamanda beklenen faz hassasiyetini karşılayabilme kabiliyeti için genel bir çerçeve ortaya konmuştur. Aynı zamanda çok taraflı dolanıklık ölçütü bulma çabalarına da katkı sağlayacak bazı bulgulara ulaşılmıştır.

Örneğin, Kuantum Fisher Bilgisi kapsamındaki çalışmalar sırasında mevcut araştırmalarda Fisher Bilgisi'nin optimize edilmeden kullanıldığı görülmüştür. Bu çalışma kapsamında Lokal Operasyon Klasik Kanal kullanarak bir optimizasyon prosedürünü KFB üzerinde uygulanmış ve elde edilen sonuçlardan yola çıkarak dolanıklık ölçütleri ile karşılaştırılmalı analizi yapılmıştır.

Tezin kapsamında öncelikle Kuantum Bilgisayar Mühendisliği'nin 10 yıllık bir projeksiyonu verilmiş ve Fisher Bilgisi ve Dolanıklık kavramlarının Bilgisayar Bilimleri ve Mühendisliği'nin ne tür problemlerini çözebileceği ile ilgili analiz çalışması yapılmıştır. Daha sonra Kuantum Bilgisayarlarının neden çalışılması gereken bir kavram olduğu irdelenmiştir. Bu kısımda Google, Nasa ve D-Wave'in çalışmaları da anlatılarak orta vadeli bir perspektifte bu teknolojide nasıl bir ilerleme beklendiği de açıklanmıştır. Daha sonraki bölümde Kuantum Mekaniği'ne ve Dolanıklık Kavramına giriş bilgileri verilmiştir. Kuantum Bilgisayarların fiziksel olarak nasıl üretildikleri ve mevcut teknolojik durumda üretilen türevlerinin nasıl üretildikleri ve neleri içerdikleri anlatılmıştır. İlerleyen bölümde, Dolanıklık Ölçütleri ile ilgili bir literatür taraması yapılmış ve literatürde yaygın olarak çalışılmış ölçütler ile ilgili bilgiler verilmiştir. Dördüncü olarak Kuantum Fisher Bilgisi ile ilgili detaylı anlatım yapılmıştır. Bir sonraki adımda ise İki Kübit ve Kübit-Kütrit Kuantum Sistemler İçin Dolanıklık Ölçütleri ve Kuantum Fisher Bilgisi'nin Analizi yapılmıştır. Bu bölümde elde ettiğimiz bulgular detaylı olarak aktarılmıştır. Son olarak elde ettiğimiz sonuçlar paylaşılmış ve alanın açık konuları açıklanmıştır.

# II. KUANTUM KURAMI VE KUANTUM BİLGİSAYARLARI

Kuantum Bilişim zorlu fiziksel deneylerin doğruluğunu kanıtlamak için kullanılan bir yoldur. Kuantum Bilgisayarları icat edilene kadar Bilgisayar Bilimcileri Kuantum Bilişim kavramından nasıl faydalanmalıdır? Gerçekte, Kuantum Bilişim tek başına yeni Hesaplama Cihazları yapmakla ilgilenmez. Bizlere dünyadaki gerçek problemlere bakış açımızı değiştirecek ve mevcut durumdan çok daha hızlı çözümler bulmamızı sağlayacak kadar radikal bir Bilimsel endüstri devriminin habercisi olmaktadır.

## II.1 Fisher Bilgisi

Kuantum Bilişimle ilgili en dikkat çeken çalışmalardan bazıları Büyük Veri ile ilgili çalışmaların *klasik Fisher Bilgisi temeliyle* açıklanmasıdır. *Fisher Bilgisi*, aslında fiziksel bir ölçüt olarak tanımlanmakla beraber bazı büyük veri ile ilgili problemlerin çözümünde oldukça ilginç sonuçlar elde edilmesini sağlamaktadır. *Eksterm Fiziksel Bilgi (Extreme Physical Information)* [31] denilen bir yaklaşımla istatistiksel olarak bu sistemler hakkında çeşitli çıkarımlar yapmak mümkündür. Bu kavramın anlaşılabilmesi için sistemdeki bileşenlerin ortak paydalarının çıkarılması ve gözlemlenmesi gerekmektedir.

Aslında tüm evren birçok hareket eden, çarpışan, birini çeken ya da farklı şekillerde etkileşime geçen varlıkların bütününden oluşmaktadır. Doğada bulunan sistemlerin bütününe ilişkin bilginin yine bu sistemlerin içinde bulunduğu bilinmektedir. Fisher Bilgisi kavramsal olarak Bilgi Teorisinin temel taşlarından birisini oluşturan Shannon Bilgisine benzetilse de ikisinin arasında bazı farklar mevcuttur. Shannon, bir mesajın içerisindeki belirsizliği olasılık kavramı ile iliskilendirerek mesajın içerisindeki bilgi miktarını tanımlamıştır. Bir mesajın içindeki toplam belirsizlik aslında o mesajın içerisindeki toplam bilgiye eşittir.

*Entropi* terimi de ilk kez *Shannon* tarafından bilgisayar bilimlerinde veri iletişiminde kullanılmıştır. Dolayısıyla literatürde *Shannon Entropisi (Shannon's Entropy)* olarak da geçen kavrama göre bir mesajı kodlamak için gereken en kısa ihtimallerin ortalama değeri alfabede bulunan sembollerin logaritmasının entropiye bölümüdür. Yani kabaca alfabemizde 256 karakter varsa bu sayının logaritmasını ($\log_2 256 = 8$) mesajın entropisine böleriz. Yani mesajdaki değişim ne kadar fazla ise o kadar fazla kodlamaya ihtiyacımız vardır. Diğer bir deyişle alfabemiz 256 karakterse ama biz sadece tek karakter yolluyorsak o zaman entropy 0 olduğundan $0/256$ farklı kodlamaya (0 bite) ihtiyacımız vardır. Veya benzer olarak her harften aynı sıklıkta yolluyorsak bu durumda $256/8 = 8$ bitlik kodlamaya ihtiyaç duyulur.

Fisher Bilgisi temel olarak bir istatiksel dağılımda bilinen bir parametreye göre bilginin miktarınının ölçütüdür. Fisher Bilgisine göre bilginin tersi belirsizliktir. *Keşifsel Veri Analizi (Exploratory Veri Analizi)* tekniği Fisher Bilgisinden yola çıkarak tanımlanmıştır. Bu teknikle açıklanabilen gerçek dünyadaki büyük veri (big data) problemlerinden bazılarını şu şekilde sıralanabilir [31]:

- Popülasyon nüfus büyüklüğünün tahmini/analizi
- Borsa hareketleri ve finansal verilerin analizi, kestirimi
- Sosyal medya uygulamalarında verilerin analizi
- İnsanlardaki çeşitli Kanser türleri için Kanserli bölgedeki büyümenin tespiti ve modellenmesi
- Optimum finansal yatırım değerlerinin tahmini ve analizi
- Moleküler biyolojideki ve biyoenfomatikteki çeşitli modelleme ve kestirim uygulamaları

Özellikle günümüz dünyasında sıkça karşımıza çıkan bu tarz problemlerin çözümünde klasik Fisher bilgisi çok yoğun olarak kullanılmaktadır. Kuantum Bilgisayarların üretilmesi ile beraber o mimaride çözülmesi öngörülen benzer ve daha karmaşık büyük veri problemlerinin çözümünde bu bilginin kuantum sistemler için geliştirilmiş türü olan kuantum Fisher bilgisinin aktif olarak kullanılacağı öngörülmektedir.

## II.2 Kuantum Fisher Bilgisi ve Dolanıklık

Kuantum sistemlerde kullanılan hesaplama yöntemlerinin temeli dolanıklık kavramına dayanmaktadır. Kuantum dolanıklığı ölçmenin birçok yöntemi bulunmaktadır ve bunlara dolanıklık monotonları veya daha özel bir ifadesi ile dolanıklık ölçütleri denmektedir. Tez kapsamında dolanıklık ölçütleri ayrı bir bölümde detaylı olarak incelenmekte ve literatürdeki güncel çalışmalar anlatılmaktadır. Tez kapsamında ağırlıklı olarak analiz edilen dolanıklık ölçütleri Eş Zamanlılık (Concurrence), Negatiflik (Negativity) ve Dolanıklığın Göreceli Entropisi (Relative Entropi of Entanglement) ölçütleridir. Bu ölçütler hem iki seviyeli (two kübit) ve kübit-kütrit kuantum sistemler açısından incelenmiş ve bu değerler hesaplanarak Sistem Durum Sıralaması problemi çerçevesinde irdelenmiştir.

Kuantum Fisher Bilgisi tek başına bir dolanıklık ölçütü değildir ancak özellikle faz hassasiyeti gerektiren durumlar için bize çok önemli yansımalar sağlamaktadır. Bu konudaki çalışmalarda açık alan olarak mevcut araştırmalarda Fisher Bilgisi'nin optimize edilmeden kullanıldığı tespit edilmiştir. Bu Tez çalışması kapsamında Lokal Operasyon Klasik Kanal kullanarak bir optimizasyon prosedürünü KFB üzerinde uygulanmış ve elde edilen sonuçlardan yola çıkarak dolanıklık ölçütleri ile karşılaştırılmalı analizi yapılmıştır. Yine Sistem Durum Sıralaması problemi çerçevesinde sıralama ilişkileri üzerinde detaylı olarak durulmuştur.

Kuantum Bilgisayarlarının 10 yıllık bir perspektifte Bilgisayar Mühendisliği'nin şu problemlerini ele alması ve klasik bilgisayardan daha performanslı bir şekilde çalışması öngörülmektedir:

- Çarpanlara ayırma tabanlı kriptografik yöntemler
- Anahtar Dağıtımı altyapıları
- Veri Arama algoritmaları
- Veri Sınıflandırma ve Makine Öğrenimi ile ilgili çalışmalar
- Olasılık ve İstatistik tabanlı hesaplama uygulamaları
- Başarıya Kadar Tekrar Et (Repeat Until Success) Tabanlı Haberleşme ve Kodlama algoritmaları
- Çizge Kuramı tabanlı problemlerin çözümleri
- Büyük Veri ve Optimizasyon problemleri

Makine öğrenimi prosedürleri hızlandırmak için bazı dolanıklık tabanlı çözümler güncel çalışmalar arasında yer almaktadır. Bu çalışmalardan bir örnek olarak 20 Mart 2015 tarihinde Physical Review Letters dergisinde yayınlanan, Çinli araştırmacılar Cai ve arkadaşlarının "Entanglement-Based Machine Learning on a Quantum" başlıklı çalışmadır [41].

Yapay Sinir Ağları'nın Kuantum Bilgi Teorisi perspektifiyle gerçeklenmesi konusunda da Japonya'dan Yamomoto ve grubunun çeşitli çalışmaları bulunmaktadır [42].

Yukarıda sayılmış problemlerin bazıları günümüz klasik kriptografi kavramlarının sonunun kuantum bilgisayarları tarafından geleceği öngörüsü ile ortaya atılan kavramlardır. Örneğin: Kuantum Kriptografi. Kriptografi alanında kuantum çağın açılması ile beraber kuantum sonrası kriptografi (post-quantum cryptograhy) kavramı da ortaya atılmıştır. Bu teorik altyapının temelinde lineer cebirsel lattice kavramı yer almaktadır. Mevcut varsayımlara göre henüz lattice tabanlı şifreleme algoritmalarının kuantum bir çözüm algoritması bulunamamıştır. Eğer günün birinde bu algoritmalar da kuantum bir yöntemle kırılabilirse kuantum sonrası kriptografi kavramı da bilim tarihinin tozlu rafları arasında yerine alacaktır.

Görüldüğü üzere Bilgisayar Bilimleri ve Mühendisliği'nin birçok güncel alanını yakından ilgilendiren konularda Kuantum Bilgi Teorisi altyapısını kullanan çalışmalar yapılmaktadır veya 10 yıllık bir perspektifte bunların yapılabileceği öngörülmektedir. Yine Gartner gibi saygın trend analizi yapan ABD menşeili araştırma kuruluşlarının raporlarına göre genel geçer bir ticari Kuantum Bilgisayarının hayatımıza girmesi 10 yıllık teknolojik perspektifte mümkün olarak gösterilmiştir. Bu sebeple bu konulardaki çalışmaların devamlılığının sağlanması ülkemizde Bilgisayar Mühendisliği camiası açısından önem arz etmektedir.

Bu bölümün daha sonraki kısımlarında öncelikli olarak Kuantum Bilgi Teorisinde bugüne kadar işlenmiş önemli bazı problemler aktarılmıştır. Bunlara Grover'ın arama algoritması, Shor'un çarpanlara ayırma algoritması ve Kuantum Fourier Dönüşümü örnek olarak verilebilir. Daha sonra büyük boyutlu, gerçek uygulamaların yapılması için kullanılan fiziksel altyapılardan bahsedilmiştir. Bunlar sırasıyla Optik Foton Tabanlı Kuantum Bilgisayarı, Optik Kavite Kuantum Elektrodinamikleri (QED), İyon Kapanları, Nükleer Manyetik Rezonans (NMR) ve Süperiletkenliktir. Bu bölümde son olarak Google, Nasa ve D-wave'in çalışmalarından bahsedilmiştir.

## II.3. Kuantum Bilgi Teorisi'nin Tarihsel Gelişimi

Geçmiş yüzyıllarda, dünyaya deterministik bakışta bulunulduğu için gerçek dünya problemleri sanki büyük bir saat benzeri sistemi çözmek gibi irdelenmiştir. Bilgisayarların hayatımıza yaygın bir şekilde girmesi ile beraber bilim, matematik ve topluma olan bakışımız da değişmiştir. Bilgisayarları artık sadece problem çözmek için kullanmıyoruz, bilgisayarları inşa edecek, programlayacak ve kullanacak şekilde ele alıyoruz.

Örneğin, DNA çözümlemesi, dil işleme ve bilişsel bilim gibi farklı şekillerde problemler için bilginin sıkıştırma ve hata düzeltmesi gibi kavramların optimize edilmesi için verilerin dönüştürülmesi gerekmektedir. Hesaplama verimliliği, Oyun Teorisi ve ekonomi problemleri gibi kavramların ele alınmasında önem arz etmektedir. Bilgisayar Bilimleri bu ve benzeri alanların amaçlarını da değiştirmiştir. Matematiksel araştırmalarda verimlilik konusuna daha fazla önem verilmektedir ve Bilgi Teorisi, Graf Teorisi ve İstatistik gibi bilgisayarla ilişkili alanlardaki çalışmalar hızlanmıştır. Clay Millenium problemleri arasında tanımlanan $P$ veya $NP$ problemlerin tanımlanması sorusu matematikteki en eski yapbozu açıklamaya çalışmaktadır: bir ispatın bulunmasını ne zorlaştırmaktadır?

Bilgisayarlar ilk defa ortaya çıktıklarında, çok az kişi dışında kimse bu kadar büyük bir ticari başarıya dönüşeceğini tahmin edememişti. Bu ticari başarı aynı zamanda entelektüel bir devrime de yol açtı. Örneğin, veri sıkıştırma veya hata düzeltme konularının teorik temeli oluşturan entropi kavramının icadı bu devrimle mümkün olmuştur. Bu kavram, Termodinamik ile 19. yüzyılda buhar makinelerinin anlaşılması için kullanıldı. Claude Shannon ise II. Dünya Savaşı zamanında Bell Laboratuvarları'nda kriptografi ile ilgili çalışmaları sırasında entropi kavramını pratikte kullandı. Bu durum sadece bilgisayar bilimleri problemlerinde oluşmamıştır. Örneğin Einstein yaptığı deneylerde saat senkronizasyonu kavramını kullanmıştır. Saat senkronizasyonu problemi trenlerin hareket saatlerini otomatize etmesi açısından o dönem için önemli endüstri problemlerin birisi olarak kurgulanmıştı. Gördüğümüz örneklerde bazı durumlarda bilim, teknolojik gelişmeleri takip etmiştir ve problemlere bakış açısını bu buluşlar sonrasında değiştirmiştir.

Kuantum Bilişim'in hikayesi de benzerdir. Kuantum mekaniği 20. Yüzyılın başlarında icat edilmiştir ve şu an kullanılan modern formu 1930 yılından beri bilinmektedir. Ancak kuantum mekaniğinin bilişimsel bir avantaj sağlayabileceği fikri çok daha sonra ortaya atılmıştır. Bu fikir fizikçilerin kuantum mekaniğini bilgisayarlar üzerinde simüle etmeyi denemeleri ile ortaya çıkmıştır. Bunu denediklerinde başka bir problemle karşı karşıya kalmışlardır. Tek bir sistem (fotonun polarizasyonu) iki karmaşık sayı (polarizasyonun dikey ve yatay bileşenlerinin genlik değerleri) ile betimlenebilirken $n$ adet sistem için bu sayı $2n$ yerine $2^n$ karmaşık sayı ile betimlenir ve buna ek olarak ölçüm işlemi sadece $n$ adet biti ortaya çıkarır. Fizikçiler bu sorunu aşmak için kapalı formda çözümler geliştirmişlerdir ve incelenen durumların sayılarının artması durumlarında çeşitli kestirim tekniklerine ihtiyaç duymuşlardır.

Kuantum mekaniğinin üstel (exponential) büyüklükte sistem durum uzayları, bu noktada, doğanın aslında hesaplama bilimleri açısında ne kadar büyük ve ilginç ortamlar barındırdıklarının farketmelerine yardımcı olmuştur. O güne kadar kuantum mekaniğinin açıklanmakta zorlanılan kavramları kısıtlayıcı öğeleri ve eksikleri olarak görülmüştü. Örneğin Heisenberg Belirsizlik İlkesi genelde ölçümler üzerinde bir kısıtlama olarak görülmekteydi. Dolanıklık kavramı "kuantumun temeli" veya kuantum mekaniğin felsefesi olarak nitelense de 1970 ve 80'lerde kuantum bilişim ve kuantum kriptografi kavramları icat edilene kadar işlemsel olarak çok detaylı incelenmemiştir.

Kuantum Bilişim veya diğer bir deyişle kuantum mekaniksel kavramlardan hesaplama bilimleri açısında faydalanma fikri, 1982 yılında Richard Feynman tarafından ortaya atılmıştır. Buradaki fikir bir kuantum bilgisayarı icat edilebilseydi kuantum mekaniğini klasik bilgisayarlara kıyasla çok daha etkin bir şekilde simüle edebileceği yönündedir. Bu model 1985 yılında David Deutsch tarafından formel bir hale getirilmiştir. Yine Deutsch tarafından ilk defa bir problem için (iki bitin *XOR* değerinin hesaplanması) kuantum mekaniksel bir bilgisayarın klasik bir bilgisayardan daha hızlı çalışacağı gösterilmiştir. Benzer çalışmalar zamanla hızlanmış örneğin Peter Shor tarafından 1994 yılında tamsayıların çarpanlara ayrılması probleminin polinamial zamanda yapılabildiği gösterilmiştir.

1970'li yıllarda o dönemde doktora öğrencisi olan Stephen Wiesner tarafından Heisenberg'in ölçüm ile ilgili kısıtları kullanılarak gizli mesajların öğrenilmesinin engellenebileceği öne sürülmüş ancak o dönemin önemli bilimsel dergileri bu çalışmayı reddetmişlerdir. Bu konu, 1984 yılında Charles Bennett ve Gilles Brassard tarafında ilk kez kuantum kriptografik bir yapı olarak yayınlanmıştır. Bu çalışma, 1991 yılına kadar yine kendileri tarafından gerçekleninceye kadar bilimsel camiada çok ciddiye alınmamıştır.

Bu noktada en önemli keşif 1950'li yıllarda tanımlanmış olan kuantum mekaniksel modellerin kuantum bilişim ve kuantum kriptografi ile ilgili problemlerde kullanılmasının altyapısının oluşmasıdır. Yine bilgi teorisi ile ilgili birçok problemin kuantum mekaniksel kavramlarla çok daha hızlı çözülebildiği gösterilmiştir. Örnek, Grover'ın arama algoritması, vs. [1][Nielsen2000]

Günümüzde Google, Nasa gibi kurumların ve yine dünyadaki birçok prestijli üniversite ve araştırma kurumlarının başını çektiği çalışmalar tüm hızıyla sürmektedir. Kuantum bilgisayarlarının fiziksel olarak nasıl üretileceği ile ilgili çalışmalar da son dönem çok hızlanmıştır. Bu konulardaki tarihsel gelişim ve kullanılan modellerle ilgili açıklamalar bu kısmın diğer paragraflarında detaylı bilgiler verilerek paylaşılmıştır.

## II.4. Kübitlerle Hesaplama, Shor ve Grover Algoritmaları, Kuantum Algoritmalar

Kuantum mekaniğine bilişsel bir bakışla yaklaşma fikri aslında kübit (kübit) dediğimiz yapılarla ortaya atılmıştır. $d$ sayıda bileşenden oluşan ve diğerlerinden mükemmel olarak ayırt edilebilen bir kuantum sistem $\mathbb{C}^d$ uzayının birim vektörleri ile açıklanabilir. En basit ve ilginç durum *n=2* için oluşmaktadır ve bu sistemler kübit olarak isimlendirilir.

$x = \binom{x_0}{x_1}$ sistemi ölçüldüğünde $|x_0|^2$ olasılıkla *0* ve $|x_1|^2$ olasılıkla *1* sonucunu vermektedir. Bu gösterimle lineer cebirsel kavramlar kullanılarak ve sistemin normu korunarak her türlü hesaplama işlemi yapılabilmektedir. Diğer bir deyişle $x$ sistemi $U$ bir üniter (unitary) matris olmak üzere $Ux$'e map edildiğinde, her zaman uzunluğu korunmaktadır. Matematiksel olarak bu denklem $U^{\dagger}U = I$ denklemine eşittir, burada $(U^{\dagger})_{ij} = \overline{U}_{ji}$ . Bu açıklamanın en güzel tarafı altında yer alan sistemden tamamen bağımsız olmasıdır. Bu sistem fotonun polarizasyonu, elektronun enerji seviyesi, çekirdeğin spin değeri veya süper iletken akımdaki çevrimin doğrultusu olabilir. Bu bakış açısıyla aslında, kübit kuantum bilginin araç-bağımsız (device-independent) gösterim yolu olmaktadır. Bize RAM, sabit disk sürücüsü veya bir abaküste kodlanmış olan bitlerin sağlamış oldukları altyapıya benzer olarak çalışma ortamı sağlamaktadırlar.

Tek kübitlik sistemler fiziksel deneyler açısından anlamlıdır ancak biz hesaplama ile ilgilendiğimiz için $n$ kübitlik sistemlerdeki olgularla ilgilenmekteyiz. Bu durumda $x$ sistem durumu $\mathbb{C}^{2^n}$ kümesindeki bir birim vektör olmaktadır ve bu vektörün elemanları $n$-bitlik karakter katarlarında oluşur. $n$-kübit sistem durumlarının çoğu dolanıktır, bu da şu anlama gelmektedir: genlikleri bir şekilde $n$ bit üzerinde birbiriyle ilişkilidir. Buradaki dinamikler üniter matrisler tarafından tanımlanırlar ve bu yapılar iki kübitlik kapıların serileri şeklinde tanımlanmaktadırlar. Bölüm 4'te kuantum kapılar daha detaylı olarak açıklanmıştır. Örneğin 3 ve 7. İndisleri üzerinde işlem yapan bir kuantum kapı varsayalım. Burada üniter matrisimizin elemanları $U_{ij}$ sıfırdan farklıdır bu durumda $n$-bitlik karakter katarları için $i$ ve $j$ 3 ve 7. İndisler hariç her yerde birbirine eşittir. Bu durumda $U_{ij}$'nin değeri sadece dört bite dayalıdır. Bunlar $i_3$, $j_3$, $i_7$ ve $j_7$ olmaktadır. Bu lineer cebirsel gösterim biraz soyut gelebilir ancak bu gösterim klasik deterministik ve rassal hesaplamalarda da kullanılmaktadır. $n$-bitlik bir karakter katarı $2^n$ uzunluğunda ve sadece bir elemanı $1$ ve diğer elemanları $0$ olan bir vektör ile tanımlanabilmektedir. Buradaki dinamikler; $M$, $0$ ve $1$'lerden oluşan bir matris her kolonunda sadece bir tane $1$ elemanı olan matrisi ifade ederken; $x$'in $Mx$'e map edilmesi ile açıklanır. Burada yine tek bir işlem iki biti ifade eder ve yine $i$ ve $j$'nin değerlerine bağlı olarak değiştiği bir matris ile tanımlanır.  Bu ifade şu anlama gelmektedir: eğer bir bit üzerinde bir işlem yapmıyorsak, onun olan bitenden etkilenmemesi ve değişmemesi gerekmektedir.

Rassal ve kuantum hesaplama arasındaki en temel farklar gerçel sayılardan karmaşık sayılara olan değişimdir. Bu değişim ile beraber sistem durumunun $l_1$ normundan $l_2$ normuna geçiş yapılırken; kuantum sistem durumları için genliklerinin mutlak değerlerinin karelerinin toplamı $1$'e eşittir. Burada olasılıkların toplamı karelerini almadan $1$'e eşittir. Buradaki bir hesaplamanın her bir dalı için farklı fazların genliklerinin birleştirilmesi toplanması (add up) (constructive interference-yapıcı girişim) veya birbirlerini iptal etmesi (cancel) (destructive interference-yokedici girişim) anlamına gelebilmektedir. Eğer *"0"*, $\begin{pmatrix}1\\0\end{pmatrix}$ vektörü tarafından ve *"1"* de $\begin{pmatrix}0\\1\end{pmatrix}$ vektörü tarafından gösterilir ise *NOT* işlemi $\begin{pmatrix}0 & 1\\1 & 0\end{pmatrix}$ matrisi ile ifade edilebilir. Geometrik olarak bu *π/2* kadar rotasyon anlamına gelmektedir. Ancak, sadece bir kuantum bilgisayar tarafından *"NOT'ın karekökü"* kapısı işlemi gerçeklenebilmektedir. Öyle ki bu işlem *π/4* kadar rotasyon anlamına gelmektedir. $\sqrt{NOT} = \begin{pmatrix}1/\sqrt{2} & -1/\sqrt{2}\\1/\sqrt{2} & 1/\sqrt{2}\end{pmatrix}$ olarak gösterilir.

Eğer *"0"* sistem durumuna $\sqrt{NOT}$ işlemini uygularsak $\begin{pmatrix}\frac{1}{\sqrt{2}}\\\frac{1}{\sqrt{2}}\end{pmatrix}$ sistem durumunu elde etmiş oluruz. Bu durumda eğer ölçüm yaparsak, *0* ve *1* sonuçları $\frac{1}{2}$ olasılıkla elde edilmektedir. Ancak, eğer $\sqrt{NOT}$ işlemini ölçüm yapmadan önce ikinci bir defa uygularsak sonuç olarak her zaman *1*'i elde etmiş oluruz. Bu sonuç bize kuantum süperpozisyonlar ve rassal karışımların arasında anahtar bir fark olduğu çıkarımını sağlar. Bir sistem durumunu süperpozisyona maruz bırakma işlemi, herhangi bir geri dönüştürülemez (irreversible) bilgi kaybına uğramadan da yapılabilmektedir. *n* kübitimizin olması durumunda süperpozisyon ve girişim bize daha büyük hesapsal avantajlar sağlamaktadır.

Bu duruma en uygun örnek Grover'ın algoritmasıdır. Algoritma, *n* bit üzerinde bir ikili fonksiyon olan *f*'nin $x \in \{0,1\}^n$ ve $f(x) = 1$ koşulunu sağlayan giriş değerini $2^{n/2}$ karşılaştırma ile bulmaktadır. Bu da diğer muadil algoritmalara göre karekök mertebesinde avantaj sağlamaktadır. Grover algoritması bu işlemi yaparken kullandığı olasılık değerleri genliklerin kareköklerididir. Böylece, $2^n$ sistem durumunun düzgün (uniform) süperpozisyonu için her bir bileşene $1/\sqrt{2^n}$ genlik değeri atanır. Ayrıca, *f*'nin değerlerinin hesaplanabilmesi için her *x* hedefinin de yaklaşık $1/\sqrt{2^n}$ ile çarpılması gerekmektedir. Bu genlikler çok büyük olduğunda yapılan bir düzeltmedir. Bu durumda, hesaplama için harcanan toplam efor $2^{n/2}$ mertebesindedir, veya genellikle *M* çözüm olan durumlarda $\sqrt{2^n/M}$ olarak bulunur.

Benzer şekilde klasik algoritmalara kıyasla dramatik bir hızda artış sağlayan bir diğer algoritma Shor'un çarpanlara ayırma algoritmasıdır. Shor'un algoritmasının içinde klasik bir parça yer almaktadır ve bu kısım çarpanlara ayırma problemini daha soyut bir problem olan periyot bulma problemine dönüştürmektedir. Bu problem girdi olarak $\{0, 1, \dots, 2^n - 1\}$ tam sayıları için $f(x) = f(y)$ ancak ve ancak $x - y$, *r* ile belli bir saklı periyotta (*r* ile) bölünebilmektedir. Amacımız *r* değerini bulmaktır. *n* sayısı büyüdükçe *r* değeri de üstel olarak büyük olabilmektedir, klasik bilgisayarlar bu değeri hesaplamak için üstel zamana (exponential time) ihtiyaç duyarlar ve *f* fonksiyonunu bir kapalı kara kutu gibi ele alırlar.

Bir kuantum bilgisayarı için, sistem durumlarının süperpozisyonunu yaklaşık $\sqrt{r/2^n}$ genliği ve her bir terimi $z, z+r, z+2r, \dots$ olacak şekilde ($z \in \{0, 1, \dots, r-1\}$ içinde rastgele seçilmiş olarak) üretmesi mümkündür. Buraya kadarki kısım aslında olasılıksal bir bilgisayarın rastgele bir *x* ve ondan hesaplanmış $f(x)$ değeri ile yaptığı işlemlere benzerdir. Bir sonraki adım sadece kuantum olarak bir kuantum Fourier dönüşümü (quantum Fourier transform-QFT) olarak adlandırılan üniter bir matrisin uygulanmasıdır. Bu matrisin *y*, *z* girişleri şu değere eşittir: $e^{2\pi i yz/2^n}/\sqrt{2^n}$. Bu işlemin etkin bir şekilde gerçekleştirilebilmesi klasik hızlı Fourier dönüşümünün kuantum uyarlaması ile mümkündür, böylece klasik FFT'nin yaptığı işlem olan bir sayı listesinin dönüştürülmesi işlemi yerine bir kuantum sistem durumunun genliklerini dönüştürür. Bu durumda QFT'nin uygulanması bizim süperpozisyonumuzu bir sistem durumuna $y = \frac{r}{2^n}(1 + e^{\pi i \frac{yr}{2^n}} + e^{\pi i \frac{2yr}{2^n}} + e^{\pi i \frac{3yr}{2^n}} + \cdots)$ değeri ile map eder. Eğer $yr$ değeri $2^n$ değerine bölünebilmekten çok uzaksa bu durumda bu toplam çok büyük olacaktır. Eğer $yr$ $2^n$ değerine bölünebilmekten çok uzaksa değişik fazlara sahip birçok karmaşık sayıyı içermektedir ve işlem dışı tutulur (cancel out) Böylece, *y*'nin ölçülmesi ile elde edilecek değer $2^n/r$ nin bir katına yakın bir değer olacaktır. Son olarak, klasik bir sürekli kesir üleştirmesi (expansion) *r*'yi yeniden elde etmemizi sağlar.

Kuantum simülasyonun pratik olarak bu kadar önemli hale gelmesinin sebebi kuantum saldırılara karşı dayanıklı yeni açık-anahtar altyapılarının oluşturulması ile daha iyi anlaşılacaktır. Shor algoritmasının en ilginç yanlarından birisi de algoritmanın çözüm yöntemi sırasında kuantum mekaniğinden tamamen bağımsız bir şekilde sonuçları elde etmesidir. Shor algoritmasının ortaya çıkması ile kuantum mekaniğin klasik bir bilgisayarda tam olarak simüle edilemeyeceği düşünülmeye başlanmıştır. Bu durum *NP-tamam (NP-complete)* problemlerin çözümündeki gelişmelere benzer, çeşitli alanlardaki NP-tamam problemlerin çözümü için benzer yaklaşımlarla çalışmalar yapılmaktadır.

Shor ve Grover algoritmaları Kuantum Bilgi Teorisi'ndeki en tanınmış algoritmalar olsalar da bu alan sadece bu iki çalışmadan ibaret değildir. Daha yakın zamanlarda keşfedilmiş bir kuantum algoritma ile [43] [HarrowPRL2009] büyük lineer denklem sistemlerini verilen bir *A* matrisi ve *b* vektörü için $Ax = b$ eşitliğini sağlayan *x* değerlerinin bulunması klasik eşleniğine göre daha hızlı çözmektedir. Klasik eşleniğine göre kuantum olan algoritmada girdi değerleri $x$ ve $b$ kuantum sistem durumlarıdır ($n$ kübitlik sistemler için $2^n$-boyutlu vektörler). Ayrıca, $A$ sparse bir matris ise ve belli bir $i$ indisi için sıfırdan farklı $A_{ij}$ giriş değerlerinin belli bir şekilde tutulduğu varsayımıyla hareket edersek üstel olarak büyük sistemlerin denklemlerini polinomial zamanda çözmek mümkündür. Burada önemli bir diğer nokta ise şudur: algoritmanın çalışma zamanı $A$'daki koşul sayısı ile ölçeklenmektedir. Bu parametre klasik bir şekilde denklem sistemlerinin çözümünün nümerik istikrarsızlığının bir ölçütüdür.

Yine yakın dönemlerde geliştirilmiş bir diğer algoritma Metropolis örnekleme (sampling) algoritmasının kuantum analoğudur [44] [TemmeNat2011]. Klasik olarak Metropolis yöntemi analizi zor dağılımların üstel olarak büyük sistem durum uzayları üzerinde örnekleme için kullanılan bir yöntemdir. Bu yöntemin uygulanabileceği alanlar çok geniş bir dağılım sergiler, örneğin istatistiksel sonuç çıkarma (inference) ve yaklaşım algoritmaları gibi. Kuantum Metropolis algoritması, termal dağılımlar için kuantum sistem durumlarını üretir. Klasik versiyonu gibi bu algoritmada düşük-sıcaklıklı sistem durumlarının üretilmesi daha uzun sürmektedir. Bu durum daha zor optimizasyon problemlerini çözerken daha kesin sınırların daha zor ancak her şartta bulunmasını sağlamaktadır. Formel bir ispat olmadan, birçok durum için klasik Metropolis algoritmasının da iyi sonuç verdiğini empirik olarak tespit edilebilir. Kuantum Metropolis algoritmasının empirik olarak sonuçlarının analiz edilmesi geniş ölçekli kuantum bilgisayarın icadı sonrası yapılabilecektir. Ancak, klasik Metropolis algoritmasını çalıştıran yapının bir altrutini olarak kuantum Metropolis algoritması konumlandırılabilirse teoride gösterilen sonuçlar pratikte de gerçeklenebilir.

Kuantum Bilgisayarları için yeni kullanım alanları oldukça çok çalışılan bilimsel konular arasında yer almaktadır. *10* veya daha fazla kübit üzerinde çalışan kuantum bilgisayarlar için birçok yeni çalışmanın klasik versiyonlarından daha hızlı çalıştıkları gösterilmiştir. Bu çeşit kuantum hesaplama cihazları genelde kuantum ölçümün duyarlılığı gibi (örneğin atom saati veya yerçekimi dalgaları gibi) alanlarda kullanılmaktadırlar. "Kuantum tekrarlayıcı (repeater)" yapıları ise kuantum anahtar dağıtımı protokollerinin gerçeklenmesinde dolanıklığın aktarımı gibi konularda kullanılırlar. Klasik bilgisayarların kullanım alanları nasıl Turing makinasının attığı temelden çok daha ileri gitmişlerse, kuantum bilgisayarlarının da birçok farklı ve potansiyele sahip alanda aktif olarak kullanılmaları kaçınılmazdır.

## II.5. Kuantum Bilgisayarları Fiziksel Olarak Nasıl Üretilebilir?

Kuantum devrelerin, algoritmaların ve iletişim sistemlerinin deneysel olarak gerçeklenmesi çoğu zaman çok zorlu prosedürlerdir. Bu bölümde optik foton tabanlı kuantum bilgisayarının deneysel olarak nasıl gerçeklendiği anlatılacaktır. Bir kuantum bilgisayarının üretilebilmesinin deneysel gereksinimleri nelerdir? Daha önceki bölümlerde de açıklandığı gibi teorinin temel birimleri kuantum bitlerdir (kübit) ve bu sistemler yine daha önce açıklandığı gibi iki seviyeli kuantum sistemlerdir. Bir kuantum bilgisayarı gerçeklemek için sadece kübitleri sağlam fiziksel gösterimlerle vermemeliyiz aynı zamanda öyle bir sistem seçmeliyiz ki istenen şekilde değişimleri sergileyebilmelidirler. Bu durumda özetle, belli ilk sistem durumları için kübitleri hazırlayabilmeliyiz ve sonuç çıktı sistem durumlarını da ölçerek elde edebilmeliyiz.

Deneysel gerçeklemenin zorluklarından birisi bu temel gereksinimlerin sadece bazılarının karşılanabilmesidir. Bir kuantum bilgisayarı kuantum özelliklerini koruyabilmek için çok iyi bir şekilde izole edilmelidir ve aynı zamanda kübitlerine kolaylıkla erişilebilmeli ve ayrıca üzerinde işlem yapılabilmesi için kolay manipüle edilebilmelidir ve çıkış değerleri kolayca okunabilmelidir. Bu durumda başlangıçta sorduğumuz soru bir kuantum bilgisayarının *nasıl* üretilebileceğinden bir kuantum bilgisayarının *ne kadar iyi* inşa edilebileceğine evrilmektedir.

Hangi fiziksel sistemler kuantum hesaplama işlemleri için iyi adaylardır? Bu noktada en çok dikkat edilmesi gereken kavram kuantum gürültü (kuantum noise) veya diğer adıyla uyum bozulması (decoherence) kavramlarıdır. Tezimizin daha sonraki bölümleri uyum bozulması altında Kuantum Fisher Bilgisi ve Dolanık Ölçütlerindeki değişimler de incelenmiş ve alandaki yakın zamanda yapılan bazı çalışmalar anlatılmıştır. Uyum bozulması yüksek olan sistemlerin toplamda stabil bir şekilde yapabilecekleri işlem sayısı azalmakta ve bu da hesaplama açısından sistemin kullanılabilirliğini azaltmaktadır.

Kuantum hesaplama için dört temel gereksinimi şu şekilde sıralayabiliriz [1]:

1-Kuantum Bilgiyi tutarlı (robust) bir şekilde ifade edebilme

2- Genel geçer bir üniter dönüşüm ailesi gerçekleyebilme

3- Uygun giriş sistem durumlarını hazırlayabilme

4- Çıkış sistem durumlarını ölçebilme

Yakın tarihte ve günümüzde yapılmış olan kuantum bilgisayarları ile ilgili fiziksel deney çalışmalarında bu dört temel gereksinimin cevaplanması ile ilgili çalışmalara odaklanılmıştır. Takip birkaç alt bölümde günümüzde kullanılan fiziksel gerçeklemeler özet olarak açıklanmaya çalışılmıştır.

## II.6. Optik Foton Tabanlı Kuantum Bilgisayarı

Bir kuantum biti ifade edebilmek için dikkat çeken fiziksel sistemlerden birisi optik fotonik sistemdir. Fotonlar yüksüz parçacıklardır ve diğer fotonlarla veya başka parçacıklarla çok kuvvetli bir şekilde etkileşim içine girmezler. Optik fiberler kullanıldığında çok uzun mesafeler için çok düşük miktarda kayıp ile yol alabilirler ve beamsplitter'lar aracılığıyla kolayca birleştirilebilirler. Fotonların kuantum özellik taşıdıklarını çift yarıklı girişim deneyi ile kolayca gösterilebilir. Temel olarak, fotonların birbiri ile lineer olmayan optik araçlarla ortalama etkileşimlere girmeleri sağlanabilmektedir. Bu anlattığımız her ne kadar mükemmel durumu anlatsa da bu bileşenlerin çalışılması ile sistem mimarisi, optik tabanlı kuantum işlemcinin kısıtları gibi konular çok kapsamlı bir şekilde anlaşılabilmektedir.

- ***Kübit gösterimi:*** Tek fotonunun yeri iki mod arasında olabilmektedir $|01\rangle$ ve $|10\rangle$ veya polarizayon
- ***Üniter dönüşüm:*** Rastgele dönüşümler faz dönüştürücüler, beam splitter'lar ve lineer olmayan Kerr ortamları sayesinde yapılabilmektedir. Bu durumda iki tekli foton faz modülasyonundan $\exp[i\chi L|11\rangle\langle 11|]$ dönüşümünü yaparak geçebilirler.
- ***Giriş sistem durumunun hazırlanması:*** Tek fotonluk sistem durumlarının lazer ışığının kırılması (attenuating) ile oluşturulması
- ***Sonucun elde edilmesi:*** tek fotonun örneğin fotoçoklayıcı tüp (photomultiplier tube) ile elde edilmesi
- ***Zorluklar:*** Yüksek oranlı çapraz faz modülasyonu gücüne sahip ve kaybı emen lineer olmayan Kerr ortamının gerçeklenmesinin zor oluşu

## II.7. Optik Kavite Kuantum Elektrodinamikleri (QED)

Kavite Kuantum Elektrodinamikleri (Cavity Quantum Electrodynamics-QED) tekil atomların (single atom) sadece birkaç optik moda eşlenmesini (coupling) sağlayan çalışma alanı olarak adlandırılır. Deneysel olarak, tekil atomların çok yüksek *kavite kalite etmenine* (cavity quality factor- Q) sahip olan optik kavitelerin içine yerleştirilmesi mümkündür, çünkü kavite içerisinde bir veya iki elektromanyetik mod bulunmaktadır ve bunların herbirisi çok yüksek elektik alan gücüne sahiptirler, böylece atom ile alan arasındaki dipol eşlenmesi çok yüksektir. Yüksek *Q* nedeniyle, kavite içerisindeki fotonların atomdan kurtulmadan önce atomla etkileşime geçme şansları vardır. Teorik olarak bu yöntem tekil kuantum sistemleri kontrol etmemizi ve üzerinde çalışma yapmamızı sağlamaktadır ve kuantum koas, kuantum geridönüş (feedback) kontrolü ve kuantum hesaplama gibi alanlarda birçok fırsat oluşturmaktadır.

Özel olarak, bu yöntem bir önceki altkısımda açıklanan optik kuantum bilgisayarına göre bazı konularda daha avantajlıdır. Tekil fotonlar kuantum bilginin taşınması için iyi taşıyıcıdırlar ama birbirleriyle etkileşime geçebilmek için başka ortamlara ihtiyaç duymaktadırlar. Diğer taraftan iyi izole edilmiş tekil atomlar her zaman uyum bozulmasına maruz kalmazlar ve ayrıca fotonlar arasında çapraz faz modülasyonu (cross phase modulation) sağlarlar. Tekil fotonun sistem durumu tekil atoma aktarılabilir ve tekil atomdan alınabilir mi sorusu bu alandaki ana çalışma konusudur.

- ***Kübit gösterimi:*** Tek fotonunun yeri iki mod arasında olabilmektedir $|01\rangle$ ve $|10\rangle$ veya polarizasyon
- ***Üniter dönüşüm:*** Rastgele dönüşümler faz dönüştürücüler, beam splitter'lar ve kavite QED sistemi (cavity QED) sayesinde yapılabilmektedir. Bu sistem birkaç atome içeren bir Fabry-Perot kavitesi içermektedir ve buna bir optik alan eşleniktir.
- ***Giriş sistem durumunun hazırlanması:*** Tek fotonluk sistem durumlarının lazer ışığının kırılması (attenuating) ile oluşturulması
- ***Sonucun elde edilmesi:*** tek fotonun örneğin fotoçoklayıcı tüp (photomultiplier tube) ile elde edilmesi
- ***Zorluklar:*** İki foton eşlenmesi bir atom üzerinde yapılmaktadır, bu durumda atom-alan eşlenmesinin (atom-field coupling) artması beklenir. Fakat bu durumda fotonun kavitenin içine ve dışına eşlenmesi zorlaşmaktadır ve basamaklanabilmeyi sınırlamaktadır.

## II.8. İyon Kapanları

Bu kısımda kübitlerin fotonlarla gösteriminden farklı olarak diğer bir ortam olan elektron ve nükleer spin üzerinde gösterimi konusuna bakacağız. Spin biraz garip ama gerçek bir kavramdır. Farklı spin durumları arasındaki enerji farkı diğer enerji ölçeklerine göre çok küçüktür. Bir atomun spin durumlarının gözlemlenmesi zordur, kontrol edilmesi ise çok daha zordur. Önceden oluşturulmuş ve kontrollü ortamlarda az sayıda yüklü atomun izole edilmesi ve kapana kıstırılması (trap) mümkündür. Bu ortamların sıcaklığı son derece düşük olmalıdır ki atomun kinetik enerjisi spin enerji katkısından daha küçük olsun. Bu koşullar sağlandıktan sonra sıradan tek renkli ışık (incident monochromatic light) spin durumlarında önceden seçilmiş dönüşümler sağlar. Bu teknik iyon kapanları dediğimiz yapıların kuantum hesaplamada nasıl kullanıldığının temelini oluşturur. Örneğin kapana kısılmış *Berilyum* atomları kullanılarak *kontrollü-NOT* kapısını gerçeklemek mümkün olmaktadır.

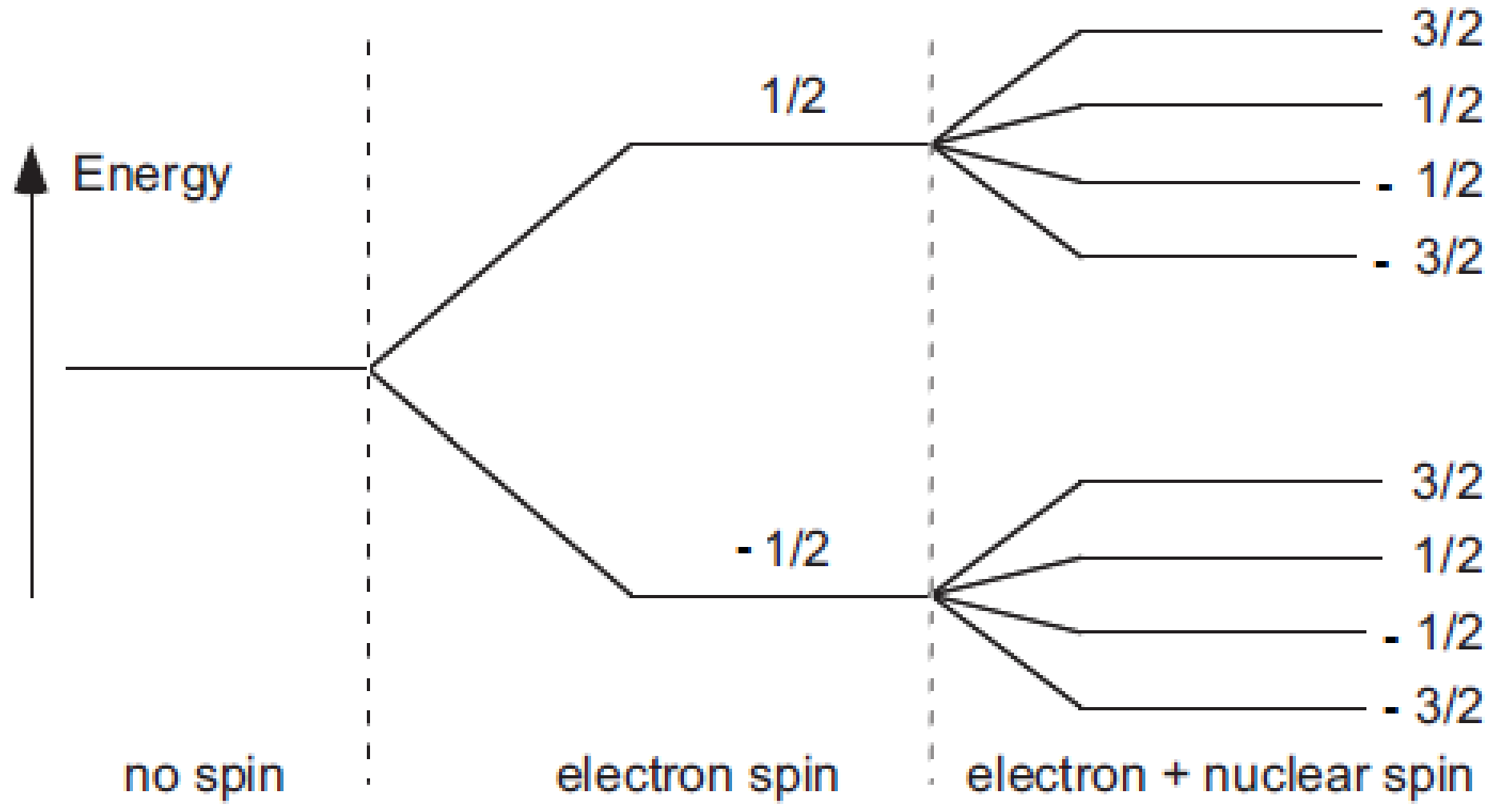

Şekil 2.1 – Spin Enerji Seviyelerinin Genel Gösterimi [1]

- ***Kübit gösterimi:*** Bir atomun *hiperiyi-nükleer spin* (hyperfine-nuclear spin) sistem durumu ve kapana kısılmış atomların *seviye titreşimsel modları-fononları* (level vibrational modes)
- ***Üniter dönüşüm:*** Rastgele dönüşümler atom durumunu Jaynes-Cummings etkileşimi ile dışarıdan manipüle eden lazer atımları (pulse) uygulamaları ile oluşturulur. Kübitler paylaşımlı fonun sistem durumları ile etkileşirler.
- ***Giriş sistem durumunun hazırlanması:*** Atomların, hareketsel yer sistem durumuna (motional ground state) ve hiperiyi yer sistem durumuna, kapan ve optik pompalama ile soğutulması
- ***Sonucun elde edilmesi:*** hiperiyi sistem durumlarının popülasyonunun ölçülmesi
- ***Zorluklar:*** Fononları stabil kalma süreleri çok kısadır ve iyonların hareketsel yer sistem durumlarının hazırlanması zordur.

## II.9. Nükleer Manyetik Rezonans (NMR)

Kavramsal olarak, tek bir molekül tek başına iyi bir kuantum bilgisayarı olabilir. Fakat bu önermemiz bir moleküller bütünü söz konusu olduğunda geçerli midir? Özel olarak, *NMR* tipik olarak bütün moleküllerden gelen sinyallerin *ortalamasının* bir ölçümüdür. Bu değer, bir kuantum bilgisayar kümesinin ortalama çıktısı olarak anlamlı bir çıktı olabilir mi? İkinci olarak, *NMR* genel olarak oda sıcaklığında denge durumunda olan fiziksel sistemlere uygulanırlar. Bu şu anlama gelmektedir ki spinleri ilk durumları neredeyse tamamıyla rassaldır. Geleneksel kuantum hesaplama sistemin saf bir sistem durumunda hazırlanmasını beklemektedir; ancak çok yüksek entropiye sahip karışık sistem durumlarında kuantum hesaplama nasıl gerçekleştirilebilir?

Yukarıda bahsedilen problemler için NMR çözümü kuantum hesaplamanın gerçeklenebilmesi için bu tür sistemlerin termal doğasından kaynaklı kısıtlamalara rağmen oldukça dikkat çeken ve umut vadeden bir yöntem olmuştur. NMR'dan birçok ders çıkarılabilir: örneğin, gerçek sistemlerdeki rastgele üniter dönüşümlerin gerçeklenmesi konusunda, sistematik hataların ve uyum bozulmasının karakterize edilebilmesi ve engellenmesi konularında, bütün sistemlerin üzerinde hesaplama ile ilgili bileşenlerin toplanırken tam kuantum algoritmaların gerçeklenmesi konusunda oluşacak durumların tespitinde vb.

- ***Kübit gösterimi:*** Bir atom çekirdeğinin spini
- ***Üniter dönüşüm:*** Rastgele dönüşümler güçlü bir manyetik alandaki spinlere uygulanan manyetik alan atımları (pulse) tarafından oluşturulur.
- ***Giriş sistem durumunun hazırlanması:*** Spinleri güçlü bir manyetik alana konularak polarize edilmesi ve daha sonra 'etkili saf sistem durumunun' (effective pure state) hazırlanma tekniklerinde kullanılması
- ***Sonucun elde edilmesi:*** manyetik momentin ölçülerek indüke edilmiş gerilim sinyalinin ölçülmesi
- ***Zorluklar:*** İlk başlangıçtaki polarizasyon yeterince yüksek değilse, etkili saf sistem durumunun hazırlanması şekli sinyalin kübit sayısında üstel olarak azalmasına neden olur.

## II.10. Süperiletkenlik

Süperiletken tabanlı kuantum hesaplama gerçeklemesi oldukça umut vadeden bir yöntemdir. Nanoteknolojik şekilde üretilen süperilerken elektrotları Josephson bağıntılarına (junction) göre eşleniktir (coupled). Süperiletken bir elektrotun içinde faz ve yük eşlenik değişkenlerdir. Yüke bağlı olan, faza bağlı olan veya ikisine de bağlı olmayan olmak üzere, üç çeşit süperiletken kübit tipi bulunmaktadır ve bunlar iyi birer kuantum sayıdır. Bunlar sırasıyla terim olarak yük kübiti, akı (flux) kübiti ve hibrit kübit olarak isimlendirilirler.

Daha önceki kısımlarda anlatılan ve genelde kübitin diğer iki seviyeli olan fiziksel gerçeklemelerinden farklı olarak süperiletken kübitleri üreten kuanrum devreler çok seviyeli sistemlerdir ve istenirse ilk iki seviyesi hesapsal temel (basis) olarak kullanılabilir. Bu gerçeklemenin temel bir gereksimi olarak enerji seviyelerinin düzgün (uniform) bir şekilde yerleşmediğini söyleyebiliriz, böylece 0 ile 1 seviyeleri arasındaki değişimleri sağlayan belli frekanslardaki fotonlar, birinci seviyeden daha yüksek seviyelere olan değişimlere sebep olmazlar. Devrenin bir kısmından diğer bir kısmına elektronik sinyalleri taşıyan sinyallerin enerji kaybına ve uyum bozulmasına uğramaması için sistemin metal parçalarının direncinin sıfır olması gerekmektedir. Bu durumdaki devreler çok düşük sıcaklıklarda işletilmelidir: bu durumda süperiletkenlik gerçekleşir ve termal oynamalar (fluctuations) enerji seviyelerinde atlamalara (transition) neden olmazlar.

## II.11. Google, NASA ve D-Wave'in Çalışmaları

Google ve Nasa ortaklaşa bir Kuantum Yapay Zeka laboratuvarı kurarak genel amaçlı ve büyük ölçütlü (large-scale) Kuantum Bilgisayarlarının üretilmesi konusunda çalışmalar yürütmektedirler. Bu laboratuvarda sadece teorik çalışmalar değil, kuantum algoritmalar, olası yeni bilgisayar mimarisi vb. konularda da çalışılmaktadır. Bu laboratuvar Kanadalı D-Wave şirketinin ürettiği 512 kübitlik kuantum işlemciyi içeren D-Wave Two isimli bilgisayarı da kullanmaktadır. D-Wave'in iddiası ilk genel amaçlı kuantum işlemciyi ürettikleri yönündedir. Özellikle Graf Teorisi ile ilgili bazı problemlerde bu işlemcinin klasik türevlerinden 10 kata kadar daha hızlı işlem yapabildiğini gösterdiklerini iddia etmektedirler. D-Wave Two cihazının olası diğer algoritmalar için klasik çözümlerden daha iyi olup olmadığının tespit edilmesi amacıyla Google ve Nasa bir Ziyaretçi Test süreci başlatmışlardır. Bu sürece katılan ve D-Wave'in ürünlerin test eden bazı araştırmacıların sonuçları Nature, Science gibi saygın bilimsel dergilerde yayınlanmıştır. Elde edilen ilk sonuçlara göre D-Wave sistemi klasik türevlerinden kuantumluk açısından daha hızlı sonuç vermeyebileceği gözlemlenmiştir. Ancak bu konuda henüz araştırmalar sona ermediğinden D-Wave'in gerçekten ilk genel amaçlı Kuantum İşlemcisinin başarısı ile ilgili tartışmalar sonuçlanmamıştır.

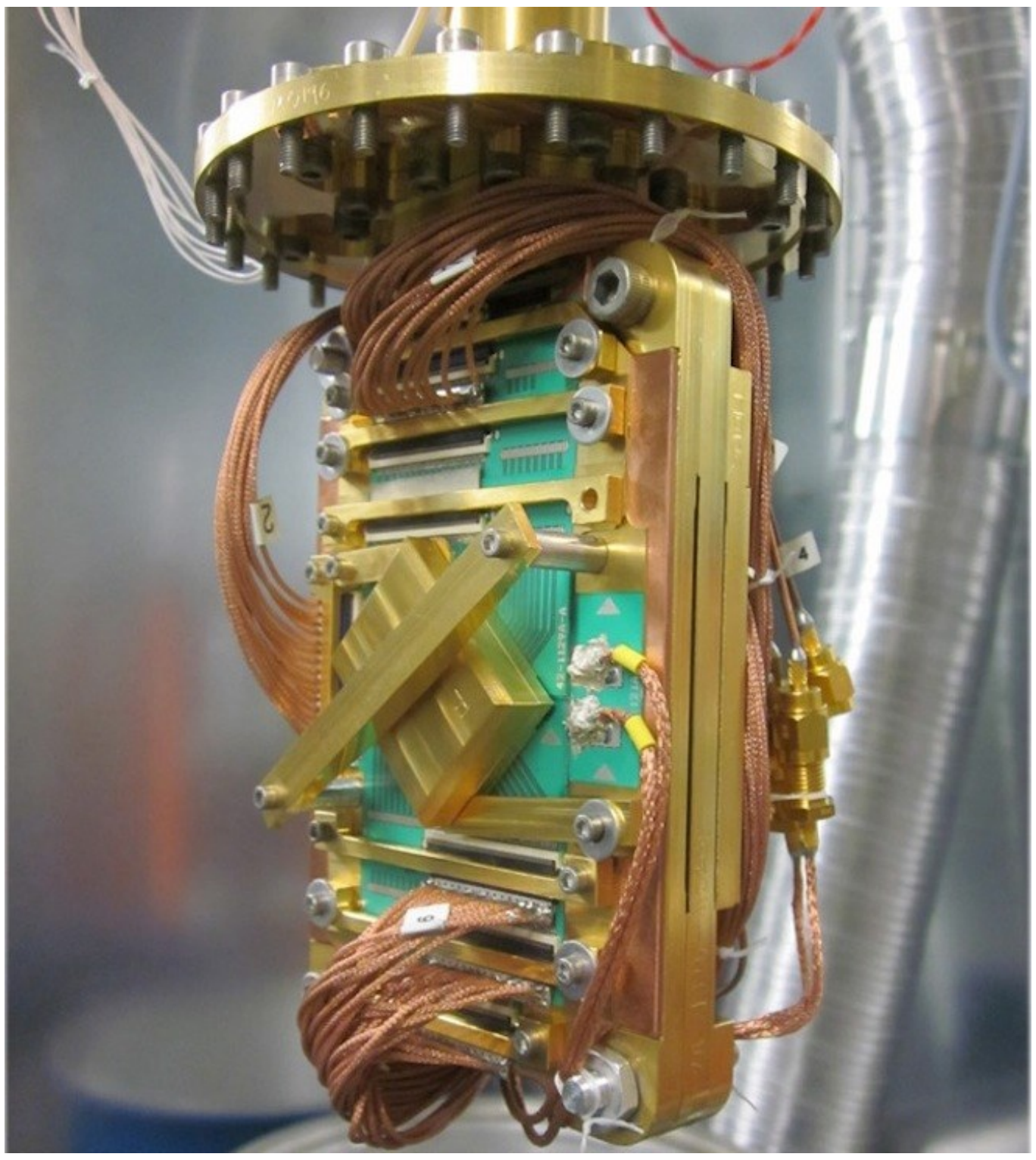

Şekil 2.2 – D-Wave'in 512 Kübitlik Kuantum İşlemcisi

Google bu konudaki çalışmalarını hızlandırmış ve ilk genel amaçlı büyük kuantum bilgisayarının üretilmesi amacıyla dünyaca ünlü araştırmacı Prof.Martinis'i kadrosuna dahil etmiştir. Martinis'in grubu bu konudaki çalışmalarına devam etmektedir. Diğer yandan D-Wave'in bu konudaki çalışmaları da devam etmektedir. Son birkaç yılda bu konu üzerine çok sayıda bilimsel çalışma yine benzer dergilerde yayınlanmıştır. Özellikle, Kuantum Anahtar Dağıtımı (Quantum Key Distribution-QKD) konusundaki çalışmalar pratik alana çok daha yakındır ve özellikle askeri Ar-Ge projelerinde aktif olarak kullanılmaktadır. Örneğin Fransız ordusu ID-Quantique isimli şirketin geliştirdiği cihazları kullanmaktadır. Ayrıca Tokyo'da çok büyük bir QKD ağı NICT tarafından işletilmektedir. Japon araştırma kurumlarının çoğunun bu konulardaki çalışmaları devam etmektedir. Özellikler NICT'nin koordine ettiği Ar-Ge projelerinde Kuantum Tekrarlayıcı (repeater) üretilmiştir. Kuantum Hafızanın üretimi konusunda ciddi yol katedilmiştir. Bazı özel durumlar için kuantum algoritmaları gerçekleyen bilgisayarlar yapılmıştır. Görüldüğü üzere gerçek kuantum bilgisayarlarının üretilmesi konusundaki çalışmalar tüm hızıyla devam etmektedir.

# III. KUANTUM BİLGİ TEORİSİ VE DOLANIKLIK

20. yüzyılda başlayan Kuantum Bilgi Teorisi yaklaşımı, bazı fiziksel olayların klasik fizik anlayışı ile çözümlenememesi sorunsalına çözüm arayarak gelişti. Özellikle atom ve radyasyona bakarken klasik kuralların pek de uygun olmadığı anlaşıldı. Kuantum mekaniği bu aşamada bilimin her aşamasında vardı, atomun yapısı, nükleer füzyon, süperi iletkenler, DNA yapısı ve doğanın her bir basit parçasını içeriyordu. Kuantum fiziğinin ortaya çıkış sebebi klasik fizikle açıklanamayan olaylardır. Bunlara Compton Saçılması (Işığın Elektronla etkileşip saçılmaları), Fotoelektrik olay, Siyah cisim ışıması vs. örnek olarak verilebilir.

Kuantum mekaniğinin gelişimine bağlı olarak kuantum bilgi teorisi de bilim adamlarının yeni hedefi olmuştu ve 1980 yılından itibaren "*kopyalanamama teorisi*", 1970 yılından itibaren "*tekil kuantum sistemlerin tamamen kontrolü*" olarak ilk adımlar atıldı.

## III.1. Kuantum Bitler "Kübitler"

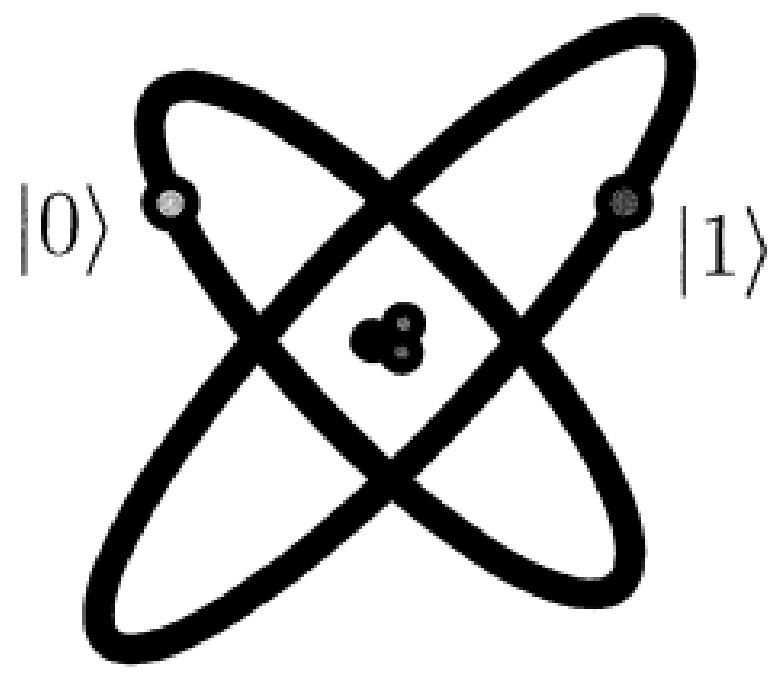

Şekil 3.1-Elektronun Enerji Seviyesine Göre Kodlanmış Bir Kübit Bilgi

Klasik bilgisayarda kullandığımız bitler bize olabilecek en küçük bilgi tanımlamasını ifade eder. Bunu bizim dünyamızda örneklendirmek gerekirse, "evet" veya "hayır" seçeneklerinden her biri bir en küçük bilgiyi ifade eder. Yalnız burada üçüncü bir ihtimal olamayan durumlarda bilgisayar sisteminin bit kavramına benzetebiliriz. Bu seçenekler, bilgisayar dünyasında "0" veya "1" olarak adlandırılır. Yani bir bitin durumunu tasvir edersek;

$$\frac{0\ durumu\ +1\ durumu}{2}$$

Klasik bilgisayarda, bilgi elektriksel gerilim yardımı ile kodlanarak taşınır ve hesaplamalar bu mantığa dayanarak yapılır. Bütün bilgisayar sistemi bu protokole göre inşa edilmiştir. Bizim kolaylık olması amacıyla kullandığımız "0" ve "1" den oluşan bit kavramı aslında 5Volt, 0Volt olmaktadır. Bilgisayar donanımının temelini oluşturan AND ve OR kapıları bu temele göre çalışmaktadır.

Geçmiş yıllarda kızılderililerin duman ile haberleştiğini tarihten öğreniyoruz, Gökyüzüne çıkan dumana göre kodlanmış bir bilgi sistemleri olduğunu kolayca anlaşılıyor, yani onların bit kavramı duman "var" veya "yok"'a göre kodlanmış bir haberleşme sistemi. Gelişmiş bir duman ile haberleşme sistemi olduğunu hayal edelim, ilk dumandan itibaren, önceden tanımlanmış küçük bir süre geçtiğinde tekrar duman çıkıp çıkmamasına bakılarak bir dizi "var", "yok" ile kodlanmış bilgi gönderilebilir. Bu "var", "yok" bilgilerini birleştirerek ve bazı kodlamalar kullanarak bir metin göndermek mümkün olabilirdi. Bu örneklemede geçen kavramları ve nesneleri günümüz bilgisayarı ile eşleştirirsek; duman var = 1 biti, duman yok = 0 biti, ilk duman = başlama biti, tanımlanmış küçük süre = saat(clock) kavramı, bazı kodlamalar = örnek, ascii.

Klasik bilgisayara geri dönersek, bir saat vuruşunda, veri yolunun 0V veya 5V olmasını kontrol etme aşamasında milyonlarca elektron devrede dolaşmaktadır. Kuantum mekaniği kullanarak kodlanmış bilgi sistemlerinde aynı bilgi bir elektronun spin durumuna göre; "spin up" veya "spin down" olmasına göre veya daha basit olarak bir atomdaki elektronun hangi enerji seviyesinde olduğuna göre kodlanabilir.

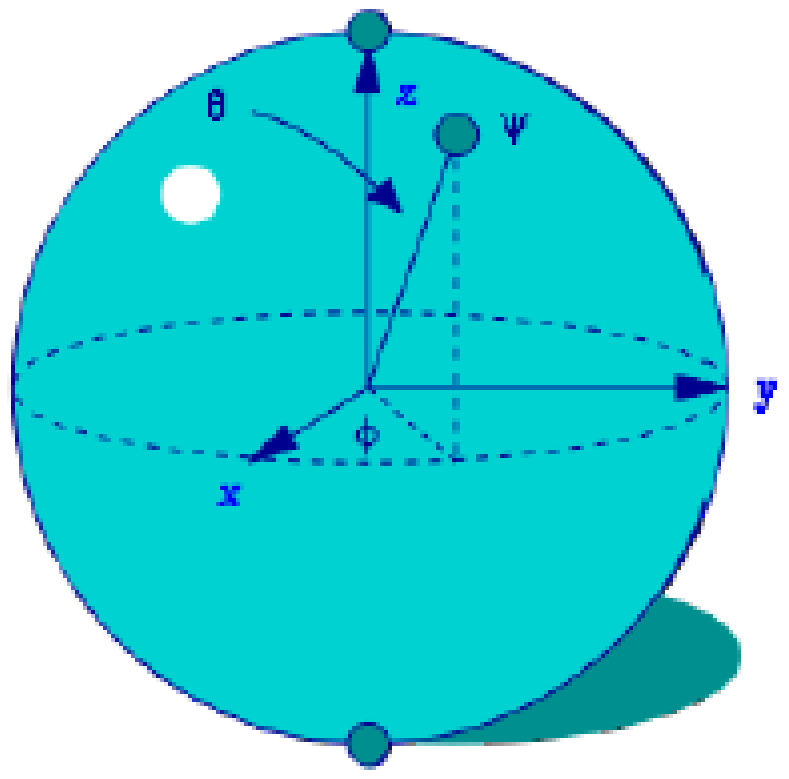

Şekil 3.2- Bir Kübitin Bloch Küresi ile Gösterimi

Bloch küresi (Şekil 3.2) kuantum mekaniğine, Felix Bloch tarafından kazandırılmış, tek kubiti göstermeye yarayan çizimdir. Üç boyutlu bir küredeki herhangi bir nokta, kubitin durumunu göstermektedir. Bu nokta, şayet kürenin yüzeyinde ise, bu durum saf durumdur (pure state) ancak nokta, kürenin içerisinde de olabilir. Bu durumda karışık durum (mixed state) olarak adlandırılır.

Klasik bit 0 veya 1 olarak kodlandığında bitin ölçülmesi durumunda olması gereken ihtimaller, eşit olarak 0 veya 1 olmaktadır. Ve tek mümkün durumlar bunlardır. Kuantum sistemlerde yani kübitlerde, bitler, ket, bra (|⟩) lardan oluşan Dirac notasyonuna göre yazılır. Klasik durumlardan farklı olarak, olabilecek iki ihtimalin lineer kombinasyonu da yer alır ve "*süperpozisyon*" olarak adlandırılır.

$$|\Psi\rangle = \alpha|0\rangle + \beta|1\rangle \quad (1)$$

Denk. 1 deki süper pozisyon sistemi tek kübit durumları tasvir etmektedir. a ve b durumların oluşma ihtimalleridir. Yani durum, $|0\rangle$ veya $|1\rangle$ durumlarının lineer kombinasyonları olarak tanımlanıyor.

## III.2. Çok Kübit Sistemler ve Dolanıklık Durumu

Elimizde iki adet kübit olduğunu varsayalım, eğer elimizdekiler klasik bitler olsaydı, olabilecek dört mümkün ihtimal mevcuttu, bunlar; 00, 01, 10 ve 11. Aynı şekilde bilgisayar tabanlı olarak iki kübit sistemler de dört mümkün ihtimal bulunmaktadır, bunlar; $|00\rangle, |01\rangle, |10\rangle\ ve\ |11\rangle$. Fakat kübit çiftleri, bu dört ihtimalin süper pozisyonu halinde de bulunabilirler. Bu durum içlerinden bir veya ikisi ile olabileceği gibi, her birini içeren yapıda da olabilir. En genel durum denk(2) deki gibidir;

$$|\Psi\rangle = \alpha_{00}|00\rangle + \alpha_{01}|01\rangle += \alpha_{10}|10\rangle + \alpha_{11}|11\rangle \quad (2)$$

Çok önemli iki kübit lik sistem olan "Bell state" veya "EPR çiftleri" durumları denk(3)-(6) da gösterilmiştir.

$$|\Psi_1\rangle = \frac{1}{\sqrt{2}}|00\rangle + |11\rangle \quad (3)$$

$$|\Psi_2\rangle = \frac{1}{\sqrt{2}}|00\rangle - |11\rangle \quad (4)$$

$$|\Psi_3\rangle = \frac{1}{\sqrt{2}}|01\rangle + |10\rangle \quad (5)$$

$$|\Psi_4\rangle = \frac{1}{\sqrt{2}}|01\rangle - |10\rangle \quad (6)$$

## III.3. Tek Kübit Kuantum Kapıları

Klasik bilgisayar devrelerinin kablo ve mantıksal kapılardan oluştuğunu önceden belirtmiştik. Kablolar bilgiyi başka bir yere taşımak için kullanılır, mantıksal kapılar ise bilgi üzerinde dönüşümler uygular. Örnek olarak tek bitlik klasik bir kapı olan *NOT* kapısının doğrulama tablosu, *1 -> 0* ve *0 -> 1* olmaktadır ve bilgi terslemesi yapılmaktadır. Örnek olarak denk(7) de süper pozisyon halinde olan bir kübit ve denk(8) de *NOT* kapısı uygulandıktan sonraki hali yer almaktadır.

$$\alpha|0\rangle + \beta|1\rangle \tag{7}$$

$$\beta|0\rangle + \alpha|1\rangle \tag{8}$$

Kapı operasyonlarını tasvir etmenin etkili bir yolu matris gösterimidir. *NOT* kapısını ele alırsak matris gösterimi aşağıdaki gibidir.

$$X \equiv \begin{bmatrix} 0 & 1 \\ 1 & 0 \end{bmatrix} \tag{9}$$

Tek kübit işlem yapam kuantum kapıları ikiye ikilik matrisler ile tanımlanabilir. Ve bu kapı matrislerinin kendi kendine uygulanması birim matrisi vermektedir, bu durum "üniter-unitary" operasyon olarak adlandırılır. Bir sisteme uygulanan kapı operasyonu tekrar uygulandığında durum ilk haline dönüyor ise operasyon "üniter" olmaktadır. Kendisine uygulandığında birim matris oluşması ise hiçbir işlem uygulanmamış gibi sonuçlandığını göstermektedir.

Kapının üniter olması önem taşımaktadır, kuantum bilgi teorisinde herhangi bir üniter operasyon geçerli olabilmektedir.Bir diğer önemli tek kübit kapı "Z" harfi ile gösterilen *Z* kapısı:

$$Z \equiv \begin{bmatrix} 1 & 0 \\ 0 & -1 \end{bmatrix} \tag{10}$$

Giriş kübiti $|0\rangle$ ise çıkış değişmemektedir, ancak giriş $|1\rangle$ olduğunda çıkış $-|1\rangle$ olmaktadır. Bir diğer kapı olan *Hadamard Kapısı :*

$$H \equiv \frac{1}{\sqrt{2}}\begin{bmatrix} 1 & 1 \\ 1 & -1 \end{bmatrix} \tag{11}$$

$|0\rangle$ durumunu $(|0\rangle + |1\rangle)/\sqrt{2}$ olan kapı matrisinin ilk kolonu şekline dönüştürür, ve $|1\rangle$ durumunu kapı matrisinin ikinci kolonuna benzetebildiğimiz, $(|0\rangle - |1\rangle)/\sqrt{2}$ durumuna dönüştürür. Hadamard kapısının giriş durumu üzerindeki etkisi Şekil 3.3'de gösterilmektedir.

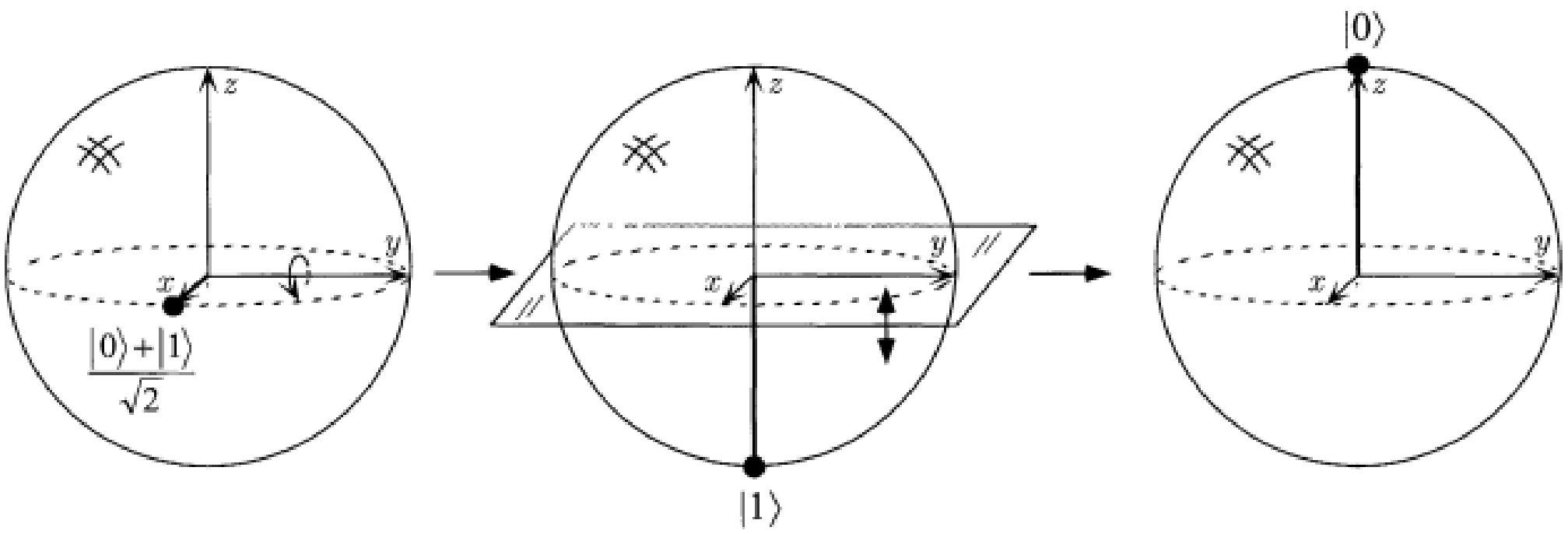

Şekil 3.3- Hadamard kapısının *x, y, z* ekseninde bir kübite etkisi. [1]

Bu mantıksal kuantum kapılarının devre elemanı olarak gösterimi ve yaptığı işlemler aşağıdaki gibi özetlenebilir.

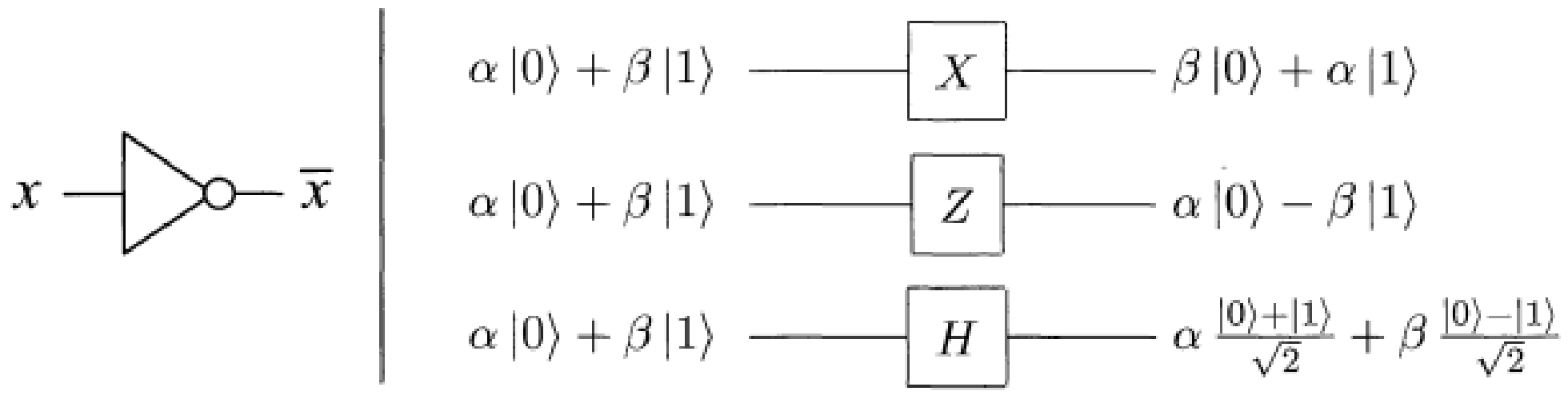

Şekil 3.4- Sol tarafta tek bit yapılan işlem, sağ tarafta ise tek kübit mantıksal kapılar. [1]

## III.4. Çok Kübit Kapılar

Klasik bilgisayar sisteminde kullanılan mantıksal kapılar Şekil 7 sol tarafta gösterilmektedir. Önemli bir teorik nokta ise, *NAND* ve *NOR* kapılarının; bilgisayar devrelerinde kullanılan bütün fonksiyonları gerçekleştirebilmesidir ve bu durum evrensel kapılar olarak adlandırılmasını beraberinde getirmiştir.

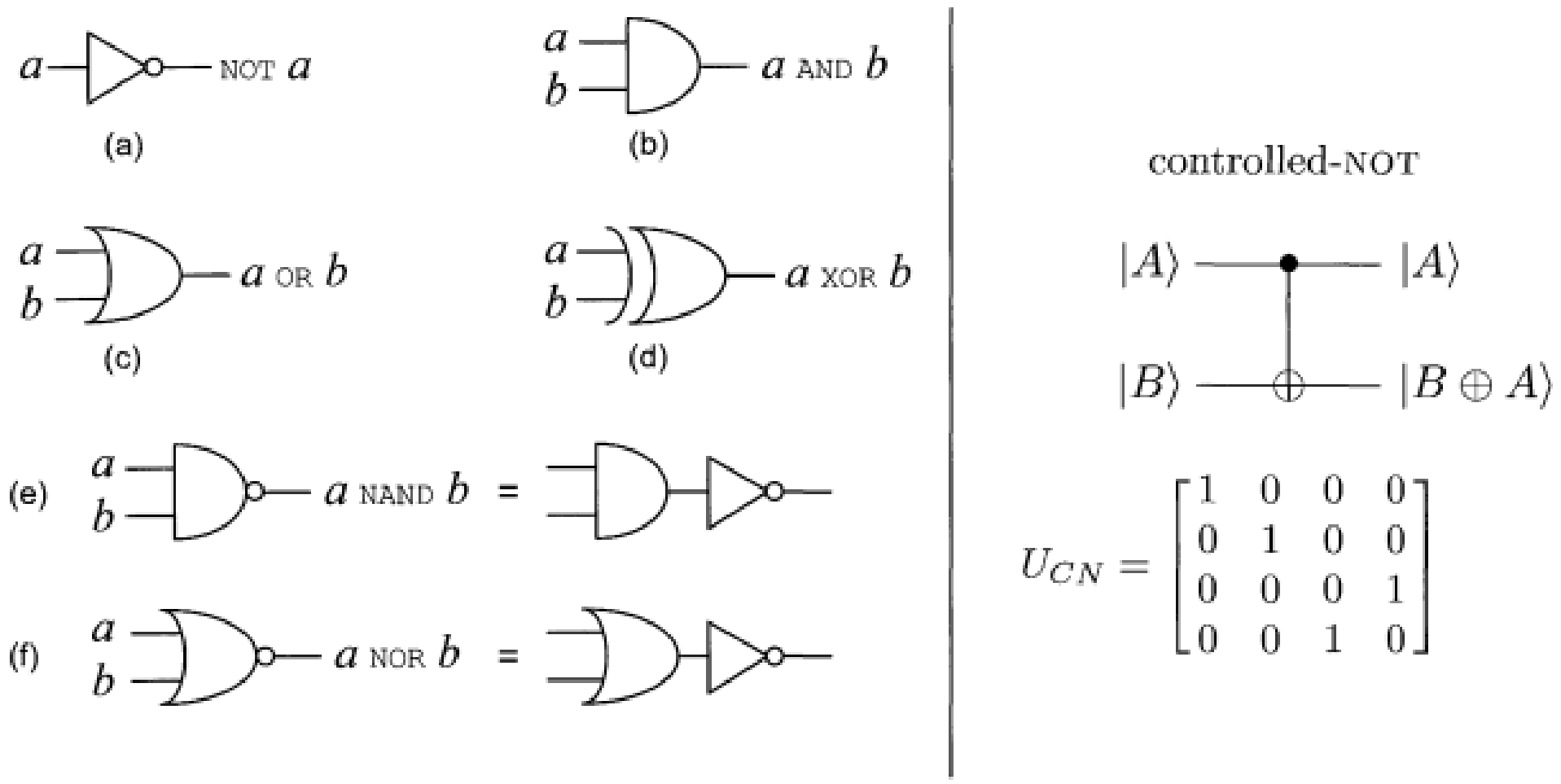

$$U_{CN} = \begin{bmatrix} 1 & 0 & 0 & 0 \\ 0 & 1 & 0 & 0 \\ 0 & 0 & 0 & 1 \\ 0 & 0 & 1 & 0 \end{bmatrix}$$

Şekil 3.5- Sol tarafta, standart iki bitlik kapılar, sağ tarafta örnek iki kübit mantıksal kapı olan *CNOT* veya *Controlled NOT* kapısı bulunmaktadır. [1]

Klasik *XOR* kapısı ile aynı etkiyi kuantum devrelerinde gösteren, kuantum devrelerin en önemli kapısı olan *Controlled NOT* veya *CNOT* kapısı Şekil 7 sağ tarafta yer almaktadır. Kapının iki girişi bulunmaktadır, girişlerden biri hedef (target) kübit olarak diğeri ise kontrol (control) kübiti olarak adlandırılır. Kontrol edilen kübit $|1\rangle$ ise hedef kübitine *NOT* işlemi uygulanır, kontrol edilen kübiti $|0\rangle$ ise hedef kübitine bir işlem uygulanmaz:

$$|00\rangle \longrightarrow |00\rangle;\ |01\rangle \longrightarrow |01\rangle;\ |10\rangle \longrightarrow |11\rangle;\ |11\rangle \longrightarrow |10\rangle.$$

Diğer önemli iki kübit kapılar *Controlled Z* kapısıdır ve aynı *CNOT* operasyonundaki gibi kontrol kübitinin durumuna göre hedef kübitine *Z* kapısı işlemini uygulamaktadır. kapının devre elemanı olarak gösterimi ve matrisi aşağıdaki gibidir.

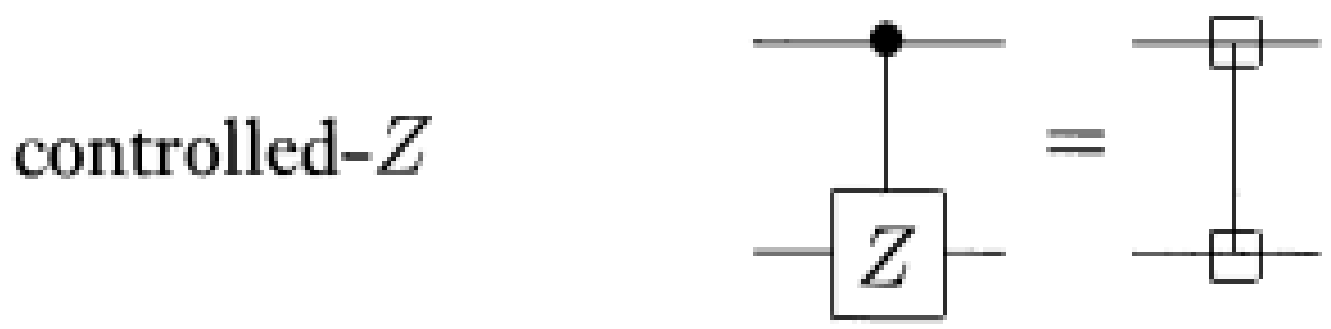

$$\begin{bmatrix} 1 & 0 & 0 & 0 \\ 0 & 1 & 0 & 0 \\ 0 & 0 & 1 & 0 \\ 0 & 0 & 0 & -1 \end{bmatrix}$$

Şekil 3.6-Kontrollü Z kapısının devre ve matris gösterimi. [1]

Üç kübit ile işlem yapma ihtiyacı duyulduğunda, tek ve iki kübit kapılar ile gerçekleştirilmesi teorik olarak yapılabilen üç kübit kapıların yaygın kullanılanları *Toffoli* ve *Fredkin* kapılarıdır. Bu kapıların devre gösterimi ve matrisleri aşağıda Şekil 3.6'te yer almaktadır. *Fredkin* kapısı; bir kontrol iki hedef kübitine işlem yapmaktadır, kontrol kübitinin durumuna göre hedef kübitlerinin yerlerini değiştirmektedir. *Toffoli* kapısı *CNOT* kapısının üç kübitlik olanı gibi düşünülebilir, kontrol kübiti iki adettir, her ikisinde kontrol edilir ikisi de $|1\rangle$ ise hedefe *NOT* kapısı uygulanmaktadır. Bu iki kapının devrede gösterimi ve matrisleri Şekil 3.7'de gösterilmektedir.

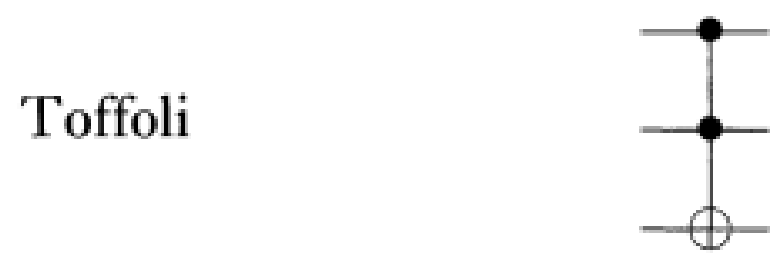

$$\begin{bmatrix} 1 & 0 & 0 & 0 & 0 & 0 & 0 & 0 \\ 0 & 1 & 0 & 0 & 0 & 0 & 0 & 0 \\ 0 & 0 & 1 & 0 & 0 & 0 & 0 & 0 \\ 0 & 0 & 0 & 1 & 0 & 0 & 0 & 0 \\ 0 & 0 & 0 & 0 & 1 & 0 & 0 & 0 \\ 0 & 0 & 0 & 0 & 0 & 1 & 0 & 0 \\ 0 & 0 & 0 & 0 & 0 & 0 & 0 & 1 \\ 0 & 0 & 0 & 0 & 0 & 0 & 1 & 0 \end{bmatrix}$$

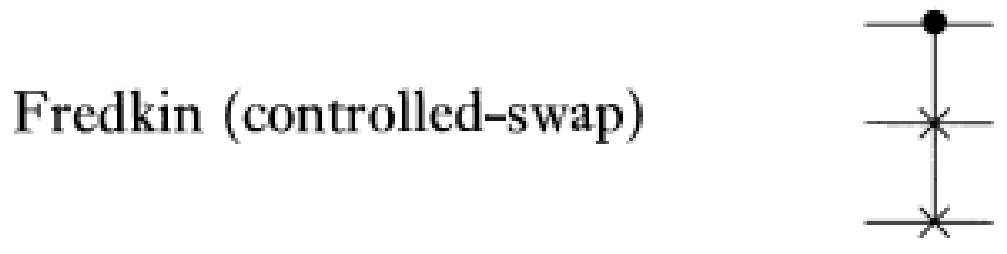

$$\begin{bmatrix} 1 & 0 & 0 & 0 & 0 & 0 & 0 & 0 \\ 0 & 1 & 0 & 0 & 0 & 0 & 0 & 0 \\ 0 & 0 & 1 & 0 & 0 & 0 & 0 & 0 \\ 0 & 0 & 0 & 1 & 0 & 0 & 0 & 0 \\ 0 & 0 & 0 & 0 & 1 & 0 & 0 & 0 \\ 0 & 0 & 0 & 0 & 0 & 0 & 1 & 0 \\ 0 & 0 & 0 & 0 & 0 & 1 & 0 & 0 \\ 0 & 0 & 0 & 0 & 0 & 0 & 0 & 1 \end{bmatrix}$$

Şekil 3.7- Toffoli ve Fredkin kapılarının devre ve matris gösterimi. [1]

## III.5. Kuantum Kapıları ile Klasik İşlemler

Kuantum devreleri ile klasik bilgisayar devreleri işlemlerini simule etmek son derece mümkün olmaktadır. Her bir klasik devre kuantum üniter kapılar ile gerçekleştirilebilir, bu da bize günümüz sistemlerini kuantum bilgisayarları ile çok daha efektif ve hızlı kıllanabileceğimizi göstermektedir. Örneğin, *Toffoli* kapısı hedef kübitine 1 gönderirsek, klasik *NAND* kapısı ile aynı özellik göstermektedir.

| Inputs | | | Outputs | | |
|---|---|---|---|---|---|
| $a$ | $b$ | $c$ | $a'$ | $b'$ | $c'$ |
| 0 | 0 | 0 | 0 | 0 | 0 |
| 0 | 0 | 1 | 0 | 0 | 1 |
| 0 | 1 | 0 | 0 | 1 | 0 |
| 0 | 1 | 1 | 0 | 1 | 1 |
| 1 | 0 | 0 | 1 | 0 | 0 |
| 1 | 0 | 1 | 1 | 0 | 1 |
| 1 | 1 | 0 | 1 | 1 | 1 |
| 1 | 1 | 1 | 1 | 1 | 0 |

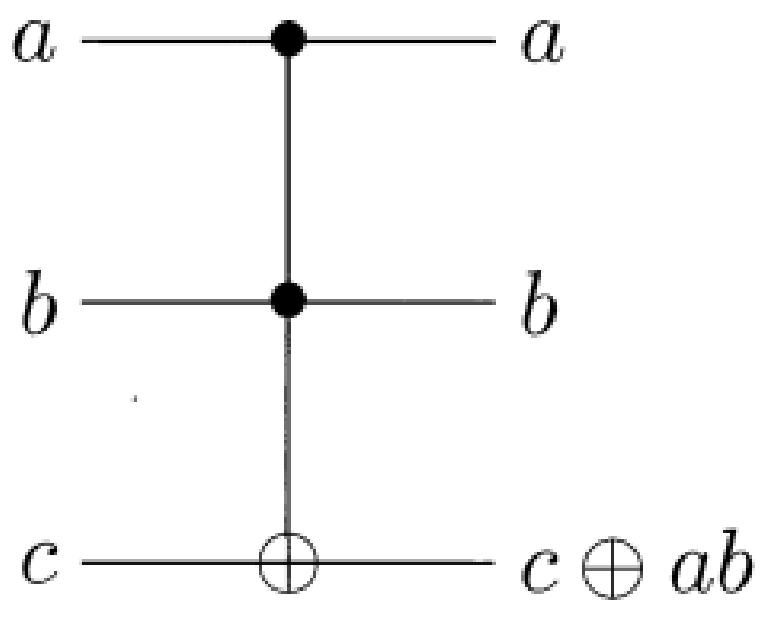

Şekil 3.8- Toffoli kapısının doğrulama tablosu ve kullanım şekli [1]

## III.6. Kuantum Işınlama

Şekil 3.9'deki kuantum devre ışınlama (teleportasyon) olayını göstermektedir. Klasik bitlerde oluşturulması imkansız olan bu yapıda amaç bir kübiti kuantum haberleşme yapmadan gönderebilmektir, yani sadece klasik haberleşme kullanılarak kuantum bilgi transfer edilmektedir.

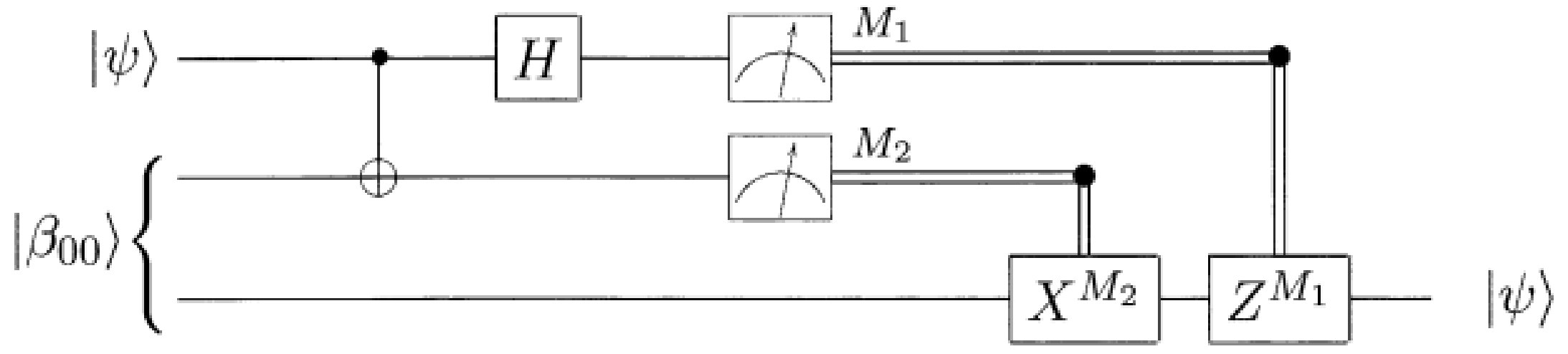

Şekil 3.9-Teleportasyon işleminin devresi, $M_1$ ve $M_2$ ölçümleri göstermektedir, ve bu sonuçlara göre *X*, *Z* kapıları uygulanmaktadır. [1]

İlk etapta Alice ve Bob'un bir Bell durumunda iki kübiti paylaştıklarını varsayalım. Sonuç olarak Bob'un elinde bir kübit bulunmaktadır. Alice ise elinde durumunu bilmediği Bob ile paylaştığı kübitin diğer çifti ve Bob'a göndermek istediği kübit bulunmaktadır. Gönderilmek istenen kübit, $|\psi\rangle = a|0\rangle + b|1\rangle$ olarak düşünelim. Bu durumu aşağıdaki gibi yazabiliriz.

$$\frac{1}{\sqrt{2}}[\alpha|0\rangle(|00\rangle + |11\rangle) + \beta|1\rangle(|00\rangle + |11\rangle)] \tag{12}$$

sol taraftaki bulunan ilk kübit *Alice*'in kübiti olmaktadır ve bu kübit gönderilmek istenmektedir. Daha önce paylaşılan *EPR* çiftleri ise ikinci ve üçüncü kübitlerdir. *Alice* elindeki iki kübite, *CNOT* operasyonu uygulamaktadır. genel durum aşağıdaki gibi olmaktadır.

$$\frac{1}{\sqrt{2}}[\alpha|0\rangle(|00\rangle + |11\rangle) + \beta|1\rangle(|10\rangle + |01\rangle)] \tag{13}$$

Sonra, *Alice* ilk kübitine *Hadamard* operasyonu uygulamaktadır ve Genel durum aşağıdaki gibi olmaktadır.

$$\frac{1}{2}[\alpha(|0\rangle + |1\rangle)(|00\rangle + |11\rangle) + \beta(|0\rangle - |1\rangle)(|10\rangle + |01\rangle)] \tag{14}$$

Bu genel durum daha basit olarak aşağıdaki gibi yazılabilir.

$$\frac{1}{2}[|00\rangle(\alpha(|0\rangle + \beta|1\rangle)) + |01\rangle(\alpha(|1\rangle + \beta|0\rangle)) + |10\rangle(\alpha(|0\rangle - \beta|1\rangle)) + |11\rangle(\alpha(|1\rangle - \beta|0\rangle))] \tag{15}$$

Bu durumda aşağıda gösterilen dört ayrı kırılım içermektedir.

$$00 \longrightarrow |00\rangle \equiv [\alpha|0\rangle + \beta|1\rangle] \;\; I\ işlemi\ uygulanmalı,$$

$$01 \longrightarrow |01\rangle \equiv [\alpha|1\rangle + \beta|0\rangle] \;\; X\ işlemi\ uygulanmalı,$$

$$10 \longrightarrow |10\rangle \equiv [\alpha|0\rangle - \beta|1\rangle] \;\; Z\ işlemi\ uygulanmalı,$$

$$11 \longrightarrow |11\rangle \equiv [\alpha|1\rangle - \beta|0\rangle] \;\; X\ ve\ Z\ işlemleri\ uygulanmalı,$$

şimdi *Alice* elindeki kübitleri ölçüyor ve ölçüm sonuçlarına bağlı olarak *Bob* ile klasik haberleşme kurarak, Şekil 11'deki diğer operasyonların hangilerini yapması gerektiğini söylüyor. Sonuç olarak yapılan işlemlerin ardından gönderilmek kübit ışınlanmış oluyor.

## III.7. Süperyoğun Kodlama (Superdense Coding)

Süperyoğun Kodlama, en basit ifade ile tek kübit göndererek iki klasik bitlik bilgiyi gönderebilir miyiz? Sorusuna cevap veren kuantum mekaniği uygulamasıdır. *Alice* ve *Bob* bu işlemi gerçekleştirmek istiyorlar ve ellerinde daha önce paylaştıkları aşağıdaki dolanık sistem durumu mevcut.

$$|\varphi\rangle = \frac{|00\rangle + |11\rangle}{\sqrt{2}}. \qquad (16)$$

Burada *Alice* elindeki kübite bir tek kübit operasyon uygulayıp *Bob*'a göndererek iki bitlik bilgi göndermek istiyor. Yapması gereken tek şey $I, Z, iY, NOT$ kapılarından her hangi birini uygulayıp kübiti Bob'a göndermek. Aşağıdaki şekilde bu yapı tasvir edilmiştir.

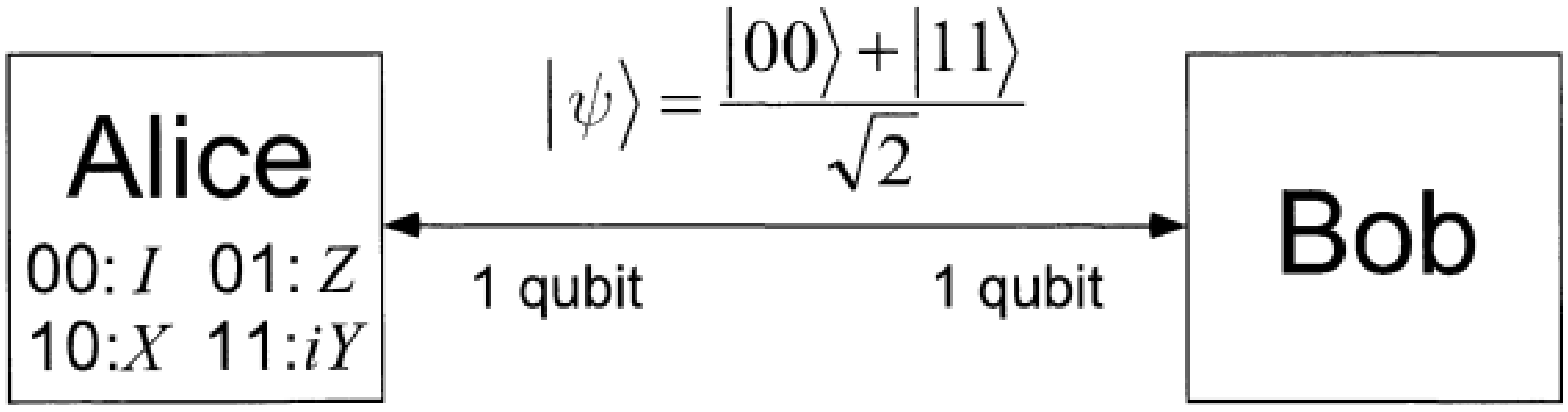

Şekil 3.10- Tek kübit ile klasik iki bit veri yollama senaryosunu niteleyen kurulum. [1]

Sonuç olarak Alice *00* göndermek isterse *I* kapısı uygulayıp gönderiyor, *01* göndermek isterse *Z* kapısı uygulayıp gönderiyor, *10* göndermek isterse, *X* kapısı uygulayıp gönderiyor, *11* göndermek isterse *iY* kapısı uygulayıp gönderiyor. Bu işlemler aşağıda gösterilmektedir.

$$00: \ |\psi_1\rangle \longrightarrow \frac{|00\rangle + |11\rangle}{\sqrt{2}} \qquad (17.1)$$

$$01 : |\psi_2\rangle \longrightarrow \frac{|00\rangle - |11\rangle}{\sqrt{2}} \quad (17.2)$$

$$10 : |\psi_3\rangle \longrightarrow \frac{|10\rangle + |01\rangle}{\sqrt{2}} \quad (17.3)$$

$$11 : |\psi_4\rangle \longrightarrow \frac{|01\rangle - |10\rangle}{\sqrt{2}} \quad (17.4)$$

## III.8 Sistem Yoğunluk Matrisi ve Kuantum Mekaniğinin Postulaları

Kuantum Sistem Durumlarını belirtmenin en kolay yöntemi onları *sistem yoğunluk matrisi (density matrix)* veya *yoğunluk operatörü (density operator)* dediğimiz matrislerle tanımlamaktır.

Bir sistem yoğunluk matrisi, durumunun tam olarak bilinmediği Kuantum Sistemleri betimlemek için kullanılır. Daha açık bir ifadeyle, bir kuantum sistemin durumu belli bir sayıdaki $|\psi_i\rangle$ sistem durumlarından birisi halindedir. Burada $i$ indisinin $p_i$ olasılıkta olduğunu varsayarsak $\{i, |\psi_i\rangle\}$ kümesini saf sistem durumlarının kümesi olarak adlandırabilirz. Bu durumda bu kuantum sistem yoğunluk operatörü-matrisi ile şu şekilde tanımlanabilir:

$$\rho = \sum_i p_i \, |\psi_i\rangle\langle\psi_i| \quad (18)$$

Yoğunluk matrisi tanımı matematiksel olarak sistem durum vektörü yaklaşımına denk olsa da yoğunluk matrisi gösterimi problem çözümlerinde ciddi kolaylıklar sağlamaktadır. Bir $\rho$ operatörü yoğunluk operatörü olarak adlandırılır ancak ve ancak şu koşulları sağlıyorsa:

1- Trace koşulu: $\rho$ operatörünün *trace-iz* değeri 1'e eşittir.
2- Pozitiflik koşulu: $\rho$ pozitif bir operatördür.

Bu koşullar altında *Kuantum Mekaniği*'nin temel postulalarını sistem yoğunluk matris-operatörü yaklaşımı ile şu şekilde tanımlamak mümkündür:

**Postula 1:** Herhangi bir izole fiziksel sistemin durumlarının uzayı bir *Hilbert Uzayı*'dır. Kuantum Sistem bir sistem yoğunluk matrisi ile betimlenir. Eğer bir sistemin $\rho_i$ durumunda olma olasılığı $p_i$ ise, tüm sistemin yoğunluk matrisi şu şekildedir:

$$\sum_i p_i \rho_i \quad (19)$$

**Postula 2:** Kapalı bir kuantum sistemin değişimi bir üniter dönüşüm (transformasyon) ile tanımlanır. Başka bir deyişle, eğer bir sistem durumu $t_1$ anında $\rho$ ve $t_2$ anında $\rho$' ise, $U$ üniter operatörü ile $t_1$'den $t_2$'ye şu şekilde tanımlanır.

$$\rho' = U\rho U^{\dagger} \quad (20)$$

**Postula 3:** Kuantum Ölçümler bir $M_m$ ile tanımlanır ve ölçüm operatörlerinin bir derlemesidir. Bu operatörler ölçülen sistemin uzayında işlem yaparlar. Buradaki $M$ indisi deney sırasında meydana gelebilecek çıktılara işaret eder. Sistem durumunun ölçüm öncesi $\rho$ olması durumunda m ölçüm sonucunun gelme olasılığı şu şekilde tanımlanır:

$$p(m) = tr(M_m^{\dagger} M_m \rho) \quad (21)$$

Ölçüm sonrası sistemin durumu ise şu şekilde tanımlanır.

$$\frac{M_m \rho M_m^{\dagger}}{tr(M_m^{\dagger} M_m \rho)} \quad (22)$$

Ölçüm öperatürü şu tamlık (completeness) koşulunu sağlamalıdır.

$$\sum_m M_m^{\dagger} M_m = I \quad (23)$$

**Postula 4:** Bileşik bir fiziksel sistemin durum uzayı fiziksel bileşen sistemlerin durum uzayının tensör çarpımıdır.

Bazı durumlarda bileşim bir kuantum sistemin sadece bir kısmı elimizde olabilir ve bizim ilgilendiğimiz kısım burası olabilir. Bu durumlarda, *indirgenmiş yoğunluk operatörüne* (reduced density matrix) ihtiyacımız bulunmaktadır. $A$ ve $B$ alt sistemlerinin olduğu bir $\rho_{AB}$ sistemimiz olduğunu varsayalım, bu durumda $A$ alt sisteminin yoğunluk operatörü

$\rho_A = tr_B(\rho_{AB})$ olarak tanımlanır. Burada $tr_B$, $B$ alt sistemi üzerinde *kısmi trace* (partial trace) adı verilen bir operatör map'idir.

*Kısmi trace* şu şekilde tanımlanır:

$$tr_B(|a_1\rangle\langle a_2|\otimes|b_1\rangle\langle b_2|) \equiv |a_1\rangle\langle a_2|tr(|b_1\rangle\langle b_2|) \quad (24)$$

Burada $|a_1\rangle$, $|a_2\rangle$, $|b_1\rangle$ $ve$ $|b_2\rangle$ *A* ve *B* durum uzayları için vektör kümeleridir.

Sağ taraftaki trace işlemi B sistemi için genel trace işlemidir:

$$tr(|b_1\rangle\langle b_2|) = \langle b_2|b_1\rangle \quad (25)$$

4. Postula'dan yola çıkarak bileşik sistemler için kuantum mekaniğinin en ilginç konularından birisi olan "dolanıklık" kavramını bulmaktayız. İki kübitlik durum için dolanık bir sistem durumu bölüm IV.2'de tanımlandığı üzere şu şekilde tanımlanabilir:

$$|\psi_1\rangle \longrightarrow \frac{|00\rangle + |11\rangle}{\sqrt{2}} \quad (26.1)$$

$$|\psi_2\rangle \longrightarrow \frac{|00\rangle - |11\rangle}{\sqrt{2}} \quad (26.2)$$

$$|\psi_3\rangle \longrightarrow \frac{|10\rangle + |01\rangle}{\sqrt{2}} \quad (26.3)$$

$$|\psi_4\rangle \longrightarrow \frac{|01\rangle - |10\rangle}{\sqrt{2}} \quad (26.4)$$

Bu sistem durumlarına Bell durumları (çiftleri) denmektedir ve bu durumlar *Lokal Operasyon Klasik İletişimde* (Local Operations and Classical Communication – LOCC) başka bir Bell durumuna dönüştürülemezler. Bu konu daha sonraki bölümlerde daha detaylı olarak incelenecektir.

## III.9 Kuantum İşlemler

Doğal olarak bir kuantum sistem çevresindeki sistemlerle etkileşim halindedir. Bir sisteminin durumunu değiştirmek istediğimizde bazı operatörlere ihtiyaç duyarız. Bu kuantum işlemlerin matematiksel olarak formülasyonu kuantum sistemlerin dinamiklerinin tanımlanması için anahtar görevi görmektedirler. Genel olarak bir Kuantum durumun transformasyonu şu şekilde tanımlanır: $\rho' = \varepsilon(\rho)$ Burada kuantum işlemin $\varepsilon$ map'iyle betimlenir.

Kuantum sistemler kapalı veya açık olabilirler. Kapalı bir kuantum sistem gürültü içermez ve çevreyle olan etkileşimi yoktur sadece üniter bir işlem ile değişmektedir. Diğer yandan, açık bir sistem, çevre ile etkileşim halindedir ve işlem hem temel sistem hem de çevre üzerinde etkilidir. Çevre ve temel sistem beraber ele alındığında değişim üniterdir; ancak çevre kapsam dışı bırakılır ve sadece temel sistem ele alınırsa temel sistemdeki değişim üniter bir değişim değildir.

Temel sistem ve çevrenin bir çarpım durumunda olduğunu varsayalım $\rho \otimes \rho_{env}$, son temel sistemi üniter işlem sonrası çevreden trace out edersek bulabiliriz:

$$\rho' = \varepsilon(\rho) = tr_{env}[U(\rho \otimes \rho_{env} U^{\dagger})] \tag{27}$$

Kuantum işlemleri *işlem-toplam gösterimi* ile gösterebiliriz. Böylece işlemlerin etkilerinin sadece temel sistem üzerinde olması sağlanabilmektedir. $|e_k\rangle$ çevrenin sonlu durum uzayının ortonormal bazı olsun ve çevre başlangıçta saf durum $\rho_{env} = |e_0\rangle\langle e_0|$ olarak tanımlansın. Herhangi bir *karışık durumdaki çevre* (mixed state environment) için saflaştırma sistemi tanımlanabilir. Bir önceki denklemi şu şekilde tanımlayabiliriz:

$$\varepsilon(\rho) = \sum_k \langle e_k | U[\rho \otimes |e_0\rangle\langle e_0|] U^{\dagger} | e_k \rangle \tag{28.1}$$

$$= \sum_k E_k \rho E_k^{\dagger} \tag{28.2}$$

Burada $E_k = \langle e_k|U|e_0\rangle$ temel sistem üzerindeki bir işlemdir. Bu işlem-toplam gösteriminde, Ek işlemleri $\varepsilon$ kuantum işleminin elemanları olmaktadır.
Trace koruyan kuantum işlemleri aşağıdaki tümlük koşulunu sağlamalıdır:

$$\sum_k E_k^{\dagger} E_k = I \tag{29}$$

# IV. DOLANIKLIK ÖLÇÜTLERİ

Bu bölümde Dolanıklık kavramı ile ilgili tanımlar verilmiş ve bu tanımlar sonrasında Dolanıklık Ölçütleri ile ilgili açıklamalar yapılmıştır. Daha sonra, bir işlemin Dolanıklık Ölçütü sayılabilmesi için gerekli koşullar tanımlanmıştır. İlerleyen kısımlarda ise literatürde sıkça çalışılmış olan Dolanıklık Ölçütleri ile detaylı anlatımda bulunulmuştur.

## IV.1. Dolanıklık Nedir?

Dolanıklık ölçütleri ile ilgili bir çalışmanın başlangıç kısımlarından birisi de dolanıklık kavramının tanımlanması olmalıdır. Bu kavramın nasıl kullanıldığının da anlatılması önem arz etmektedir. Dolanıklık kavramının kullanışlılığı *LOCC* kısıtı dediğimiz ve daha sonra detaylandıracağımız bir kısıttan yola çıkmasındandır. Bu kısıt hem teknolojik hem de temel motivasyonumuzu, incelediğimiz sistemler üzerinde uzun mesafeli kuantum haberleşmeyi doğrudan etkilediği için önemli hale getirmektedir.

Herhangi bir kuantum haberleşme deneyinde, kuantum parçacıkları birbirinden uzak laboratuvarlar arasında dağıtmak isteriz. Mükemmel kuantum haberleşme yapılabilmesi için mükemmel dolanıklık dağıtımı yapılması gerekmektedir [45]. Eğer bir kübiti uyum bozulması (decoherence) olmadan dağıtabilirsek, onun paylaştığı dolanıklığı da mükemmel bir şekilde dağıtabiliriz. Ters durum olarak, dolanık haldeki sistem durumlarını mükemmel bir şekilde dağıtabilirsek, az miktarda klasik haberleşme ile kuantum sistem durumlarını yayınlayabilmemiz için ışınlamayı kullanabiliriz. Bununla beraber, bu süreçleri uygulayabildiğimiz yapılabilir deneylerde, gürültünün etkisi kuantum sistem durumlarını uzun mesafelerde göndermemizi engelleyecektir.

Bu problemin çözümü için kuantum sistemlerinin dağıtımının zaten varolan gürültülü kuantum kanallar üzerinden yapılmalıdır; daha sonra gürültünün etkilerinin engellenmesi için gerekli işlemlerin birbirinden uzak mesafede yer alan laboratuvarlardaki yüksek kalitedeki lokal kuantum süreçlerle yapılması uygun olacaktır. Bu lokal kuantum işlemler *('Local Operations-LO')* çok-kontrollü ortamlarda yapıldığı için ideal duruma yakın almaktadır ve bu şekilde çok uzun mesafedeki iletişimin etkileri engellenmektedir. Bu sistemlerin çoğu zaman birbirinden tamamen bağımsız ortamlarda çalıştılması uygun olmayabilir. Bu durumda şu an var olan klasik iletişim *(classical communication-CC)* hal-i hazırda var olan standart iletişim teknolojileri ile gerçekleştirilebilr. Şekil 4.1'de de gösterildiği gibi bu şekildeki bir iletişimi farklı laboratuvarlardaki işlemleri koordine etmek için kullanabiliriz.

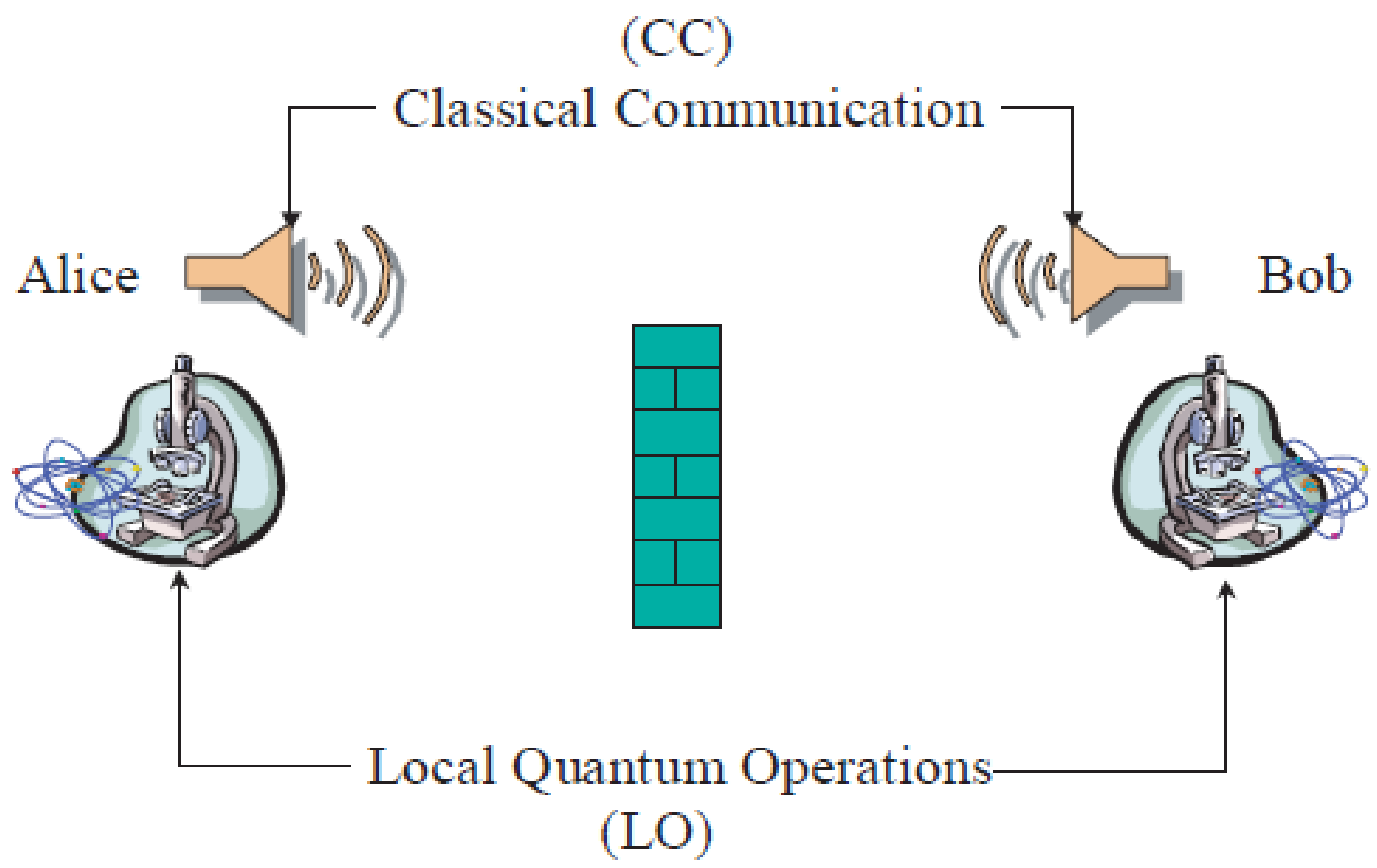

Şekil 4.1- Lokal Operasyon (LO) Klasik İletişim (CC) yönteminin gösterimi [45]

Birçok Kuantum Bilgi Teorisi çalışmasında klasik iletişimin kullanılabiliyor olması hayati bir önem taşımaktadır, örneğin kuantum ışınlama. Şu an yaptığımız kabuller mevcut teknolojik durumla ilgili olmakla beraber ışınlama çalışmalarında *LOCC* önem arz eden bir kavram olmaktadır [45].

Dolanıklığı, çoklu kuantum sistemler arasındaki *kuantum korelasyonlar* olarak tanımlayabiliriz. Bu durumda şu soru akla gelebilir kuantum korelasyon ne demektir ve *klasik korelasyon*dan ne farkı vardır? 'Kuantum' ve 'klasik' etkiler üzerindeki tartışma oldukça sıcak bir konudur. Kuantum bilgi kapsamındaki klasik korelasyonları *LOCC* kullanımında ortaya çıkanlar olarak tanımlayabiliriz. Bir kuantum sistemi gözlersek ve bunları klasik olarak simüle edemiyorsak, bunlara genel olarak *kuantum korelasyonlar* deriz. Gürültü bir kuantum sistem durumumuz olduğunu ve *LOCC* üzerinde bunun üzerinde işlem yaptığımızı varsayalım. Bu süreçte öyle bir sistem durumu elde edebiliriz ki bununla klasik korelasyonlar ile elde edemeyeceğimiz bazı işlemler yapabiliriz, örneğin Bell eşitsizliğinin çiğnenmesi gibi (violating Bell inequality). Bu durumda bu etkileri, *LOCC* işlemler sonrasında değil kaynak konumunda zaten varolan (çok gürültülü bir sistem durumu olsa bile) ilk sistem durumundaki kuantum korelasyonlar ile elde edebiliriz. Bu dolanıklık çalışmalarının en önemli noktasını oluşturmaktadır.

*LOCC* işlemleri kısıtı, kaynak sistemin durumunu dolanık hale yükseltmesidir. Sadece lokal olmayan ikili veya çoklu kuantum sistemler üzerinde işlemler yapılabildiği için dolanıklığın diğer bir tanımı da sadece *LOCC* tarafından oluşan bir korelasyonu olmayabileceği şeklinde olabilir. *LOCC* işlemlerin daha detaylıca anlaşılabilmesi için Kuantum İşlemler konusu bir de dolanıklık perspektifiyle anlatılmıştır.

## IV.2. Dolanıklık Perspektifiyle Kuantum İşlemler

Kuantum Bilgi Teorisi ile ilgili çalışmalarda genellikle 'genelleştirilmiş ölçümler' (generalized measurements) kullanılır. Bahsedilen bu genelleştirilmiş ölçümler standart kuantum mekaniğinin ötesine geçmemektedir. Kuantum değişimlere ait genel yaklaşımda, bir sistem üniter işlemlere göre değişim gösterir veya projektif ölçümlerle (projective measurements) çöker. Bir sistemin diğer kuantum sistemlerle etkileşimi sırasında oluşan yapıyı üç adımda tanımlayabiliriz: (1) öncelikle ek parçacık ekleriz (2) sonra hem sistem hem de ek parçacık üzerinde eş zamanlı üniter ve ölçüm işlemlerini uygularız ve son olarak (3) ölçüm sonuçlarının bazındaki bazı parçacıkları göz ardı ederiz. Eğer bu süreçteki ek parçacıklar bahsedilen sistemden orijinal olarak ilgisiz ise bu etkileşim *Kraus operatörleri* ile açıklanabilir. Herhangi bir ölçümün sonucunda çıkan toplam bilginin hesabını yapmak istersek $\rho_i = tr\{A_i \rho_{in} A_i^\dagger\}$ olasılığı ile oluşan ölçüm sonucu şu şekilde hesaplanır:

$$\rho_i = \frac{A_i \rho_{in} A_i^\dagger}{tr\{A_i \rho_{in} A_i^\dagger\}} \tag{30}$$

Burada $\rho_{in}$ ilk sistem durumunu ve $A_i$ Kraus operatörleri olarak bilinen matrisleri temsil etmektedir.

Olasılıkların normalizasyonu Kraus operatörlerinin şu koşulu sağlamasını gerektirir:

$$\sum_i A_i^\dagger A_i = \mathbb{1} \tag{31}$$

Bazı durumlarda, örneğin bir sistem çevre ile etkileşim içinde ise ölçüm sonuçlarının tamamına veya bir kısmına ulaşılamayabilir. Bu kapsamdaki en uç durumda ise ek parçacıklar trace out olur. Bu durumda harita (map) şu formül ile verilir: $\sigma = \sum_i A_i \rho_{in} A_i^\dagger$ ve Şekil 18 b kısmında gösterilmiştir. Bu şekildeki haritalara trace koruyan kuantum işlem adı verilir ve çoğu zaman *ölçen* (measuring) kuantum işlem denir. Bunun tersi olarak, $\sum_i A_i^\dagger A_i = \mathbb{1}$ koşulunu sağlayan herhangi bir $A_i$ lineer işlem kümesi için, *ek parçacık*, *eş üniter işlem* ve *van Neumann ölçümleri*nden oluşan bir süreç bulabiliriz. *İz (Trace) koruyan işlemler* için $A_i$ bütün matrisleri aynı boyutlarda olmalıdırlar ancak sonuç bilgisi korunuyorsa $A_i$ farklı boyutlara sahip olabilir. Genel kuantum işlemleri ile ilgili temel yapıtaşları tanımladıktan sonra artık hangi işlemleri *LOCC* altında uygulanabilir olduğunu tanımlayabiliriz. *LOCC* kısıtı Şekil 4.2'de görselleştirilmiştir. Genel olarak bu çeşit işlemler oldukça karmaşıktır. *Alice* ve *Bob* belli sayıdaki lokal hareket öncesinde veya sonrasında klasik olarak iletişim içerisinde olabilirler, bu durumda herhangi bir tur sonrası hareketler bir önceki ölçümlerin sonuçlarına bağımlı olur. Bu karmaşıklığın bir sonucu olarak, *LOCC* işlemlerin basit bir açıklaması mevcut değildir. Bu durum, daha kolay açıklanabilen ve daha büyük işlem sınıflarının geliştirilmesini motive etmiştir ve *LOCC* yapılabilme konusunun önemli bir parçası olarak kalmaktadır. Bu önemli sınıflardan birisi *ayrılabilir işlemlerdir* (separable operations). Bu şekildeki işlemler Kraus işlemlerinin cinsinden bir *çarpım* (product) gösterimi şeklinde yazılabilirler:

$$\rho_k = \frac{A_k \otimes B_k \rho_{in} A_k^\dagger \otimes B_k^\dagger}{tr A_k \otimes B_k \rho_{in} A_k^\dagger \otimes B_k^\dagger} \qquad (32)$$

Burada $\sum_k A_k^\dagger A_k \otimes B_k^\dagger B_k = \mathbb{1} \otimes \mathbb{1}$ koşulu sağlanmaktadır.

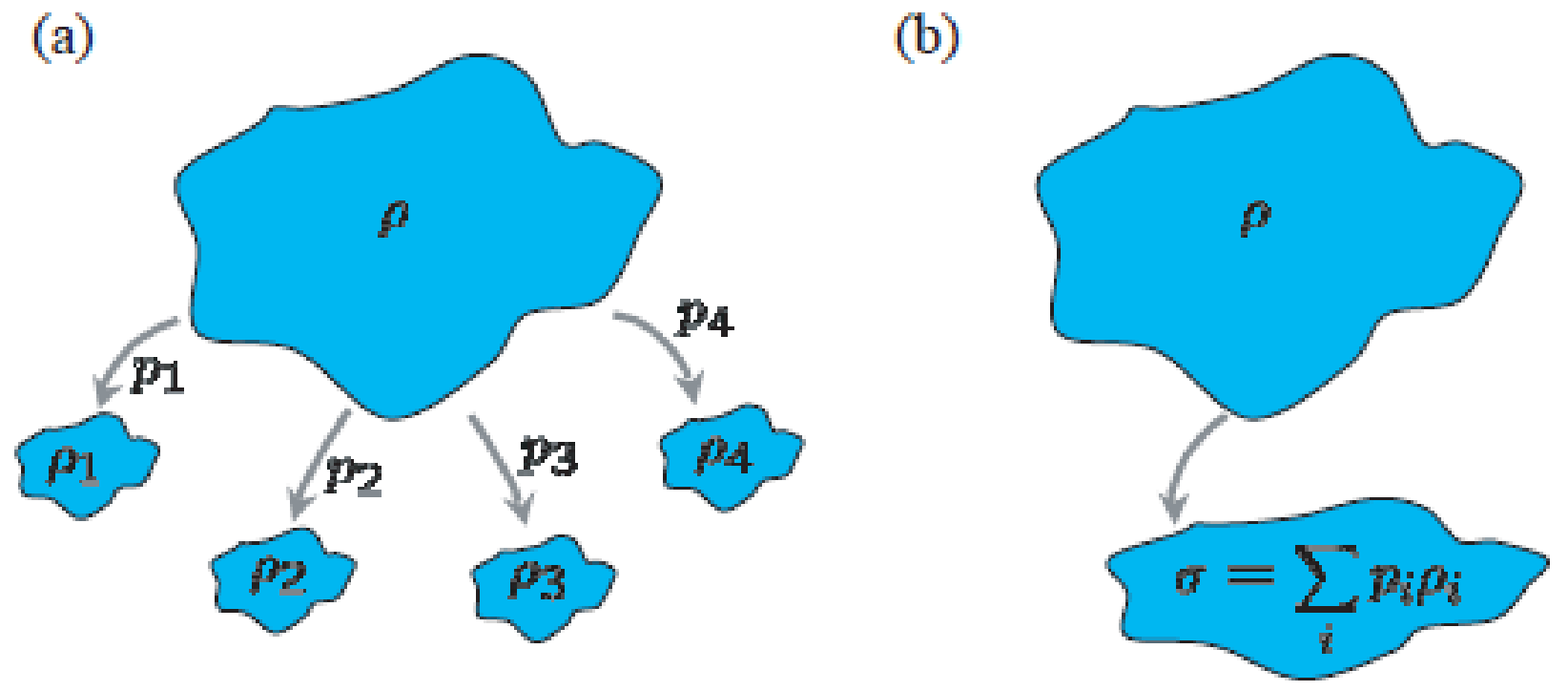

Şekil 4.2-Kuantum İşlemlerin alt seçmeyi içeren (a) ve içermeyen (b) şekillerde şematik gösterimi [45]

Açıkça, herhangi bir LOCC işlem bir ayrılabilir işleme dönüştürülebilir, öyle ki beraber yapılan işlem Alice ve Bob'un bireysel lokal Kraus işlemlerinin çarpımlarından oluşan işleme karşılık gelir. Ancak, bu durumun tersi geçerli değildir. Bir ayrılabilir işlem LOCC işlemler kullanılarak elde edilemeyebilir.

Ayrılabilir işlemler matematiksel bir bakış açısıyla irdelenirse, verilen bir görevin LOCC kullanılarak ciddi sınırlamalarla karşılaşmasına karşın, ayrılabilir işlemler kullanılarak optimize edilebilmektedir. Bazı durumlarda bu süreç, bazı zor sonuçlara sebep olabilmektedir: Simetrilerin olması durumunda optimal ayrılabilir işlemlerin LOCC ile de elde edilebileceğini not etmek isteriz. *Pozitif kısmi transpoze koruyan* (positive partial transpose (PPT) preserving operations) genel işlem sınıfları dolanıklığın anlaşılması için oldukça avantajlı bir matematiksel model oluşturmaktadır.

## IV.3. Dolanıklığın Temel Özellikleri

Bu kısımda Dolanıklığın Temel Özelliklerinin tanımlamak istiyoruz.

- Ayrılabilir sistem durumları dolanık değildirler:

*A, B, C* alt parçalarından oluşan ve şu şekilde yazılabilen $\rho_{ABC}$ sistem durumu *ayrılabilirdir* (separable)

$$\rho_{ABC\ldots} = \sum_i p_i \rho_A^i \otimes \rho_B^i \otimes \rho_C^i \otimes \ldots \tag{33}$$

Burada $p_i$ bir olasılık dağılımıdır. Bu sistem durumları LOCC ile kolayca oluşturulabilmektedir. Alice $p_i$ dağılımından diğer kısımlarında yer alanlar kullanıcılara $i$'nin sonucu hakkındaki bilgileri paylaşır ve her kısımdaki kullanıcı *X* lokal olarak $p_X^i$ değerini hesaplar ve $i$'nin sonucunda gelen bilgiyi göz ardı ederler. Bu sistem durumları lokal gizli değişkenler modelinden LOCC ile sağladığı için doğrudan oluşturulabilir ve bunların bütün korelasyonları klasik olarak açıklanabilir. Bu şekilde, mantıken ayrılabilir sistem durumlarının hiç dolanık olmadığı sonucuna ulaşabiliriz.

- Ayrılabilir olmayan sistem durumları (non-separable state) bazı görevlerin LOCC'ye göre daha iyi bir şekilde yapılmasını sağlarlar, bu durumda ayrılabilir olmayan tüm sistem durumları dolanıktır:

Herhangi bir ayrılabilir olmayan sistem durumu $\rho$ için başka öyle bir $\sigma$ sistem durumu bulunabilir ki bunun ışınlama doğruluğu (teleportation fidelity) eğer $\rho$ varsa geliştirilebilir olduğu sonucu yakın zamanda gösterilmiştir. Bu ilginç sonuç ayrılabilir sistem durumlarında olmayan olumlu bir sonucu elde etmemizi sağlamıştır. Bu durum aynı zamanda ayrılabilir olmayan ve dolanık terimlerinin birbirlerinin eşanlamlısı olarak kullanımını da desteklemektedir.

- LOCC dönüşümler altında sistem durumlarının dolanıklığı artmaz
- Lokal üniter işlemlerin altında dolanıklık değişmez
- Maksimal olarak dolanık sistem durumları mevcuttur:

Bir sistem durumunun dolanık olması kavramı, bazı durumlarda bir sistem durumunun diğerinin daha dolanık olduğu şeklindeki gerçeği de tanımlayabilmemizi sağlar. Bu da *maksimal dolanık sistem durumunun* (maximally entangled state) varlığı sorusunu doğurur. Maksimal dolanık sistem diğerlerinin tümünden daha dolanıktır. Bu durumda iki parçacıklı iki seviye sistemler olabileceği gibi, qudit dediğimiz iki parçacıklı *d-boyutlu-seviye* alt sistemler de olabilir. Bir saf sistem durumu için aşağıdaki denklem maksimal dolanık sistem durumunu tanımlar:

$$|\psi_d^+\rangle = \frac{|0,0\rangle + |1,1\rangle + \cdots + \ |d-1, d-1\rangle}{\sqrt{d}} \tag{34}$$

Bu tanımların tamamından yola çıkarak; *"sistem durum sıralaması mümkün müdür?"* veya *"sistem durum sıralaması problemi bir kısmi sıralama mıdır yoksa tam sıralama mıdır?"* gibi sorular sorulabilir. Bu soruların cevaplarının sorgulanabilmesi için bir sistem durumunun diğerine LOCC işlemler altında dönüştürülüp sorusunun cevaplanması gerekmektedir.

## IV.4. Dolanıklık Ölçütlerinin Postulaları

Bu bölümde, herhangi bir dolanıklık ölçütünün sağlaması gereken birkaç temel aksiyom tanımlayacağız. Tam olarak iyi bir dolanıklık ölçütünün uyması gereken özellikler nelerdir? Bir dolanıklık ölçütü öyle bir matematiksel büyüklüktür ki dolanıklıkla ilişkili temel özellikleri sağlamalıdır ve ideal olarak bazı operasyonel prosedürlere göre çalışmalıdır. Amaçlarımıza göre, bu durum bize bazı olası istenen özellikleri tanımlamamızı sağlar. Aşağıdaki liste dolanıklık ölçütlerinin uyması gereken postulaları tanımlamaktadır [45]. Bazı çokluklarda bu özellikleri tamamı sağlanmamaktadır:

1- Bir iki taraflı (bipartite) dolanıklık ölçütü $E(\rho)$ sistem yoğunluk matrislerinden pozitif sayılara bir eşlemedir (mapping): $\rho \rightarrow E(\rho) \in \mathbb{R}^+$

   Bu ölçüt herhangi bir iki taraflı sistemin durumu için tanımlanabilir. Bir normalizasyon değeri/çarpanı (factor) de/da genelde kullanılmaktadır, örneğin iki küdit(qudit)lik maksimal dolanık sistem durumu

$$|\psi_d^+\rangle = \frac{|0,0\rangle + |1,1\rangle + \cdots + |d-1, d-1\rangle}{\sqrt{d}} \tag{35}$$

   İçin bu değer $E(|\psi_d^+\rangle) = \log d$ değerine eşittir.

2- Eğer sistem durumu ayrılabilirse $E(\rho) = 0$'dır.

3- *E*, *LOCC* altında ortalamada artmaz, başka bir deyişle

$$E(\rho) \geq \sum_i p_i E\left(\frac{A_i \rho A_i^\dagger}{tr A_i \rho A_i^\dagger}\right) \tag{36}$$

   Burada $A_i$ matrisleri bazı *LOCC* protokolünü tanımlayan *Kraus* operatörlerini temsil eder ve *i* sonucunun olasılığı şu denklem ile hesaplanabilir: $p_i = tr A_i \rho A_i^\dagger$

4- $|\psi\rangle\langle\psi|$ Saf sistem durumu için ölçüt dolanıklığın entropisine iner:

$$E(|\psi\rangle\langle\psi|) = (S \circ tr_B)(|\psi\rangle\langle\psi|) \tag{37}$$

İlk 3 koşulu sağlayan herhangi bir *E* fonksiyonuna *dolanıklık monotonu (entanglement monotone)* denir. 1,2 ve 4. Koşulları sağlayan ve deterministik LOCC dönüşümleri altında artmayan fonksiyonlara ise *dolanıklık ölçütü (entanglement measure)* denir. Literatürde bu iki terim birbirinin yerine kullanılabilmektedir. Yaygın olarak, bu özelliklere ek olarak dolanıklık ölçütleri için bazı diğer gereksinimler tanımlanmıştır:

- Konvekslik (Convexity)

Bir dolanıklık ölçütünün konveksliği sağlayabilmesi için aşağıdaki eşitsizliği sağlaması gerekmektedir:

$$E(\sum_i p_i \rho_i) \leq \sum_i p_i E(\rho_i) \tag{38}$$

- Toplanabilirlik (Additivity)

Bir dolanıklık ölçütü ve $\sigma$ sistem durumu tanımlandığında her *n* tamsayısı için $E(\sigma^{\otimes n}) = nE(\sigma)$ eşitliği sağlanıyorsa bu ölçüt toplanabilir (additive) denir. Bu özelliği sağlamayan birçok dolanıklık ölçütü olduğu için bu özellik dolanıklık ölçütleri için temel bir özellik olarak tanımlanamamaktadır. Bu eşitliğin daha düzenlenmiş (regularized) veya asimptotik versiyonu şu şekilde tanımlanabilir:

$$E^{\infty}(\sigma) := \lim_{n\to\infty} \frac{E(\sigma^{\otimes n})}{n} \tag{39}$$

Bu durumda bir ölçüt bunu otomatik olarak sağlamaktadır. Daha kuvvetli bir gereksinim olarak, herhangi bir $\sigma$ ve $\rho$ sistem durum çifti için $E(\sigma \otimes \rho) = E(\sigma) + E(\rho)$ eşitliğimiz varsa bu ölçüt için tam toplanabilir (full additive) denmektedir.

- Süreklilik (Continuity)

Bir $L$ dolanıklık monotonu saf sistem durumları için toplanabilir ise şu eşitsizliği sağlamaktadır:

$$n(L(|\phi\rangle) = L(|\phi\rangle^{\otimes n}) \geq L(\rho_n) \tag{40}$$

Bu eşitsizlik dolanıklık monotonları için 3. Koşuldan yola çıkarak yazılabilmektedir. Bu $L$ monotonu için şu eşitlik varsa

$$L(\rho_n) = L\big(|\psi^-\rangle^{\otimes nE(\phi\rangle)}\big) + \delta(\epsilon) = nE(\phi\rangle) + \delta(\epsilon) \tag{41}$$

$L$ için “yeterince sürekli” denebilir. Burada $\delta(\epsilon)$ küçük bir değer olacaktır ve şu eşitliği elde etmiş oluruz:

$$L(|\phi\rangle) \geq E(|\phi\rangle) + \frac{\delta(\epsilon)}{n} \tag{42}$$

Asimptotik sürekli terimi ise şu özellik ile tanımlanmaktadır:

$$\frac{L(|\phi\rangle_n) - L(|\psi\rangle_n)}{1 + \log(dimH_n)} \rightarrow 0 \tag{43}$$

Burada iki sistem durumu akışı $|\phi\rangle_n, |\psi\rangle_n$ arasındaki trace norm $tr||\phi\rangle\langle\phi|_n - ||\varphi\rangle\langle\varphi|_n|$işlemi $n \rightarrow 0$’a gitmektedir. L’nin saf sistem durumları yeter ve gerekli kısıtları olduğunu gözlemlenmektedir.

## IV.5. Dolanıklık Monotonları, Dolanıklık Ölçütleri

Bu kısımda iki taraflı sistemler için literatürde tanımlanmış olan birkaç dolanıklık ölçütü ve monotonunu tanımlayacağız. Burada tanımlanan bazı ölçütlerin fiziksel anlamlı diğerlerinde daha değerlidir. Öncelikle *distile edilebilir* (distilable) dolanıklık kavramını tanımlamakla başlayacağız:

- Oluşum Entropisi (Entanglement of Formation):
  Bir karışık sistem durumu (mixed state) $\rho$’nun, *Oluşum Entropisi (Entaglement of Formation)* $E_F$ değeri Bennett ve arkadaşları tarafından herhangi bir saf sistem durumunun (pure state) $\rho$ ile şu eşitliği sağlayan ortalama en küçük dolanıklığı olarak bulunmuştur [46,47] [BennettPRA1996] [BennettPRL1996]:

$$E_F(\rho) = inf \sum_i E(|\varphi_i\rangle\langle\varphi_i|) \tag{44}$$

Burada infimum değeri

$$\rho = \sum_i p_i |\varphi_i\rangle\langle\varphi_i| \tag{45}$$

tüm saf-durum dağılımlarını ve $E(|\varphi_i\rangle\langle\varphi_i|)$ ise *von Neumann entropisi* olarak kolayca tanımlanan dolanıklığın entropisini göstermektedir.

İki parçacıklı iki seviyeli sistemlerin özel durumunda ise, Wootters bir sistem durumu $\rho$ nun Oluşum Entropisinin şu formülle hesaplandığını göstermiştir [48] [WoottersPRL1998]:

$$E_F(\rho) = H(\frac{1}{2}\left[1 + \sqrt{1 - C^2(\rho)}\right]) \tag{46}$$

Burada, $H(x) = -\mathrm{x}\log_2 x - (1-x)\log_2(1-x)$ tanımı *ikili entropiyi* tanımlamaktadır. Wootters'ın *Eş Zamanlılık (Concurrence)* olarak tanımladığı ölçütün formülü ise şu şekilde verilmektedir [48] [WoottersPRL1998]:

$$C(\rho) = \max\{0, \lambda_1 - \lambda_2 - \lambda_3 - \lambda_4\} \tag{47}$$

Burada $\lambda_i$ 'ler azalan sırada

$$\rho(\sigma_y \otimes \sigma_y)\rho^*(\sigma_y \otimes \sigma_y) \tag{48}$$

Matrisinin özdeğerlerinin karekök değerlerini vermektedirler. $\sigma_y$ ise Pauli spin matrisini ve * operatörü ise kompleks konjuge operatörünü temsil etmektedir.

$E(\rho)$ ve $C(\rho)$ *ayrılabilir* bir sistem durumu (separable state) için *0* ve *tam dolanık* bir sistem durumu (maximally entangled state) için *1* değerleri arasında yer almaktadır.

- *Konveks Çatı Kurulumlarından (Convex Roof Constructions)* yola çıkan dolanıklık ölçütleri:

Oluşum Entropisi $E_F$ konveks çatı kurulumu kavramını anlamamız için güzel bir örnektir. Konveks çatı bir $f$ fonksiyonun $\hat{f}$ konveks çatısı, $f$ fonksiyonun bütün değerini üstten sınırlayan en büyük konveks fonksiyon olarak tanımlanır.

Herhangi bir saf sistem durumu dolanıklık monotonunun konveks çatısının, karışık sistem durumlarından karışık sistem durumlarına LOCC dönüşümleri için dolanıklık monotonu olduğunu söyleyebiliriz:

- Dolanıklığın Göreceli Entropisi (Relative Entropy of Entanglement):

*Dolanıklığın Göreceli Entropisi (REE)* sistemin, dolanık olmayan sistemler arasından kendisine en yakın olan sisteme uzaklığına dayanan bir ölçüttür. Matematiksel olarak şu şekilde tanımlanmaktadır: Tüm ayrılabilir sistem durumlarının kümesi olan *D* içerisindeki $\sigma$ sistem durumları için Kuantum Göreceli Entropi $S(\rho||\sigma) = Tr(\rho log\rho - \rho log\sigma)$ denklemi ile hesaplanır. Bu durumda REE tanımı şu şekilde olmaktadır:

$$E(\rho) = \min_{\rho \epsilon D} S(\rho||\sigma) = S(\rho||\overline{\sigma}) \quad (49)$$

Burada $\overline{\sigma}$, $\rho$'ya en yakın olan sistemi ifade etmektedir.

Bu ölçütün *2-seviyeli (two-level) sistemlerde* veya daha çok seviyeli sistemlerde genel geçer bir kapalı formülü henüz bulunamamıştır. Çok seviyeli sistemlerde bazı özel durumlar için formül önerileri ve 2-seviyeli sistemlerde ise optimizasyon yöntemleriyle kestirim yöntemleri mevcuttur. Bu kestirim yöntemleri içerisinden reel sistem yoğunluk matrisleri için bilinen en hızlı yöntem bu alanın tanınmış araştırmacıları Zinchenko, Friedland ve Gour tarafından yayınlanan *"Numerical Estimation of Relative Entropy of Entanglement"* makalesinde tanımlanmıştır [49] [ZinchenkoPRA2010] .

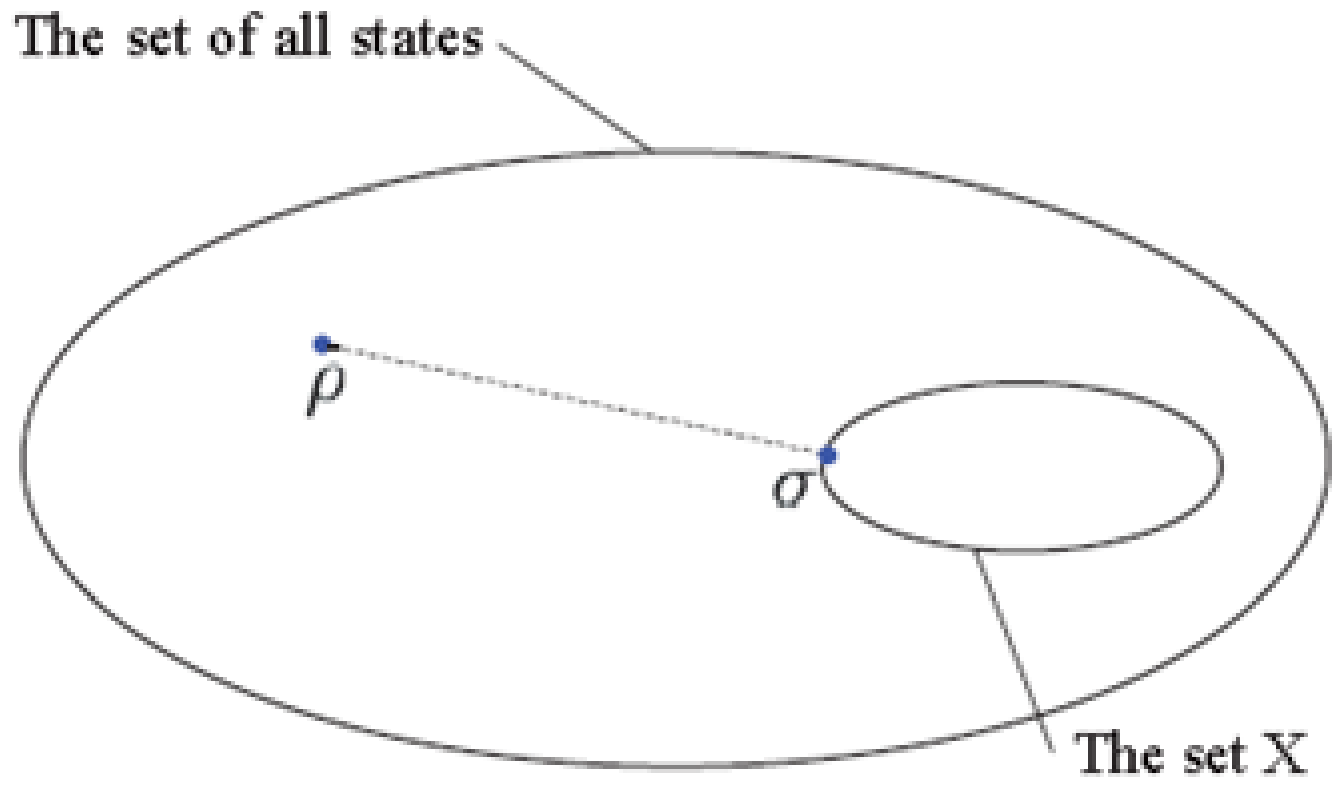

Şekil 4.3: Dolanıklığın Göreceli Entropisinin gösterimi [45]

- Negatiflik ve Logaritmik Negatiflik (Logarithmic negativity):

Peres-Horodecki kriterinin [62,63] kantitatif bir versiyonu olan ve *Negatiflik* ismi verilen diğer bir dolanıklık ölçütünü irdeleyeceğiz. İki parçacıklı iki seviyeli bir sistem için Negatiflik ölçütü şu şekilde tanımlanmaktadır [50-52]:

$$N(\rho) = \max\{0, -2\mu_{min}\} \tag{50}$$

Burada $\mu_{min}$ değeri $\rho$'nun kısmi transpozesinin minimum özdeğerini ifade eder. Yukarıdaki formülde tanımlanan *Negatiflik*, *Eş Zamanlılık* gibi 0 ve 1 değerleri arasında değişmektedir. 1 değeri yine aynı şekilde maksimum derecede dolanıklığa sahip sistemler için elde edilmektedir. Vidal ve Werner tarafından Negatifliğin dolanıklık için monoton bir fonksiyon olduğu gösterilmiştir [50].

Logaritmik Negatiflik ise $E_N(\rho) = \log_2(2N(\rho) + 1)$ eşitliği ile hesaplanır [2].

Negatiflik ve logaritmik negatiflik ölçütleri kolayca hesaplanabilen ölçütler oldukları ile için sıklıkla kullanılan ölçütler olarak literatürlerde sıkça çalışılmış olan ölçütlerdir.

# V. KUANTUM FISHER BİLGİSİ

Fisher Bilgisi; Bilgisayar Mühendisliği açısından önemli olan bazı problemlerin çözümünde; Örneğin büyük miktarda ve çok değişkenli verinin işlenmesi (Exploratory Data Analysis-Tahminsel Veri Analizi), veri madenciliği, vb. konularının araştırılmasında kullanılan bir kavramdır [31]. Kuantum Bilgisayarları üretildiğinde benzer veya daha karmaşık problemlerin çözümünde Kuantum Fisher Bilgisi'nin kullanılacağı öngörülmektedir.

## V.1. Kuantum Fisher Bilgisi ile İlgili Temel Tanımlar

Kuantum Fisher Bilgisi (KFB), Faz hassasiyeti gerektiren durumların analizde oldukça kullanışlı bir kavramdır. Bu özelliği ile dikkat çekmiştir ve klasik Fisher Bilgisini genişletmektedir. Özel olarak KFB değeri daha yüksek olan sistemlerin, kesinliği daha net bir şekilde elde edilir; örneğin saat senkronizasyonu [53] ve kuantum frekans standartlarını [54] verebiliriz. Saf dolanık sistemlerin bazıları klasik limiti geçebilse de bu durum bütün dolanık sistemler için geçerli değildir [34]. Kuantum sistem ve çevre arasındaki etkileşim sadece dolanıklığı azaltmaz aynı zamanda sistemin Kuantum Fisher Bilgisini de, genelde azaltır. Böylece şunu söyleyebiliriz ki kuantum sistemlerin KFB konusunda araştırma yapmak kuantum teknolojilerin ilerlemesi için önem arz etmektedir. Yakın zamandaki çalışmalarda, tek bir parametre, $\chi^2$ parametresi, faz hassasiyeti konusu irdelenirken incelenen sistemin sadece kendisinden gelen bilginin ölçülmesi amacıyla eklenmiştir [21]. Genel bir kuantum sistem için $\chi^2 < 1$ koşulu sağlanmadığı için sistemin çoklu dolanıklığa sahip olduğu anlaşılır ve bu sistem ayrılabilir bir sistemden daha iyi bir faz hassasiyeti sağlar. Bu kuantum sistemlere literatürde "kullanışlı" (useful) sistemler adı verilir. İki seviyeli N-parçacıklı kuanrum sistemler için Cramer-Rao limiti aşağıdaki formül ile tanımlanır [38,39]:

$$\Delta\phi_{QCB} \equiv \frac{1}{\sqrt{N_m F}} \tag{51}$$

Burada $N_m$ değeri ölçüm yapılan sistem üzerindeki deneylerin sayısını ve $F$ ise Kuantum Fisher Bilgisi değerini tanımlamaktadır. Açısal momentum operatörlerinin $n$. doğrultudaki normalize edilmiş 3-boyutlu vektörlerini, $J_n$ , Pauli matrisleri şu şekilde yazabiliriz:

$$J_{\vec{n}} = \sum_{\alpha=x,y,z} \frac{1}{2} n_\alpha \sigma_\alpha \quad (52)$$

$J_n$ için $\rho$ kuantum sisteminin Fisher Bilgisi simetrik bir matris olan $C$ cinsinden şu şekilde ifade edilebilir [21]:

$$F(\rho, J_{\vec{n}}) = \sum_{i \neq j} \frac{2(p_i - p_j)^2}{p_i + p_j} |\langle i|J_{\vec{n}}|j\rangle|^2 = \vec{n} C \vec{n}^T \quad (53)$$

Burada $p_i$ ve $|i\rangle$ sırasıyla $\rho$ sisteminin özdeğer ve özvektörlerini temsil etmektedir ve $C$ matrisi de şu şekilde tanımlanır

$$C_{kl} = \sum_{i \neq j} \frac{(p_i - p_j)^2}{p_i + p_j} [\langle i|J_k|j\rangle\langle j|J_l|i\rangle + \langle i|J_l|j\rangle\langle j|J_k|i\rangle] \quad (54)$$

$N$ seçenek arasında en büyük $F$ değeri seçilir ve $N$ parçacık üzerinde ortalaması alınır. Fisher Bilgisi değeri $C$ matrisinin en büyük özdeğeri olarak hesaplanır. Bu tanım denklem ile şu şekilde ifade edilir:

$$\bar{F}_{\max} = \frac{1}{N} \max_{\vec{n}} F(\rho, J_{\vec{n}}) = \frac{\lambda_{\max}}{N}. \quad (55)$$

## V.2. Kuantum Metroloji ve Kuantum Fisher Bilgisinin Uygulamaları

Kuantum mekaniksel sistemler, fiziksel dünya ile ilgili birçok konuda bilgi edinmemiz konusunda uygun ortamı sağlarlar. Bu bilgiler arasında ölçüm cihazlarından elde edilen bilgiler oldukça önemli bir yer tutmaktadır. Kuantum mekaniği bu ölçüm cihazlarının netliğini Heisenberg Belirsizlik İlkesi ve Margolus-Levitin Teoremi ile sınırlamaktadırlar. Kuantum mekaniği, Yarı-Klasik limitler olan standart kuantum limit ve vuruş gürültüsü sınırı geçmek için çeşitli stratejiler sağlamaktadır. Bilim adamları ve mühendisler, girişim cihazlarının ve konum ölçümlerinin hassasiyetinin artırılması ile başlayarak, sıkıştırma (squeezing) ve dolanıklık kavramlarının etkilerini çok farklı tiplerdeki ölçümlerin doğruluğunu arttırmak amacıyla stratejiler kurgulamaya çalıştılar.

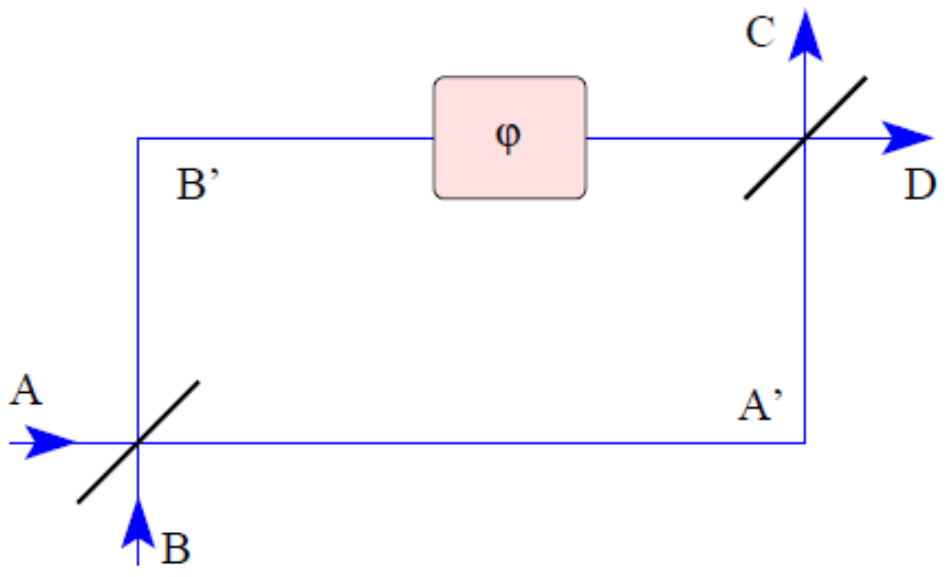

Şekil 5.1- Mach-Zehnder girişim cihazı [33]

Bazı Kuantum tekniklerin uygulanabilirliği halen çok fütüristik olsa da, mevcut durumda, dolanık sistem durumlarının türetilmesi ve üzerinde çeşitli işlemlerin yapılması başlangıç aşamasında olsa da kısmen başarılabilmektedir. Literatürdeki örneklere bakılırsa, kuantum mekaniksel sistemler ölçüm doğruluğu açısında kullanılan N parçacığın karekökü kadar kat daha başarılı sonuçlar vermektedirler. Teknik olarak şu anda N = 5 veya 6 parçacıklı dolanık sistemlerin türetilmesi bile oldukça karmaşıktır. Buna karşın, milyonlarca parçacıktan oluşan klasik sistemlerin oluşturulması ve bunlarla ölçüm yapılması teknik olarak mümkündür. Kuantum Teknolojiler geliştikçe, dolanıklık ve sıkıştırma kavramlarının da ölçüm doğruluğunun ve hassasiyetinin geliştirilmesi konusundaki etkilerinin daha belirgin olacağı aşikardır.

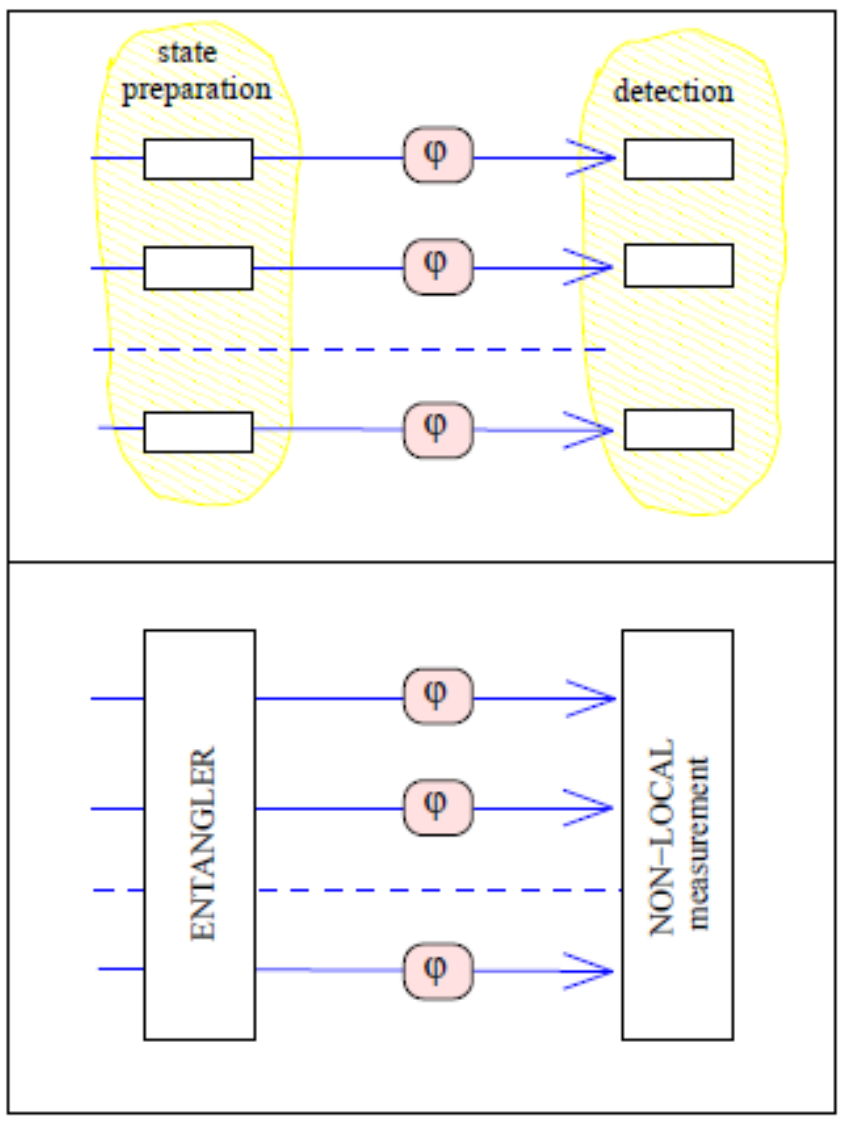

Şekil 5.2- Klasik ve Kuantum Stratejilerin karşılaştırılması [33]

Özellikle uzay-zaman geometrisinin kuantum sistemlerle ölçülmesi konusundaki çalışmaları incelediğimizde bilinen evren ile ilgili gerçekleri anlamamız konusunda kuantum mekaniksel sistemlerin eşsiz araçlar olarak yararlı olması kaçınılmazdır.

Güncel çalışmalarda [33] bilim adamları Kuantum Metroloji altyapısı üzerinde çalışırken tek parametre belirlenmesi işlemleri için bir alt sınır bulduklarını sunmuşlardır. Bu alt sınırın hem üniter (unitary) hem de üniter olmayan (non-unitary) süreçler için her zaman erişilebilir bir sınır olduğunu göstermişlerdir. Gürültü olması durumunda en iyi sistem durumu için esas Kuantum sınırın hesaplanması oldukça zorlu bir işlemdir. Hesaplanacak kaynakların artması durumunda yapılacak olan sayısal analizin zorluğu artmaktadır. Buradaki hesaplamalar ilk sistem durumuna bağlı olmayacak ve olası bütün Kraus gösterimlerine göre optimizasyon yapılmasına gerek kalmayacaktır. Bunun yerine, incelenen sistem durumu için Kraus operatörlerinden fiziksel duruma en uygun olan sınıfın seçilmesi yeterli olacaktır.

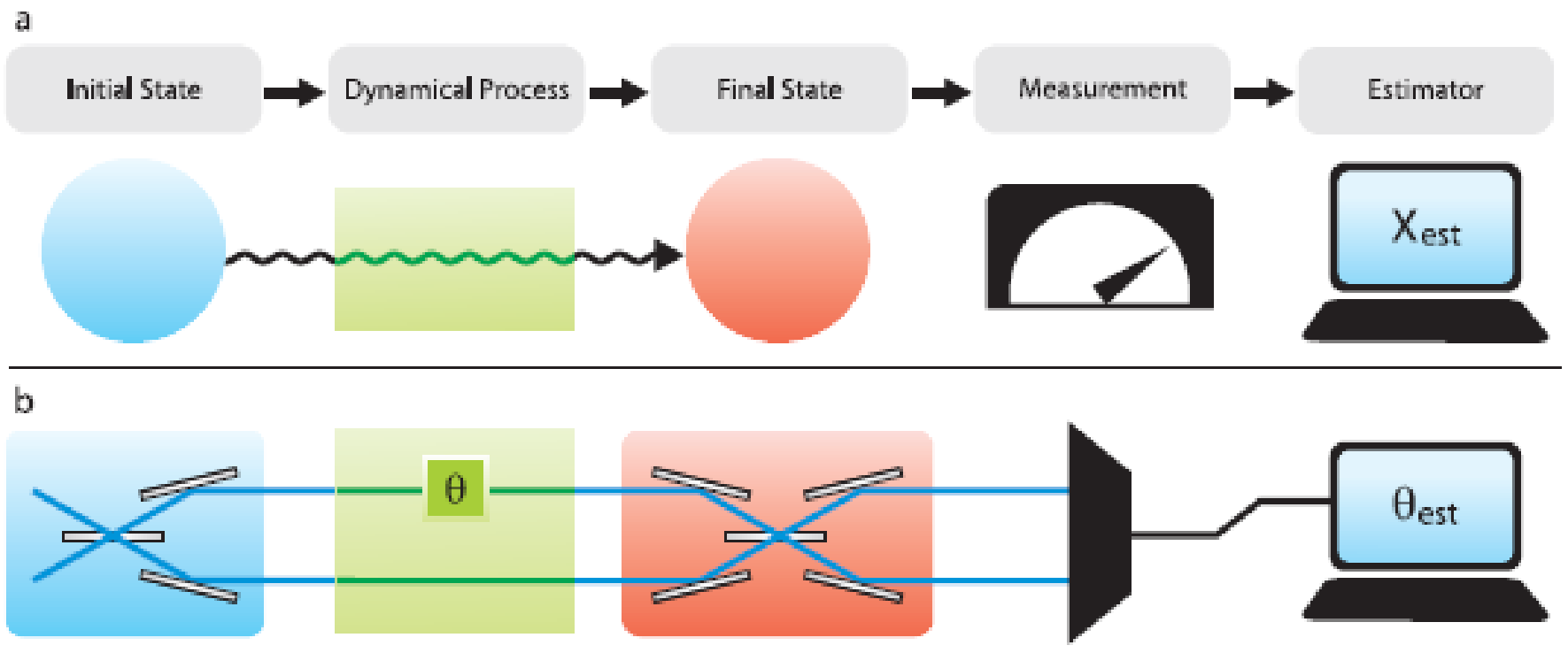

Şekil 5.3- Kuantum Parametre Tahmini için kurulan sistemler [33]

(a) Bilinmeyen bir parametre olan x için rastlantısal bir dinamik sürecin hazırlanışı

(b) Optik girişim aracında ϴ kadar bir faz kaymasının tespiti için kurulan sistem

Burada anlatılan yöntemin gücü Kuantum Metroloji kapsamında tanımlanan iki problem örneği ile de açıklanabilir: Optik Girişim Aracı fazın tahmin edilmesi ve atomik tayf araştırma için dönüşüm frekansının belirlenmesi vb.

Sunulan yöntem, optik girişim aracı uygulamasında, Heisenberg limitinden asimptotik vuruş gürültüsü benzeri davranışa dönüşümü sağlamaktadır. Bunun yanında, bu yöntemin kati bir etkisi de şudur ki: çok az miktarda gürültü olması durumunda bile, vuruş gürültüsü sınırındaki gelişim belli bir çarpan mertebesinde kalmaktadır ve $\frac{1}{\sqrt{N}}$ olan davranışı değiştirmemektedir.

Açıklanan yöntem, uygun olan her yerde ve her türlü problemde oluşabilecek bir durum olan bütünsel olmayan dinamikler ve parametre tahmini için genel geçer olabilecek bir hesaplama aracı olarak açıklanabilmektedir.

## V.3. Kuantum Girişim (Interferometry)

Faz tahmini işlerinde işe yarayabilecek, mümkün olabilecek çok fotonlu Fock sistem durumlarının zerk edilmesiyle oluşturulan çok kapılı cihazların kullanıldığı ölçüm protokolleri teorik olarak oluşturulmuştur. Ulaşılan sonuçlar şu anda var olan gerçek çok kapılı cihazlar üzerinde test edilmiş ve oluşturulması olası çok fotonlu kuantum sistemler üzerinde elde edilmiştir.

Bazı çalışmalarda [15] Fock sistem durumlarını girdi olarak alıp foton-sayma tespitini içeren bir protokol tanımlanmış ve bu protokol ile Standart Kuantum Limitin altına inen ölçümleri simüle edilmiştir[15].

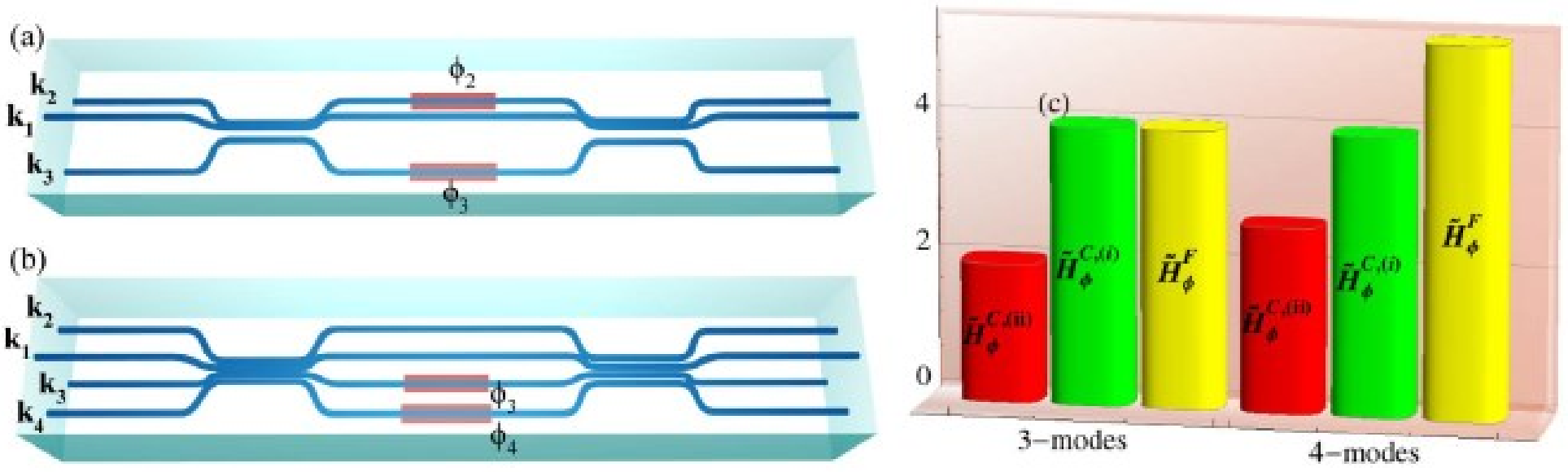

Şekil 5.4- İki adet 3 ve 4 kapılı girişim cihazlarının Kuantum Fisher Bilgisi sonuçları [15]

Diğer güncel çalışmalar iki-modlu girişim araçlarını Kuantum Bilgi Teorisi perspektifinden incelemeyi hedeflemektedir. Çözümlenmeye çalışılan problem, $N$ parçacıklı bütün saf dolanık sistem durumlarının bu girişim araçlarında parçacıklar üzerindeki lokal işlemlerle optimize edildiklerinde alt vuruş gürültüsü hassasiyetine ulaşabilip ulaşamadıklarının tespitidir. Kuantum Fisher Bilgisi $F_Q$ sayesinde optimal hassasiyet üzerinde bir sınır tanımlayan Cramér-Rao teoremi çalışmada kullanılmıştır. $F_Q > N$ için, alt vuruş gürültüsü hassasiyeti bir merkezi limitle erişilebilmektedir.

Mach-Zehnder girişim aracı gibi bir genel iki sistemli lineer girişim aracının maksimum Fisher bilgisi üzerinde çalışılmıştır.

Bu çalışmalarda ilk kez Kuantum Fisher Bilgisinin optimize edilmesi gerektiği üzerine literatüre katkıda bulunulmuştur. Optimizasyon işlemlerinin doğrudan karışık sistem durumları üzerinde ve deneysel olarak yapılabileceği belirtilmiştir[15].

# VI. KUANTUM FISHER BİLGİSİ OPTİMİZASYONU ÖNERİSİ VE DOLANIKLIK ÖLÇÜTLERİ İLE İLİŞKİSİ

Bu kısımda Dolanıklık Ölçütlerinden Eş Zamanlılık, Negatiflik ve Dolanıklığın Göreceli Entropisi seçilmiş ve bu ölçütlerin Kuantum Fisher Bilgisi ile birkaç farklı Kuantum Sistem Durumu kümesi üzerinde karşılıklı analiz çalışmaları yapılmıştır. Öncelikle Kuantum Fisher Bilgisinin Uyum Bozulması Kanalları altında değişimleri gözlemlenmiştir, ikinci olarak bugüne kadar literatürde bulgusu tespit edilememiş olan Kuantum Fisher Bilgisi'nin Dolanıklık Ölçütü gibi davranmasının koşulları iredelenmiş ve bir Optimizasyon Yöntemi Önerisi ile iki kübit Kuantum Sistemler için KFB'nin değerleri Dolanıklık Ölçütleri ile kıyaslanmış ve bu yolda elde edilen önemli sonuçlar burada aktarılmıştır. Üçüncü olarak da kübit-kütrit Kuantum Sistem Durumları için Negatiflik ve Dolanıklığın Göreceli Entropisi ölçütleri kıyaslanmış ve bu sonuçların ilginç bir şekilde iki kübit sistemlerdeki sonuçlarla korelasyon gösterdikleri tespit edilmiştir. Son olarak literatürdeki çalışılabilecek açık konular araştırmacılara aktarılmış ve bu tezin açtığı perspektifte gelecekte yapılabilecek çalışmalar için bazı fikirler aktarılmıştır.

## VI.1. Üç Farklı Uyum Bozulması Kanalındaki Değişimlere Göre Kuantum Fisher Bilgisindeki Değişimlerin Gözlemlenmesi

Güncel çalışmalarda [21] GHZ sistem durumu için maksimal KFB değerinin üç farklı uyum bozulması (decoherence) kanalındaki değişimleri incelenmektedir. Bu kanallar sırasıyla genlik azaltan kanal (Amplitude Damping Channel-ADC), faz azaltan kanal (Phase Damping Channel-PDC) ve depolarize olan kanal (Depolarising Channel-DPC) olarak sıralanır. Ani kayıp durumları hem ölçüm hem de spin sıkıştırmada görülebilmektedir ancak maksimal *KFB* için özel bir durumu vardır. *ADC* için elde edilen değer aşağıdaki grafikte de görülebileceği gibi bir bir $p$ değeri sonrası ani bir artış göstermektedir.

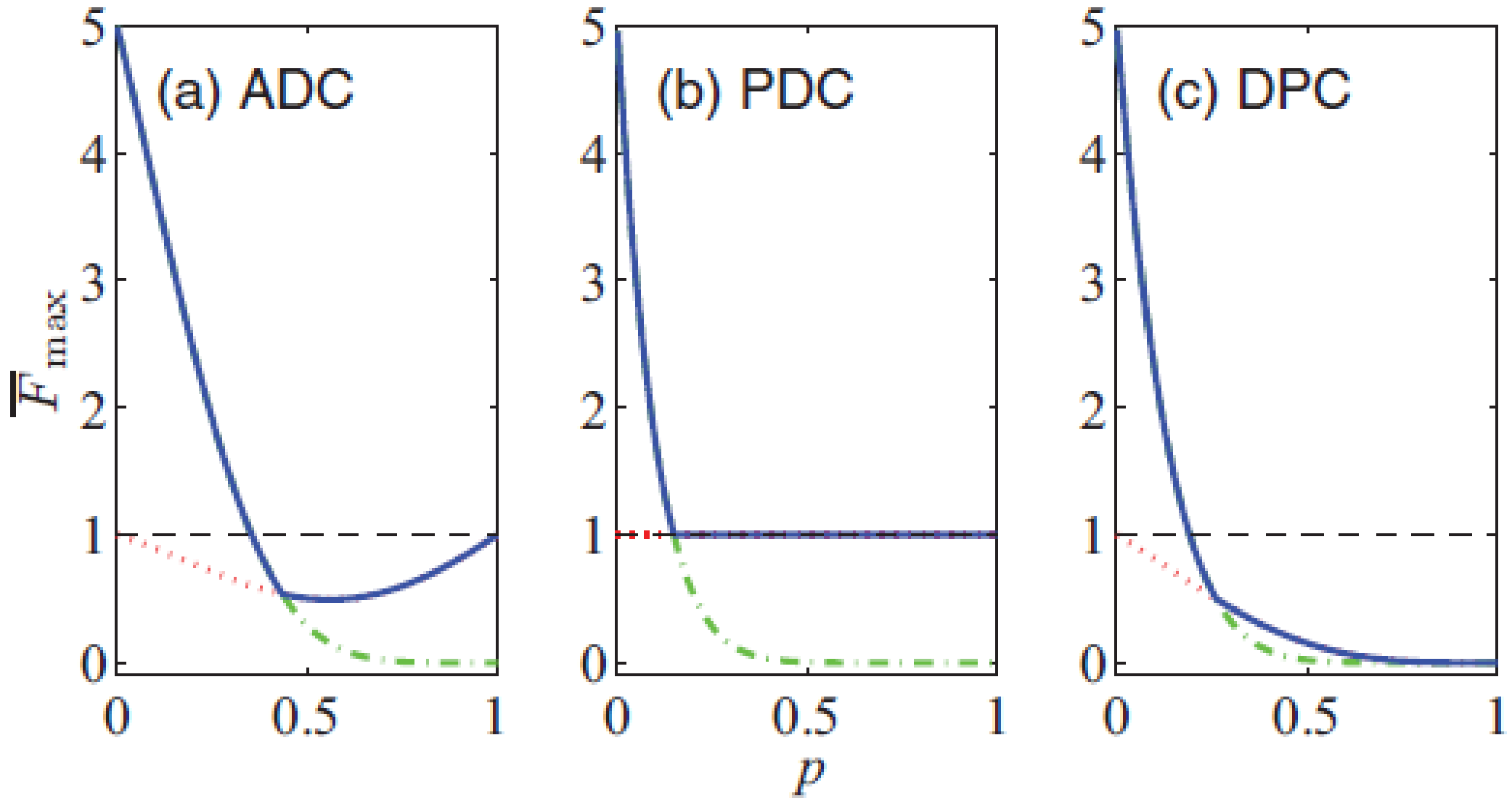

Şekil 6.1 – ADC, PDC ve DPC değerlerinin p'ye göre değişimleri

Başlangıçtaki *GHZ* sistem durumu çiftlik (parity) operatörü kullanılarak, Heisenberg limitinden daha iyi bir parametre tahmin netliği ile kullanılabilmektedir. Gürültü (Decoherence) olması durumunda ve dolanıklığın da kaybıyla sistem durumunun hassasiyeti *SU(2)* rotasyonlarının daha zayıf olması nedeniyle azalmaktadır. Yeterince fazla miktardaki $p$ gürültüsü olması durumunda ADC için KFB değeri belli bir noktadan sonra artmaktadır ve 1'e eşitlenir. PDC için belli bir $p$ gürültüsü değeri sonrası 1'e eşitlenip sabit kalmaktadır. DPC için kırılma olmasına rağmen azalma devam etmektedir. Üstteki şekilde bu durum açıklanmıştır. Bu yayının en önemli katkılarından birisi KFB hesabının açık bir şekilde anlatılmasıdır.

Diğer güncel çalışmalarda [17] $N$ parçacıklı çoklu dolanık $W$ ve *GHZ* sistem durumlarının bir süperpozisyonu olan bir sistem durumu için Kuantum Fisher Bilgisi değerlerini hesaplamışlar ve değerlerdeki değişiklikleri incelemişlerdir. Parçacık başı ortalama (mean) *KFB (RMQFI)* için elde edilen değerlerin parçacık sayısına ters orantılı olarak azaldıklarını ve 0.6 ve 0.8 değerleri arasında keskin bir pik noktası olduğunu bulmuşlardır. *RMQFI* değerinin davranışının $N$'nin 2'den 10'a kadar olan değerlerinin için değişimini vermişlerdir.

Değerlendirilen sistem durumu aşağıdaki denklemler ile tanımlanır.

$$|GHZ_N\rangle = \frac{|0\rangle^{\otimes N} + |1\rangle^{\otimes N}}{\sqrt{2}} \quad (56)$$

$$|GHZ_2\rangle = \frac{|00\rangle + |11\rangle}{\sqrt{2}} \quad (57)$$

$$|W_N\rangle = \frac{1}{\sqrt{N}}\left( \left|0_{(N-1)}\right\rangle|1_1\rangle + \sqrt{N-1}\left|W_{(N-1)}\right\rangle|0_1\rangle\right) \quad (58)$$

$n \geq 2$ $için$ $|0_N\rangle, |0_1 0_2 \dots 0_N\rangle$ ; $|1_N\rangle$ $de,$ $|1_1 1_2 \dots 1_N\rangle$ anlamına gelmektedir.

$$|W_2\rangle = \frac{|01\rangle + |10\rangle}{\sqrt{2}} \quad (59)$$

$$|\psi_N\rangle = \alpha|W_N\rangle + \beta|GHZ_N\rangle \quad (60)$$

Burada $\alpha$ ile $\beta$ arasında bağıntı şu şekildedir:

$$|\alpha|^2 + |\beta|^2 = 1 \quad (61)$$

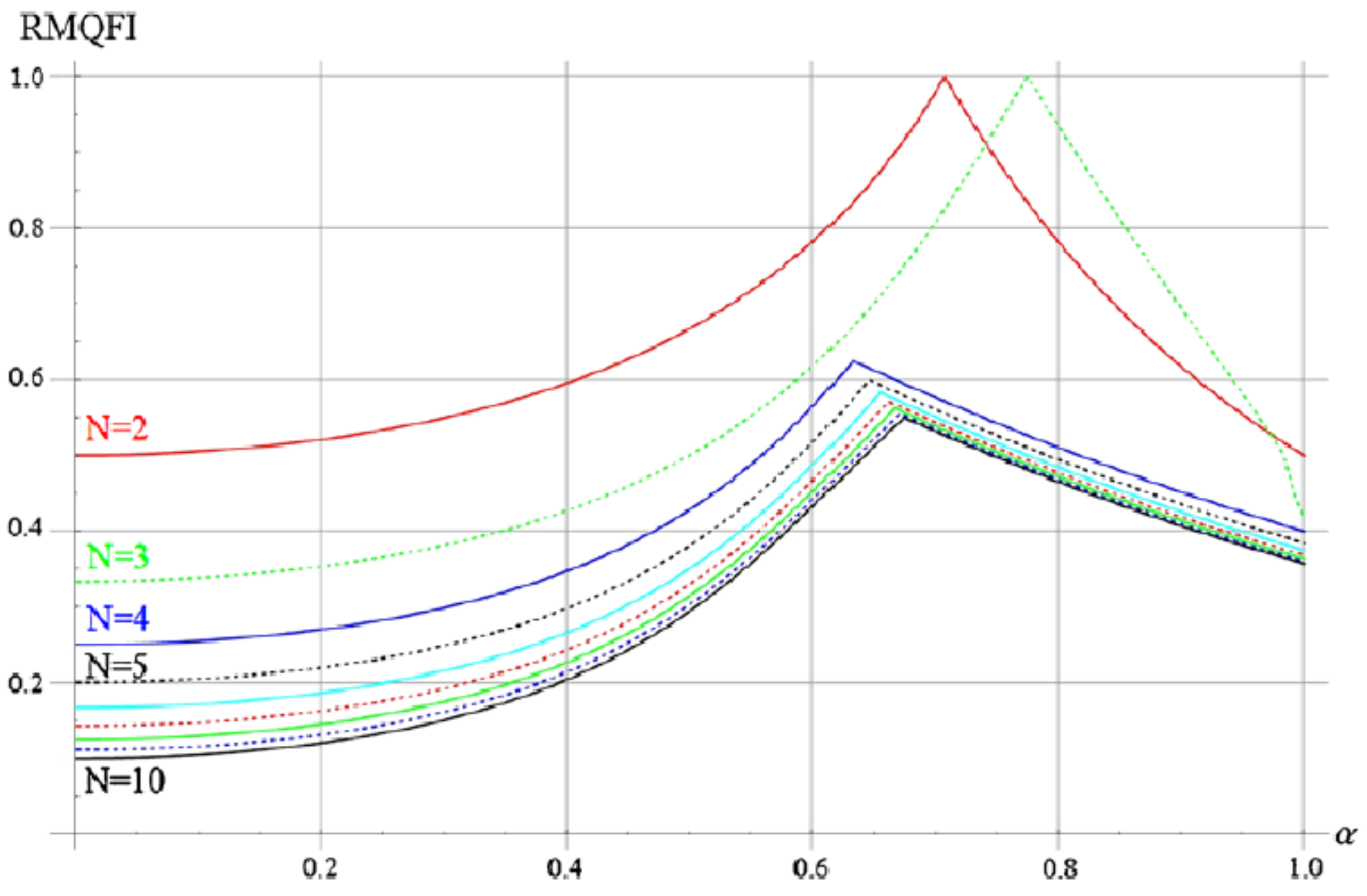

Şekil 6.2 - Açıklaması yapılan değişimler [17] [OzaydinIJTP2013]

Yine güncel çalışmalarda [18] birden fazla parçacıktan oluşan ve *GHZ* ve 2 *W* sistem durumunun süperpozisyonu olan ve rastgele bir göreli faza sahip olan bir sistem durumu için KFB değerlerinin nasıl değiştiğini incelenmiştir. Parçacık sayısının 3'ten 4'e çıktığı durumda KFB değerinde olan olağan dışı değişim not edilebilir. Ayrıca KFB bilgisine olan bağımlılık parçacık sayısı arttıkça azalmıştır. Bahsedilen sistem durumu için göreli fazlardaki değişime göre KFB değerindeki değişim analiz edilmiştir.

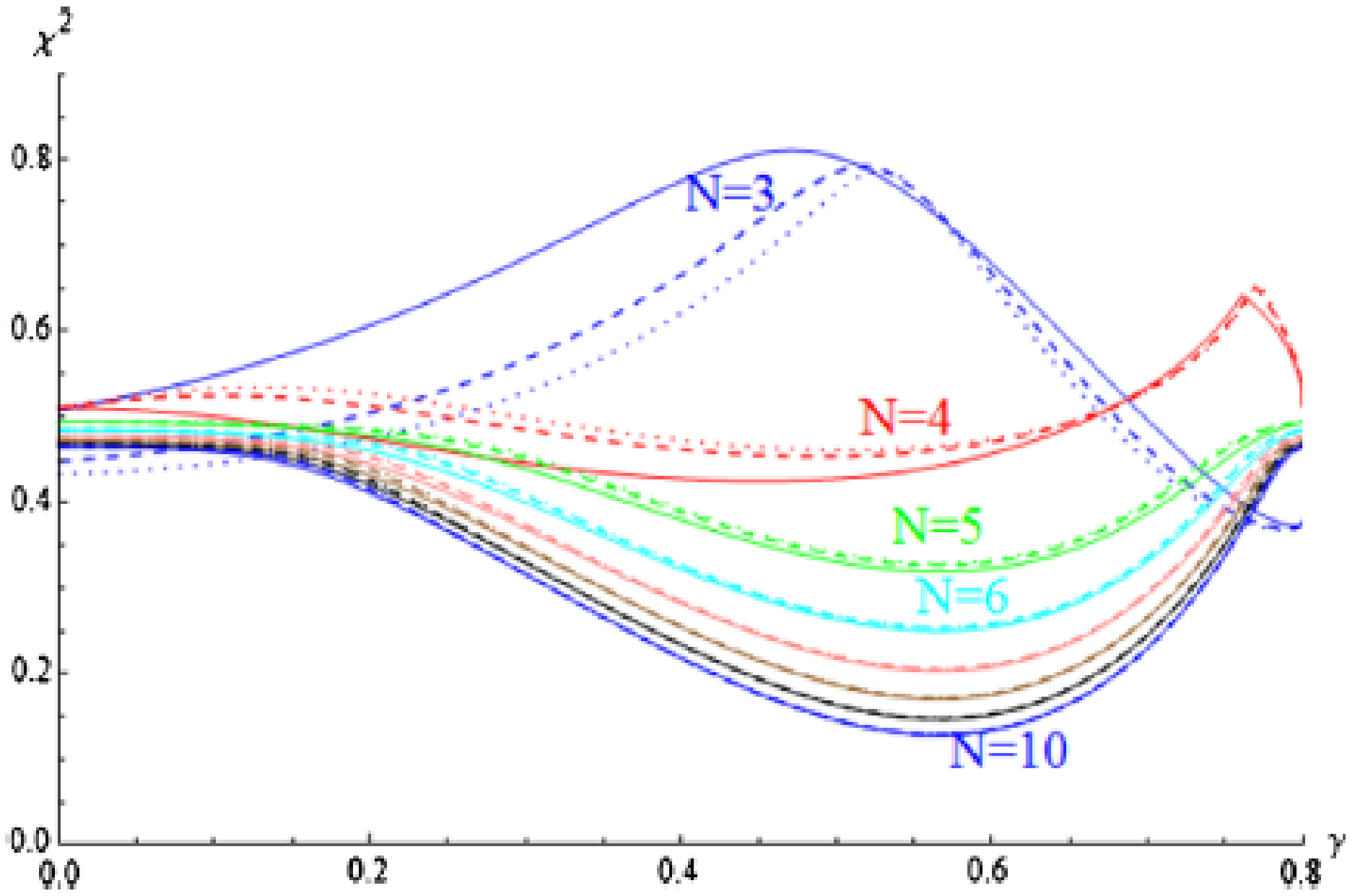

Şekil 6.3 - α = 0.6 değeri için hesaplaması yapılan ve daha önceki paragraflarda açıklanan değişimler [18]

Tez Kapsamında önceki paragraflarda tanımlanan problemlerin benzeri ve devamı niteliğinde şu problem de çözümlenmiştir: *KFB* değerinin değişimi *W* sistem durumlarının bir parçacık kaybetmesi ve uyum bozulmasına maruz kalması durumunda incelenmiştir [55]. Özel olarak, 3 parçacıklı orijinal bir W sistem durumunun ve 3 parçacıklı "W benzeri" sistem durumu olarak isimlendirdiğimiz ve orijinal bir 4 parçacıklı W sistem durumundan 1 parçacığı yoksayarak elde ettiğimiz diğer bir sistem durumununun değişimini inceledik. Bu iki sistem durumundan her parçacığı aynı uyum bozulması kanalına maruz bıraktığımızda genlik azaltan ve depolarize kanal altında orijinal W sistem durumunun diğerinden daha iyi sonuç verdiğini ancak genlik genişleten veya faz azaltan kanal altında ilginç bir şekilde W benzeri sistem durumunun daha iyi sonuç verdiğini tespit ettik.

$p_i$ özdeğerleri ve onlara bağlı $|i\rangle$ özvektörlerine sahip $N$ parçacıklı bir karışık $\rho$ sistem durumunun maksimum ortalama kuantum Fisher bilgisi $\lambda_{max}/N$ değeri ile ifade edilir.

Burada $\lambda_{max}$ değeri aşağıdaki elemanlara simetrik $C$ matrisi ile hesaplanır. Bu hesabın nasıl yapıldığın denklem (54)'te tanımlanmıştır.

$J_n$ açısal momentum operatörleri $\sigma_{x,y,z}$ pauli operatörleri cinsinden n. doğrultuda tanımlanır:

$$J_n = \sum_{\alpha=x,y,z} \frac{1}{2} n_\alpha \sigma_\alpha. \tag{62}$$

Gerçek bir 3 parçacıklı W sistem durumu, $\rho^G = |W_3\rangle\langle W_3|$ şu denklemle verilebilir:

$$|W_3\rangle = \frac{|001\rangle + |010\rangle + |100\rangle}{\sqrt{3}}, \tag{63}$$

Ve W-benzeri 3 parçacıklı system durumu $\rho^L$ 4 parçacıklı orijinal bir W system durumundan 1 parçacık trace out edilerek şu şekilde hesaplanır:

$$\rho^L = \frac{3}{4}|W_3\rangle\langle W_3| + \frac{1}{4}|000\rangle\langle 000|. \tag{64}$$

Bu çalışmada değerlendirilen uyum bozulması kanalları aşağıdaki tabloda verilen Kraus işlemleri ile tanımlanır [55]:

| | | |
|---|---|---|
| Genlik azaltan | $\begin{pmatrix}1 & 0\\ 0 & \sqrt{1-p}\end{pmatrix}, \begin{pmatrix}0 & \sqrt{p}\\ 0 & 0\end{pmatrix}$ | (65) |
| Genlik genişleten | $\begin{pmatrix}\sqrt{1-p} & 0\\ 0 & 1\end{pmatrix}, \begin{pmatrix}0 & 0\\ \sqrt{p} & 0\end{pmatrix}$ | (66) |
| Depolarize kanal | $\begin{pmatrix}\sqrt{1-\frac{3p}{4}} & 0\\ 0 & \sqrt{1-\frac{3p}{4}}\end{pmatrix}, \begin{pmatrix}0 & \frac{\sqrt{p}}{2}\\ \frac{\sqrt{p}}{2} & 0\end{pmatrix}, \begin{pmatrix}0 & -\frac{i\sqrt{p}}{2}\\ \frac{i\sqrt{p}}{2} & 0\end{pmatrix}, \begin{pmatrix}\frac{\sqrt{p}}{2} & 0\\ 0 & -\frac{\sqrt{p}}{2}\end{pmatrix}$ | (67) |
| Phase Damping | $\begin{pmatrix}\sqrt{p} & 0\\ 0 & 0\end{pmatrix}, \begin{pmatrix}0 & 0\\ 0 & \sqrt{p}\end{pmatrix}, \begin{pmatrix}\sqrt{1-p} & 0\\ 0 & \sqrt{1-p}\end{pmatrix}$ | (68) |

$\rho^G$ ve $\rho^L$ sistem durumlarının her kübitine uyum bozulması kanallarının Kraus operatörlerini uyguladığımızda etkilenmiş sistem durumlarının yoğunluk matrislerini elde ederiz. (54) nolu denkleme öz değerleri, onlara bağlı özvektörleri ve açısal momentum operatörlerini koyduğumuzda her uyum bozulması kanalı altındaki KFB değerini hesaplamak çok basit işleme dönüşür. Elde ettiğimiz sonuçlara gore genlik azaltan ve depolarize kanal için $\rho^G$, $\rho^L$'ye gore daha dayanıklıdır; ancak ilginç bir şekilde genlik genişleten ve faz azaltan kanallar için $\rho^L$, $\rho^G$'den daha dayanıklı çıkmıştır. Elde ettiğimiz sonuçları aşağıdaki şekilde bulabilirsiniz:

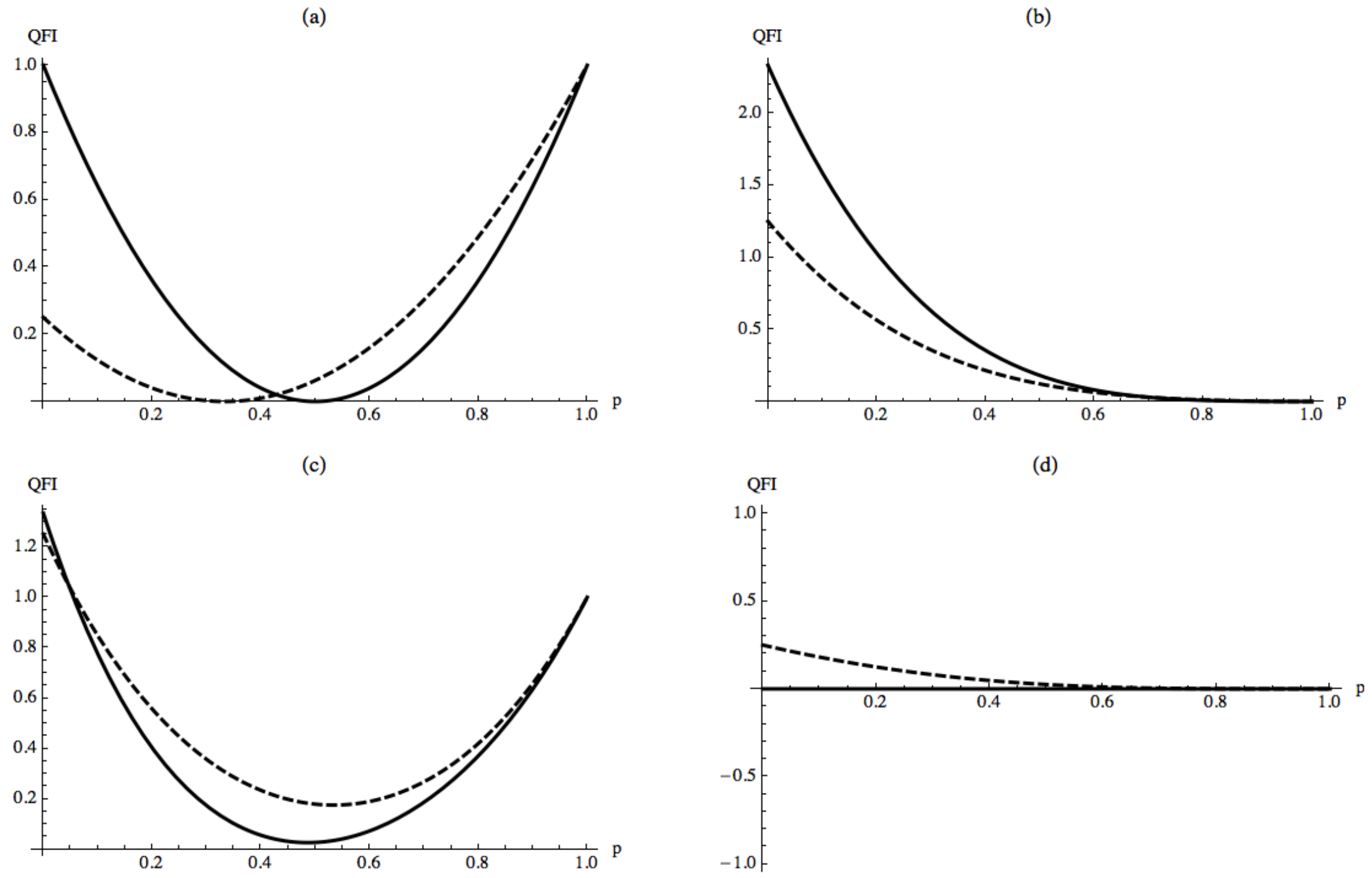

Şekil 6.4- 3 parçacıklı orijinal bir W sistem durumu (düz çizgiler) ve 4 parçacıklı original bir W sistem durumunda 1 parçacığın çıkarılması ile elde edilen 3 parçacıklı W-benzeri sistem durumunun (kesikli çizgiler) (a) genlik azaltan kanal, (b) depolarize kanal, (c) genlik genişleten kanal, (d) faz azaltan kanal altında p uyum bozulması şiddetine göre KFB değerlerindeki değişimler verilmiştir [55]

## VI.2. Sistem Durum Sıralaması Probleminin Tanımlanması

Sistem Durum Sıralaması problemi bir süredir literatürde farklı araştırmacılar tarafından çalışılmış bir konudur [4,35,36,56]. İki spin parçacıkların A ve B isimli alt sisteme bölündüğü ve dolanık olmadığı $\rho$ sistem durumları çarpım halindeki sistem durumlarının konveks toplamı olarak şu şekilde yazılabilirler.

$$\rho = \sum_i p_i \rho_i^A \otimes \rho_i^B \tag{69}$$

Buradan yola çıkarak negatif özdeğer ölçütü ($E_N$) şu şekilde tanımlanabilir.

$$E_N(\rho) = |\min\{0, \lambda_1^{T_B}, \lambda_2^{T_B}, \lambda_3^{T_B}, \lambda_4^{T_B}\}| \tag{70}$$

Buradaki özdeğerler $\rho$'nun hesapsal temelde kısmi transpozesinin özdeğerleridir.

Eisert-Plenio bir çalışmalarında kapsamında [referans], 10 bin adet sistem durumu için Oluşum Entropisi ($E_F$)-Negatif Özdeğer Ölçütü ($E_N$) ve Oluşum Entropisi ($E_F$)-Eş Zamanlılık/Concurrence ($C$) kıyaslamalı grafiklerini aşağıdaki şekilde sunmuşlardır.

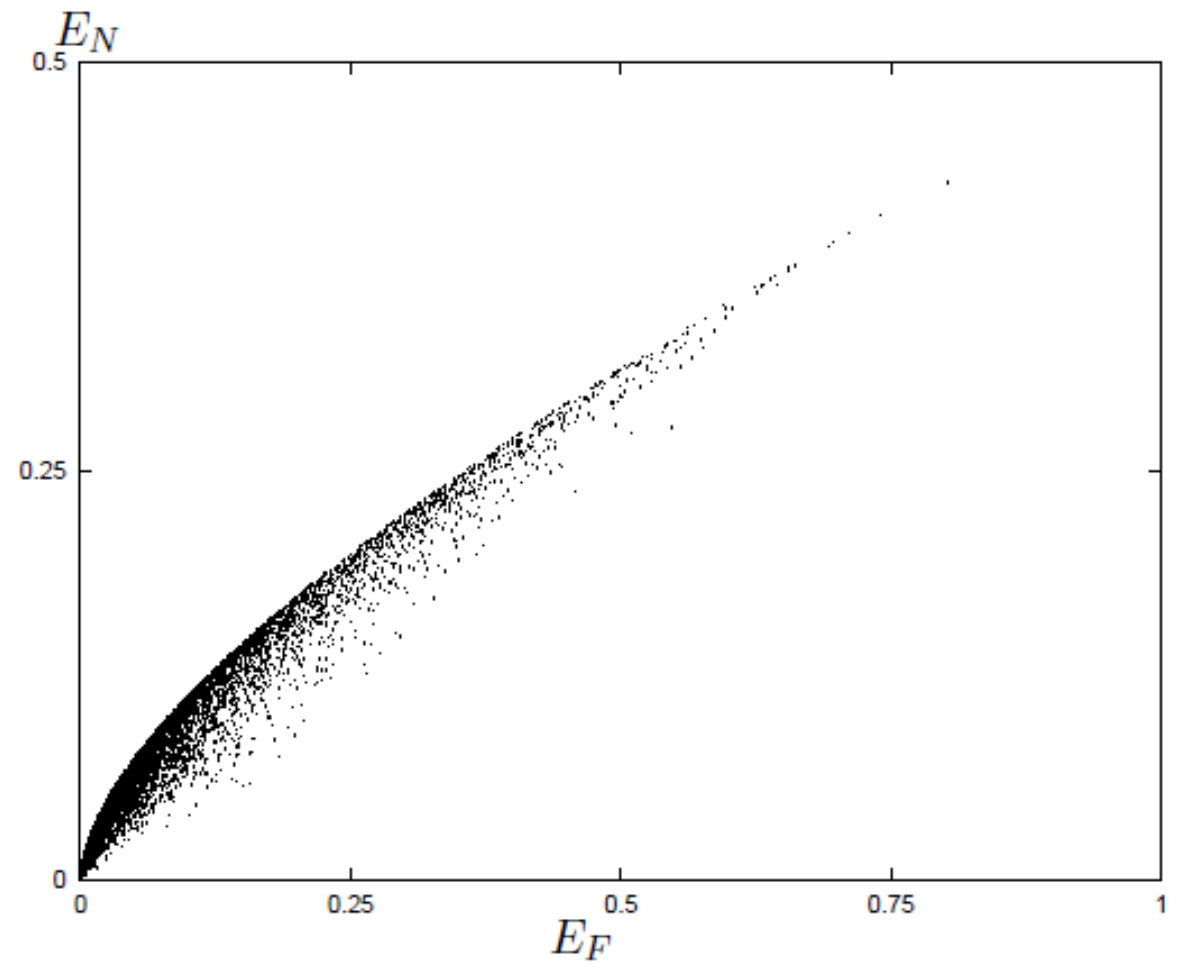

Şekil 6.5 – Negatif Özdeğer Ölçütü ($E_N$)-Oluşum Entropisi ($E_F$) karşılaştırılması [4]

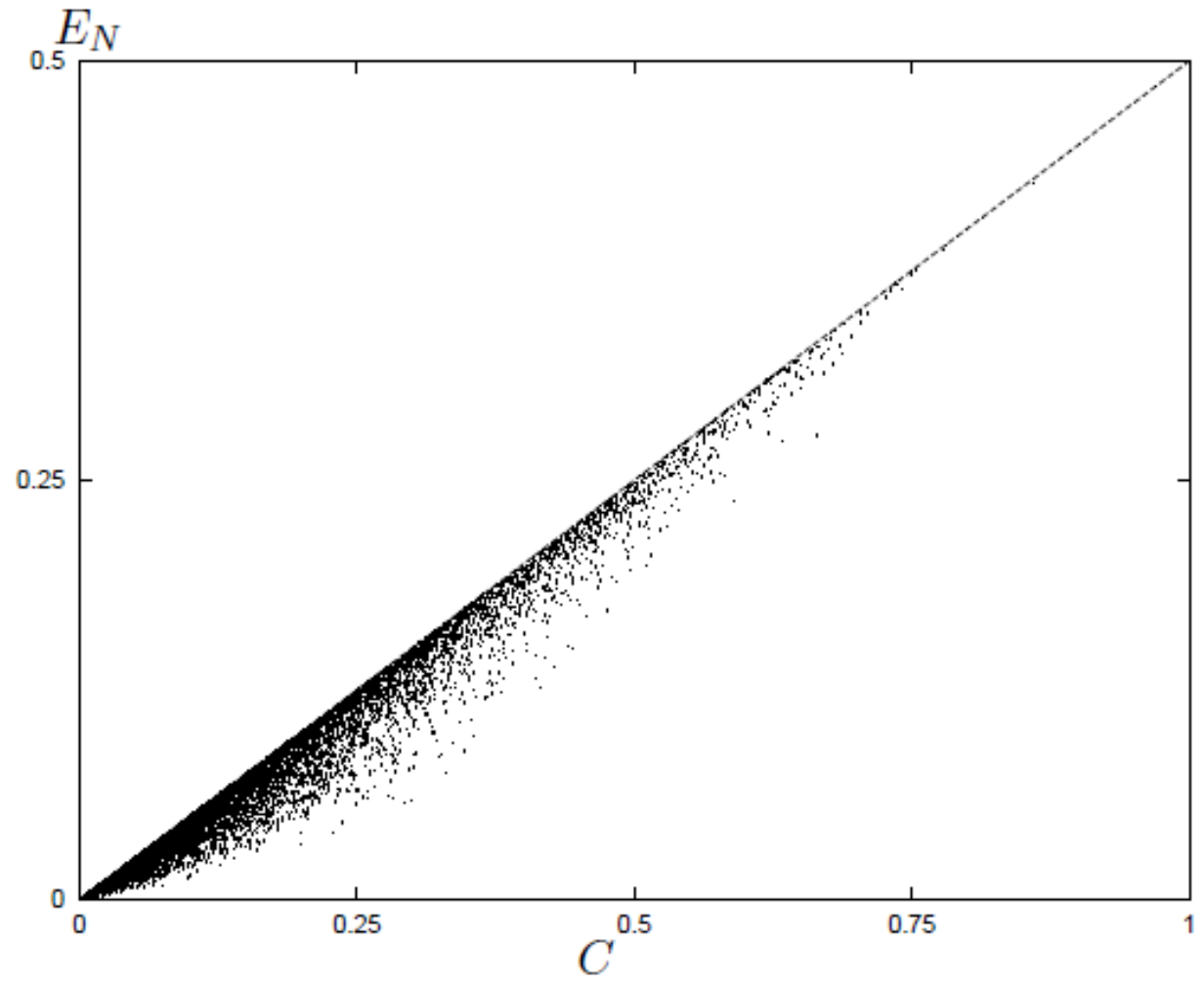

Şekil 6.6 – Negatif Özdeğer Ölçütü ($E_N$)-Eş Zamanlılık (C) karşılaştırılması [4]

Buradaki sonuçlardan yola çıkarak sistem durumlarının bu ölçütlere göre sıralanabilirliği fikri ilk defa ortaya atıldığı için çalışmanın tarihsel önemi büyüktür.

$$E'(\rho_1) < E'(\rho_2) \Leftrightarrow E''(\rho_1) < E''(\rho_2) \quad (71)$$

$E_F$ ve $C$ tanımlarının üzerine sistemin yoğunluk matrisinin kısmi transpozisyonunun özdeğerlerinin negatifliğine ve niceliğine dayanan Negatiflik (Negativity) ölçütü tanımlanmış ve Eisert-Plenio tarafından ortaya atılan sistem durumlarının sıralanması fikri bu ölçütler açısından irdelenmiştir.

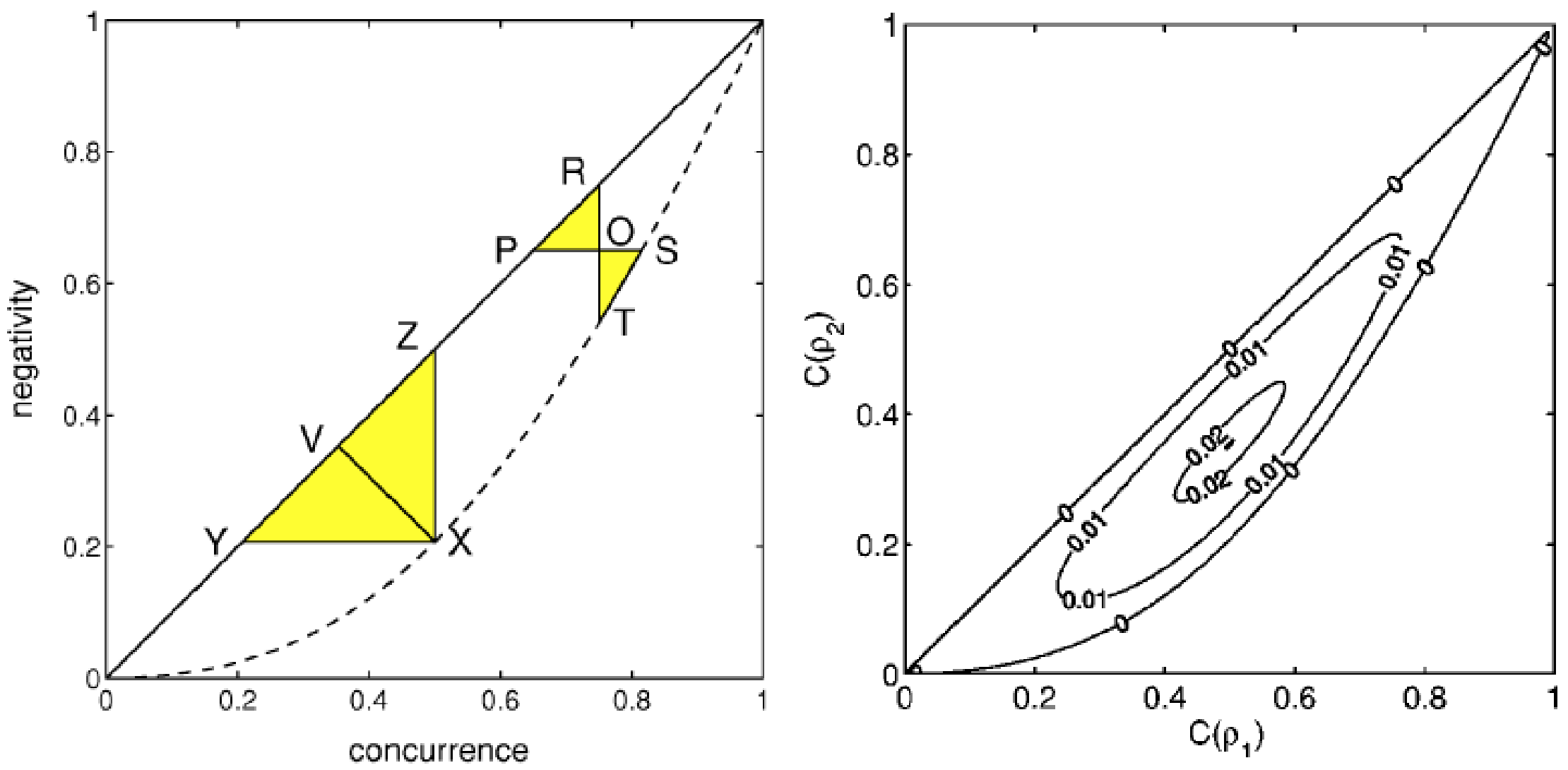

Şekil 6.7- Eş Zamanlılık/Concurrence ve Negatiflik değerlerinin grafiksel gösterimleri[35]

Eş Zamanlılık/Concurrence ve Negatiflik değerlerinin özel olarak irdelendiği taranmış alanların sınıflandırması açıklaması [35]’te detaylıca verilmiştir.

Yine aşağıda verilen grafiklerde ise Negatifliği sabit kaldığı durumlar, Eş Zamanlılığın sabit kaldığı durumlar ve iki ölçütün ters orantılı olduğu durumlar gösterilmiştir.

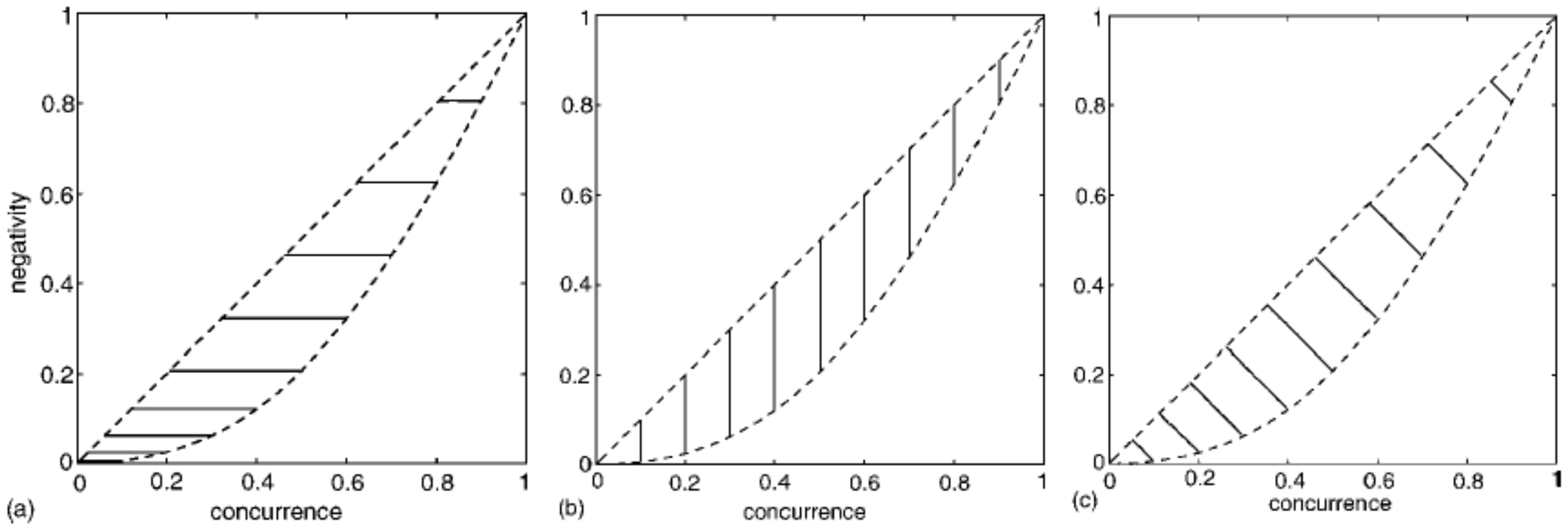

Şekil 6.8 – Negatiflik ölçütünün sabit kaldığı durumlar (a), Eş Zamanlılık ölçütünün sabit kaldığı durumlar (b), iki ölçütün ters orantılı olduğu durumlar [35]

Daha önceki paragraflarda anlatılan çalışmalara ek olarak $E_F$, $C$ tanımlarına ek olarak Dolanıklığın Göreceli Entropisi (Relative Entropy of Entanglement-*REE*) tanımını vererek sonrasında da bu sistem durum sıralama mantığı üçlü olarak geliştiren bir sınıflar kümesi tanımlanabilir.

Dolanıklığın Göreceli Entropisi (REE) sistemin, dolanık olmayan sistemler arasından kendisine en yakın olan sisteme uzaklığına dayanan bir ölçüttür. Matematiksel olarak şu şekilde tanımlanmaktadır: Tüm ayrılabilir sistem durumlarının kümesi olan $D$ içerisindeki σ sistem durumları için Kuantum Göreceli Entropi $S(\rho||\sigma) = Tr(\rho log\rho - \rho log\sigma)$ denklemi ile hesaplanır. Bu durumda REE tanımı şu şekilde olmaktadır:

$$E(\rho) = \min_{\rho \in D} S(\rho \parallel \sigma) = S(\rho \parallel \bar{\sigma}) \qquad (72)$$

Burada $\bar{\sigma}$ , $\rho$'ya en yakın olan sistemi ifade etmektedir.

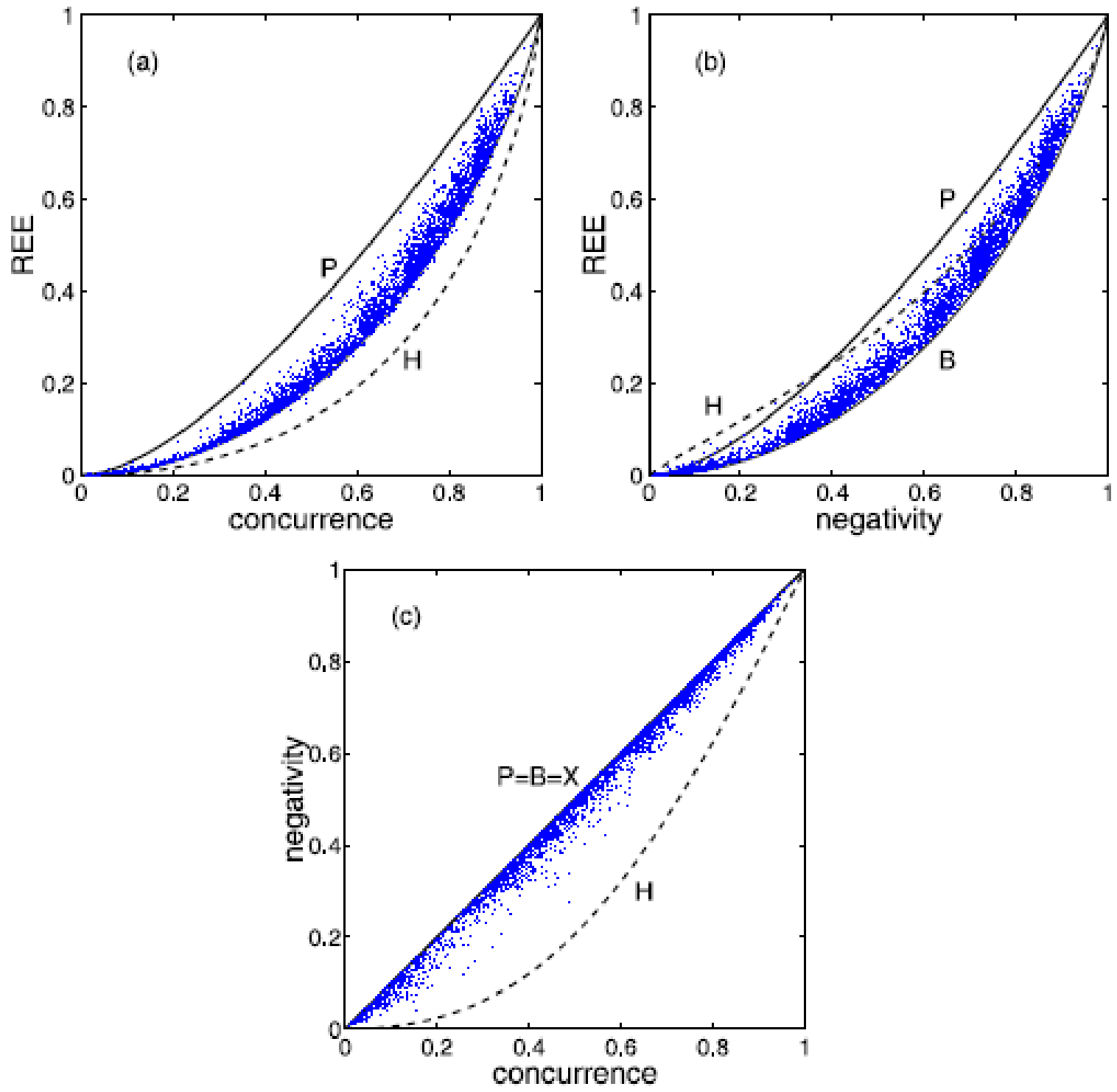

Şekil 6.9 – Yukarıda tanımlanan ölçütlerden Negatiflik, Eş Zamanlılık ve Dolanıklığın Göreceli Entropisi'nin 100bin rastgele sistem durumu için karşılaştırılması [35]

Bu karşılaştırmalara ek olarak aşağıdaki sınıfları bularak sistem durum sıralaması konusunda önemli bulgulara ulaşılmıştır[35,36].

| Class | Concurrences | Negativities | REEs |
|---|---|---|---|
| 1 | $C(\sigma') < C(\sigma'')$, | $N(\sigma') < N(\sigma'')$, | $E(\sigma') < E(\sigma'')$ |
| 2 | $C(\sigma') < C(\sigma'')$, | $N(\sigma') > N(\sigma'')$, | $E(\sigma') < E(\sigma'')$ |
| 3 | $C(\sigma') > C(\sigma'')$, | $N(\sigma') < N(\sigma'')$, | $E(\sigma') < E(\sigma'')$ |
| 4 | $C(\sigma') < C(\sigma'')$, | $N(\sigma') < N(\sigma'')$, | $E(\sigma') > E(\sigma'')$ |
| 5 | $C(\sigma') = C(\sigma'')$, | $N(\sigma') = N(\sigma'')$, | $E(\sigma') = E(\sigma'')$ |
| 6 | $C(\sigma') < C(\sigma'')$, | $N(\sigma') = N(\sigma'')$, | $E(\sigma') < E(\sigma'')$ |
| 7 | $C(\sigma') = C(\sigma'')$, | $N(\sigma') < N(\sigma'')$, | $E(\sigma') < E(\sigma'')$ |
| 8 | $C(\sigma') < C(\sigma'')$, | $N(\sigma') < N(\sigma'')$, | $E(\sigma') = E(\sigma'')$ |
| 9 | $C(\sigma') = C(\sigma'')$, | $N(\sigma') = N(\sigma'')$, | $E(\sigma') < E(\sigma'')$ |
| 10 | $C(\sigma') < C(\sigma'')$, | $N(\sigma') = N(\sigma'')$, | $E(\sigma') = E(\sigma'')$ |
| *11 | $C(\sigma') = C(\sigma'')$, | $N(\sigma') < N(\sigma'')$, | $E(\sigma') = E(\sigma'')$ |
| *12 | $C(\sigma') > C(\sigma'')$, | $N(\sigma') = N(\sigma'')$, | $E(\sigma') < E(\sigma'')$ |
| *13 | $C(\sigma') = C(\sigma'')$, | $N(\sigma') > N(\sigma'')$, | $E(\sigma') < E(\sigma'')$ |
| 14 | $C(\sigma') < C(\sigma'')$, | $N(\sigma') > N(\sigma'')$, | $E(\sigma') = E(\sigma'')$ |

Tablo 6.1 – Elde edilen sınıflar [35]

Buldukları genel sonuca göre, örneğin kimi sistem çiftlerinde bir ölçüt değeri birinci sistemde büyük ikincide küçükken, başka bir ölçütten gelen değer bunun tam tersi davranış sergilemektedir. Bu sonuç, kuantum dolanıklık ölçütlerinin, sistemlerin farklı özelliklerini yansıttığını önermektedir.

## VI.3. LOCC Altında Kuantum Fisher Bilgisi İçin Bir Optimizasyon Yöntemi Önerisi

Bölüm VI.1'deki orijinal bulgularımıza ek olarak yapılan çalışmalar bu kısımda ve bir sonraki kısımda anlatılmıştır. Bölüm VI.2'de tanımlanan Sistem Durum Sıralaması Problemi Tezimiz kapsamında Kuantum Fisher Bilgisi açısından ele alınmıştır. İlk olarak 1000 adet rastgele türetilmiş iki kübit sistemi simüle edilmiştir. Bu sistemlerin her biri için ayrı ayrı, Eş Zamanlılık dolanıklık ölçütleri ve Kuantum Fisher Bilgisi değerleri hesaplanmıştır. Bu niceliklere dair elde edilen karşılaştırma Şekil 6.10'de gösterilmiştir.

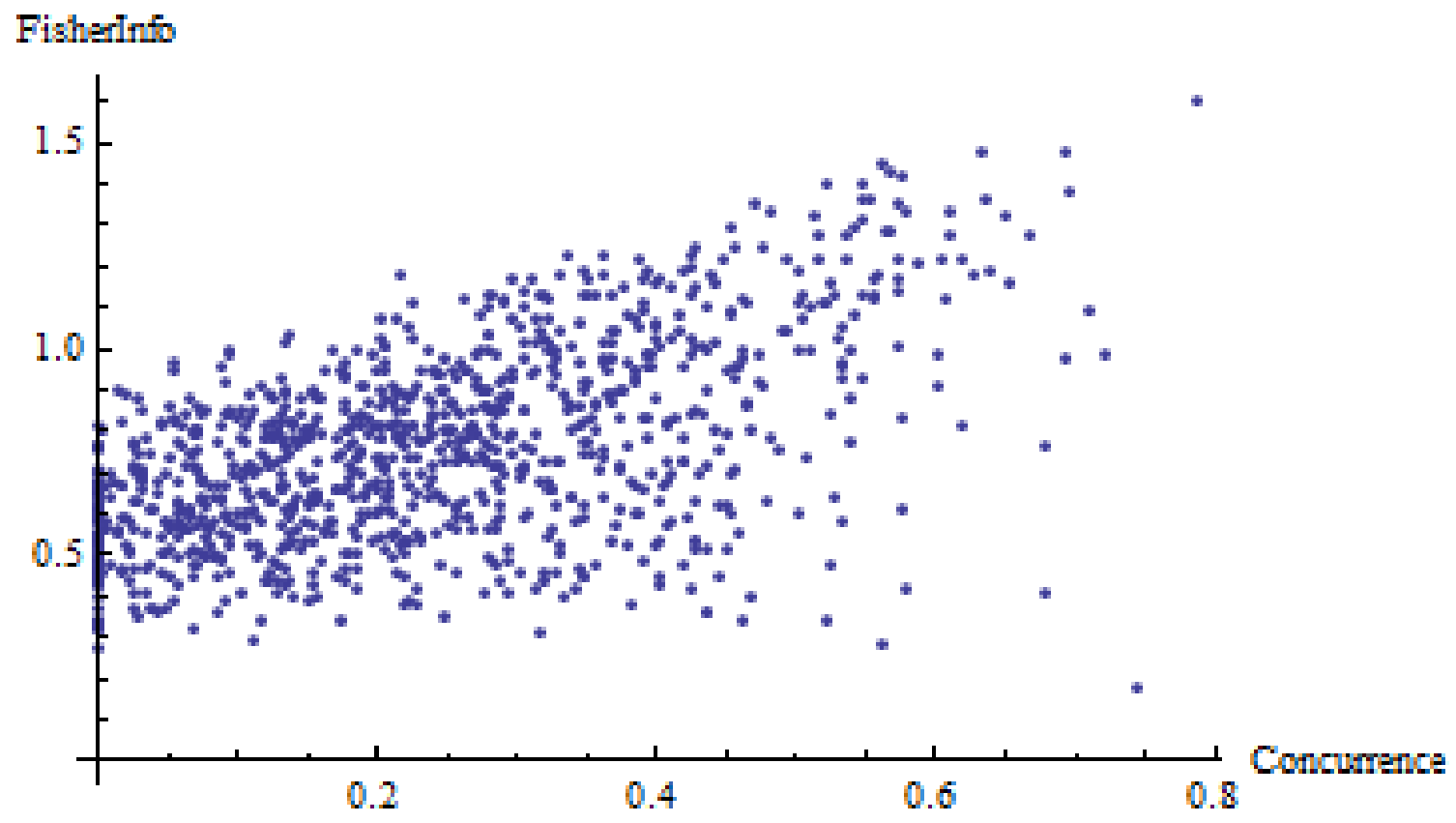

Şekil 6.10-1000 adet rastgele türetilmiş iki kübit sistem durumlarının Eş Zamanlılık dolanıklık ölçütü ve Kuantum Fisher Bilgisi değerlerinin karşılaştırılması [57].

Burada elde ettiğimiz sonuçlara göre Eş Zamanlılık değeri 0 olup Kuantum Fisher Bilgisi değeri sıfırdan farklı olan birçok durum olduğu gözlemlenmiştir.

Yine benzer şekilde Eş Zamanlılık değeri Kuantum Fisher Bilgisi değerinden büyük veya tam tersi Kuantum Fisher Bilgisi değeri Eş Zamanlılık değerinden büyük olan çok sayıda durum olduğu gözlemlenmiştir. Bu durum aşağıdaki tabloda özetlenmiş ve elde edilen sonuçlara göre sistem durumların sıralanması durumunda üç sınıf olduğu söylenebilir.

| Sınıf | Kuantum Fisher Bilgisi – Eş Zamanlılık Karşılaştırması |
|---|---|
| 1 | *F(S1)* > 0, *C(S1)* = 0 |
| 2 | *F(S2)* < *C(S2)* |
| 3 | *F(S3)* > *C(S3)* |

Tablo 6.2- Kuantum Fisher Bilgisi – Eş Zamanlılık Karşılaştırma Tablosu [57]

Özellikle 1. sınıfın keşfi, yalnızca [35,36] çerçevesinde değil, aynı zamanda Kuantum Fisher Bilgisinin dolanıklık kavramıyla olan ilişkisine dair, kuantum bilgi teorisi ve kuantum haberleşme açısından oldukça önemlidir.

Diğer çalışmalarımızda [57,58] KFB değerinin sistem durum sıralaması problemi açısından anlamlı olması için Lokal Operasyon Klasik Kanal'a (LOCC) göre Maksimize edilmesi gerektiğini bulduk. Bu çalışmadaki optimizasyon (maksimizasyon) prosedürü şu şekilde çalışmaktadır.

Her kübit için Euler gösterimine göre genel rotasyonlar uyguladık. Genel rotasyonları Hilbert uzayında bütün aralıkları tarayabilmek amacıyla önce *X*, sonra *Z* ve en son *X* eksenine göre uyguladık. Buna göre rotasyon denklemi şu şekilde tanımlanmaktadır:

$$U_{Rot}(\alpha,\beta,\gamma) = U_x(\alpha)U_y(\beta)U_x(\gamma) \tag{73}$$

Buradaki rotasyonlar akslara göre şu şekilde tanımlanmaktadır

$$U_j = \exp\left(-i\alpha\frac{\sigma_j}{2}\right), j \in \{x, z\} \tag{74}$$

Her kübit için her üç rotasyon açısında $[0, 2\pi]$ arasında $\theta$ derece aralığında rotasyonlar uygulayarak KFB değerini hesaplattık. Bu durumda her kübit için $O(\left(\frac{2\pi}{\theta}\right)^6)$ mertebesinde KFB hesabı yapılması gerekmektedir. $\theta = \frac{\pi}{2}$ lik adımların 1000 rastgele sistem durumdan %98'i için istediğimiz oranda optimizasyon yapılmasını sağladığını farkettik. Kalan yüzde %2 için rotasyon açısını π/3'e düşürdük ve bütün sistemler için KFB değerinin optimize olduğunu gördük.

KFB optimizasyon prosedürünün pseudokodunu şu şekilde bulabilirsiniz:

```
{SubtimizedQFI, QFI, OptimizedQFI} = OptimizeQFI (ρ, α)
Begin
QFI = OptimizedQFI = SubtimizedQFI = QFI(ρ)
For (a = 0; a < 2 π; a = a + α)
   For (b = 0; b < 2 π; b = b + α)
         For (c = 0; c < 2 π; c = c + α)
         Rotation1 = URot(a, b, c)
         For (d = 0; d < 2 π; d = d + α)
               For (e = 0; e < 2 π; e = e + α)
                     For (f = 0; f < 2 π; f = f + α)
                           Rotation2 = URot(d, e, f)
                           Rotation =
                                 KroneckerProduct(Rotation1,Rotation2)
                           Rotated ρ = Rotation.ρ. Rotation†
                           QFIRotated ρ = QFI(Rotated ρ)
                           If [QFIRotatedρ > OptimizedQFI]
                                 OptimizedQFI = QFIRotated ρ
                           If [QFIRotatedρ < SubtimizedQFI]
                                 SubtimizedQFI = QFIRotated ρ
                           End
                     End
               End
         End
   End
```

Burada *URot(a,b,c) = UR(σ, a).UR(β, b).UR(σ, c)* olarak tanımlanır ve *UR(σ, a) σ* matrisinin -*ia σ/2* ile üstel çarpımıdır. (Euler rotasyonunun psödokoddaki gösterimi)

Buradaki algoritmanın karmaşıklığı $O(N^6)$ mertebesindedir.

Burada yapılan işlemleri şu şekilde sıralayabiliriz:

- Çalışma kapsamında öncelikle bin adet iki kübitlik rastgele sistem durumu türettik ve bunlar 625'inin ayrılabilir sistem durumu diğerlerinin ise dolanık olduğunu [4-59]'un sonuçlarıyla da uyumlu olarak not ettik.
- Daha sonra her sistem durumu için dolanıklık ölçütleri ve maksimize KFB değerleri hesaplanmıştır, $\{\rho_1, \rho_2\}$ sistem durumları için elde edilen sıralama ilişkileri Tablo 3'te listelenmiştir.
- 625 ayrılabilir sistem durumu için bütün dolanıklık ölçütlerinin değerleri 0'a eşittir ve bunları *Dolanıklık*($\rho_1$)=*Dolanıklık*($\rho_2$)=0 durumunda gruplamış olduk.
- Dolanık olan sistem durumları için Eş Zamanlılık, Dolanıklığın Göreceli Entropisi ve Negatiflik dolanıklık ölçütleri için üç farklı olası sıralama sınıfı için gösterimleri yine Tablo 6.3'te yapılmıştır.

Şekil 6.11'de Eş Zamanlılık, Negatiflik ve Dolanıklığın Göreceli Entropisi sonuçlarının Maksimize KFB (kırmızı noktalar), KFB (mavi noktalar) ve Minimize KFB (yeşil noktalar) değerleri ile sistem durum sıralamalarını görebilirsiniz.

Buradan elde edilen en temel sonuç KFB değerlerinin maksimize edilmesi durumunda özellikle REE değerleri için oldukça anlamlı sıralama ilişkilerinin elde edilebildiği gerçeğidir. Bu sonuca gore LOCC'ye gore maksimize edilmiş KFB değerlerinin dolanıklık açısından anlamlı olduğu ve sistem durum sıralaması problem için daha kullanışlıdır. Tablo 3'te bu sıralama ilişkileri sistematik bir şekilde listelenmiştir.

Buradan elde edilen sıralama sonuçlarından yola çıkarak özellikle çoklu dolanık sistemlerin incelenmesi durumunda çok daha ilginç sonuçlar elde edilebileceği öngörülmektedir.

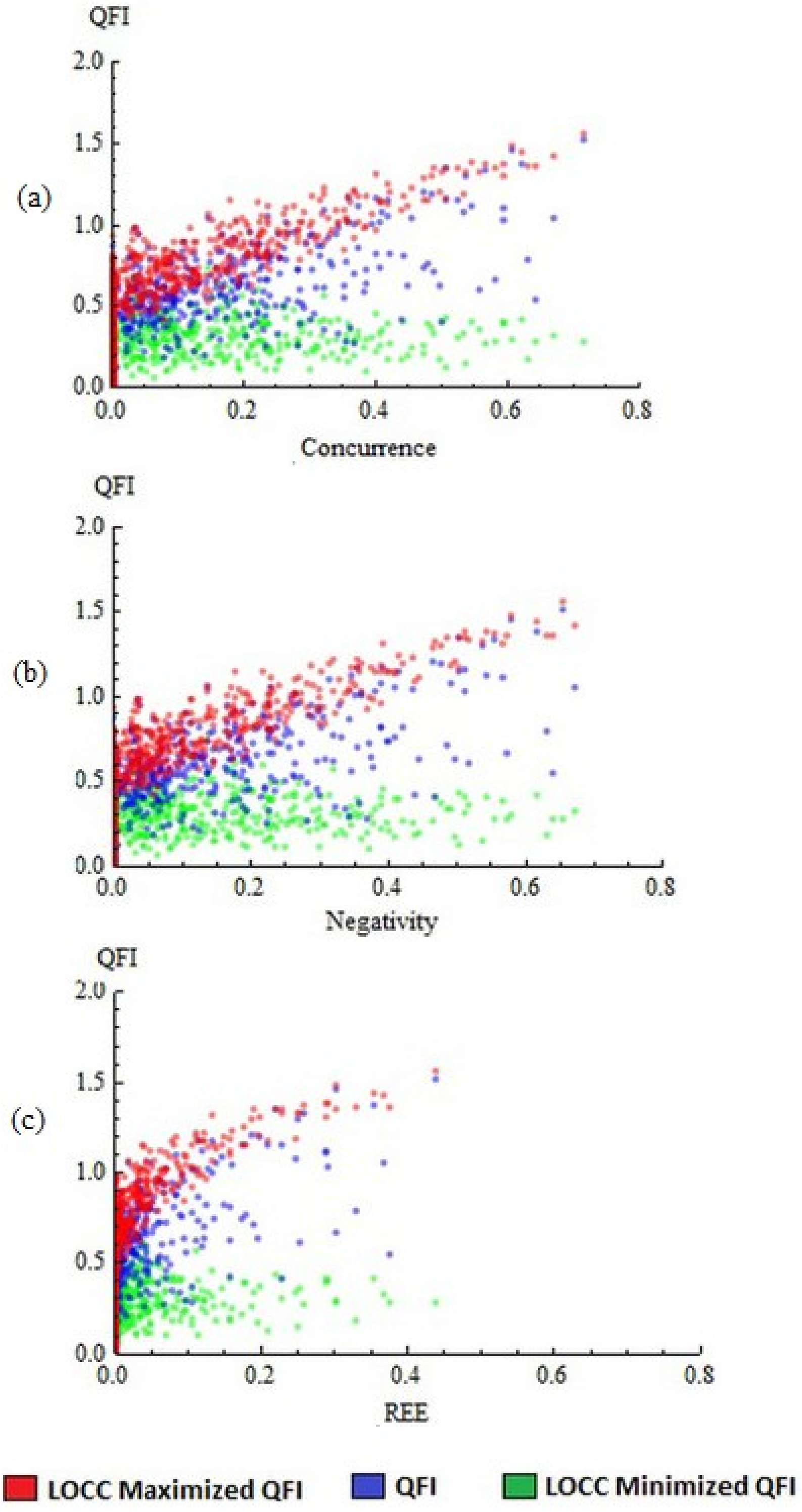

Şekil 6.11 – LOCC Maksimize KFB (QFI) değerlerinin (kırmızı), KFB değerlerinin (mavi), LOCC Minimize KFB değerlerinin üç dolanıklık ölçütü (a) Eş Zamanlılık, (b) Negatiflik, (c) Dolanıklığın Göreceli Entropisi ile karşılaştırılması [58]

| Sınıf | Maksimize KFB değeriyle olan kıyaslama |
|---|---|
| Dolanıklık$(\rho_1)$ = Dolanıklık$(\rho_2) = 0$ | $MKFB\ (\rho_1) > MKFB\ (\rho_2)$ |
| | $MKFB\ (\rho_1) = MKFB\ (\rho_2)$ |
| | $MKFB\ (\rho_1) < MKFB\ (\rho_2)$ |
| Dolanıklık$(\rho_2)$ > Dolanıklık$(\rho_1) = 0$ | $MKFB\ (\rho_1) > MKFB\ (\rho_2)$ |
| | $MKFB\ (\rho_1) = MKFB\ (\rho_2)$ |
| | $MKFB\ (\rho_1) < MKFB\ (\rho_2)$ |
| Dolanıklık$(\rho_1)$ = Dolanıklık$(\rho_2) > 0$ | $MKFB\ (\rho_1) > MKFB\ (\rho_2)$ |
| | $MKFB\ (\rho_1) = MKFB\ (\rho_2)$ |
| | $MKFB\ (\rho_1) < MKFB\ (\rho_2)$ |
| Dolanıklık$(\rho_1)$ < Dolanıklık$(\rho_2) = 0$ | $MKFB\ (\rho_1) > MKFB\ (\rho_2)$ |
| | $MKFB\ (\rho_1) = MKFB\ (\rho_2)$ |
| | $MKFB\ (\rho_1) < MKFB\ (\rho_2)$ |

Tablo 6.3- İki kübit sistem durumları için Maksimize KFB değerlerinin sıralama ilişkileri

## VI.4 İki Kübit ve Kübit-Kütrit Sistem Durumları için Dolanıklık Ölçütlerinin Analizinde Elde Ettiğimiz Diğer Sonuçlar

Bu kısımda daha önce verilmiş olan dolanıklık ölçütü tanımlara ek olarak iki kübit sistemler için eş zamanlılık ölçütünün bir türevi olan spektral orbit üzerindeki maksimal eş zamanlılık tanımını şu şekilde verebiliriz [60] :

$$C_{max}(\rho) = \max\{0, \lambda_1 - \lambda_2 - 2\sqrt{\lambda_2\lambda_4}\} \qquad (75)$$

Burada özdeğerler şu şekilde sıralanmaktadır:

$$\lambda_1 \geq \lambda_2 \geq \lambda_3 \geq \lambda_4 \qquad (76)$$

İki Kübit sistemler için elde edilen diğer sonuçlarımız Şekil 6.12-6.16'da gösterilmiştir. Bunların sistem durum sıralaması açısından anlamı da Tablolarla açıklanmıştır.

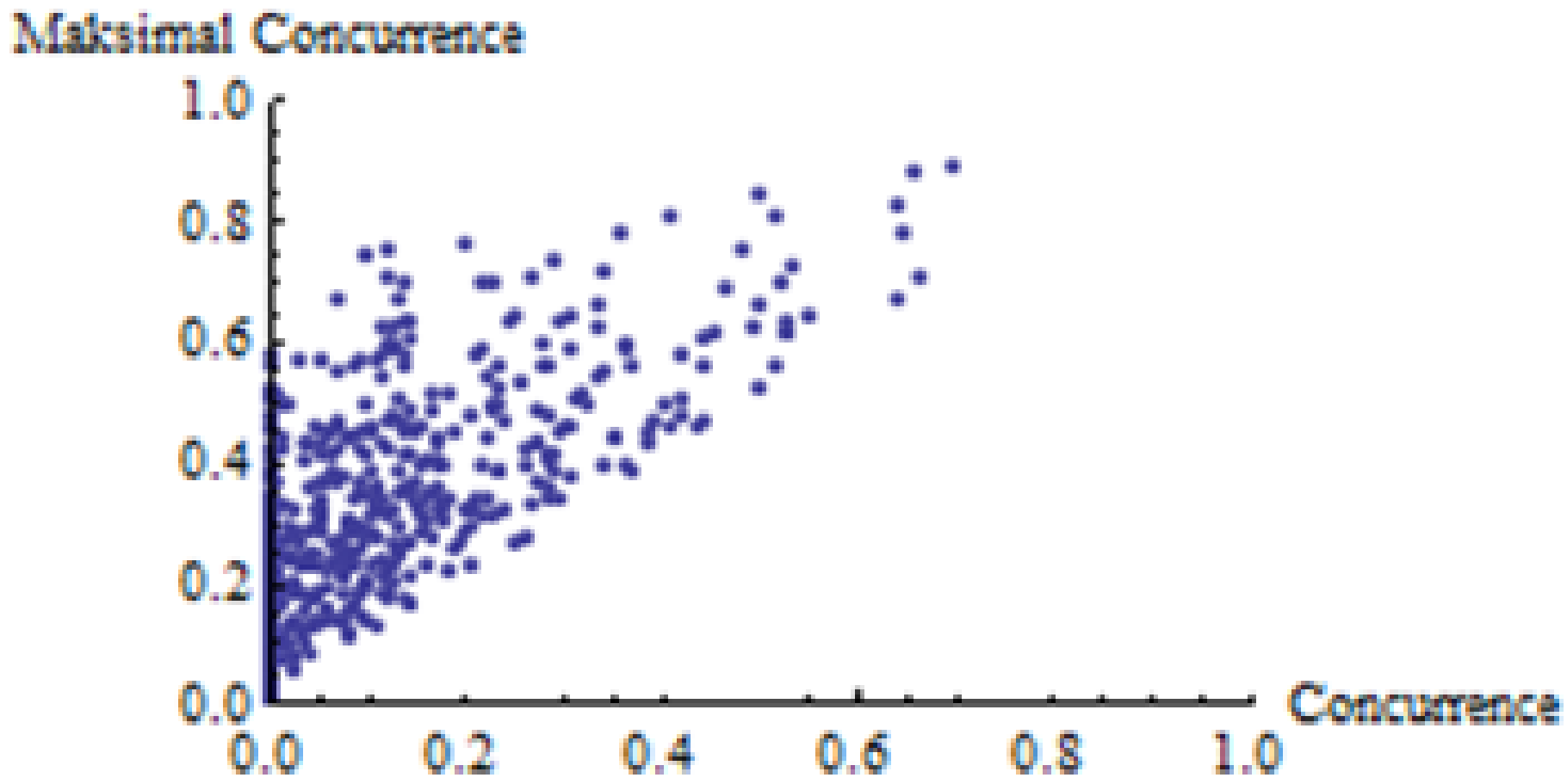

Şekil 6.12- Maksimal Eş Zamanlılık sonuçlarının karşılaştırılması

| Sınıf | Eş Zamanlılık – Maksimal Eş Zamanlılık Karşılaştırması |
|---|---|
| 1 | $C_{max}(S1) > 0$, $C(S1) = 0$ |
| 2 | $C_{max}(S2) > C(S2)$ |

Tablo 6.4 – Şekil 6.12'deki karşılaştırma sonucunda elde edilen sınıflar

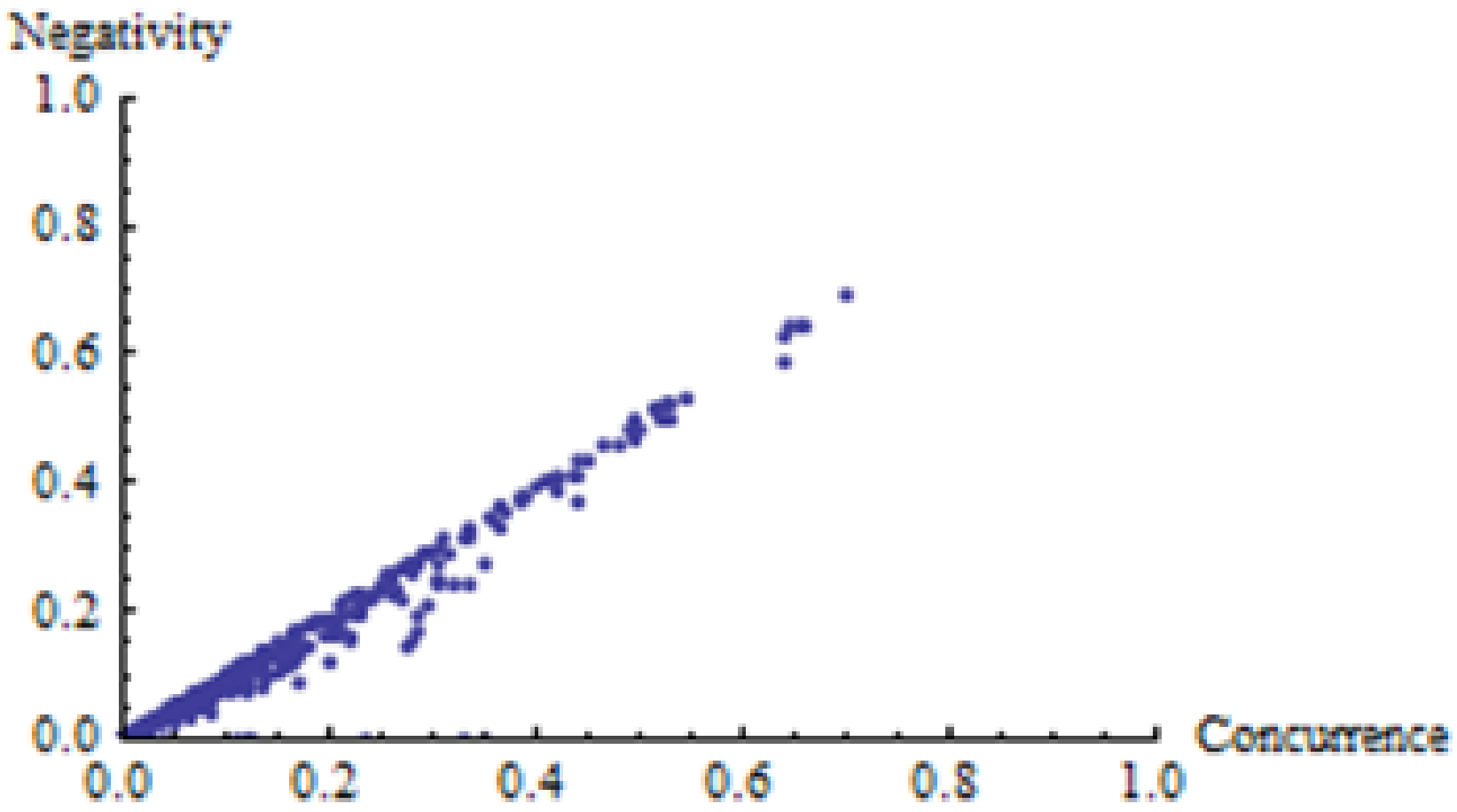

Şekil 6.13-İki Kübit 1000 adet rastgele türetilmiş sistem durumu için Eş Zamanlılık ve Negatiflik sonuçlarının karşılaştırılması

| **Sınıf** | **Eş Zamanlılık – Negatiflik Karşılaştırması** |
|---|---|
| 1 | *C(S1) > N(S1)* |

Tablo 6.5 – Şekil 6.13'deki karşılaştırma sonucunda elde edilen sınıf

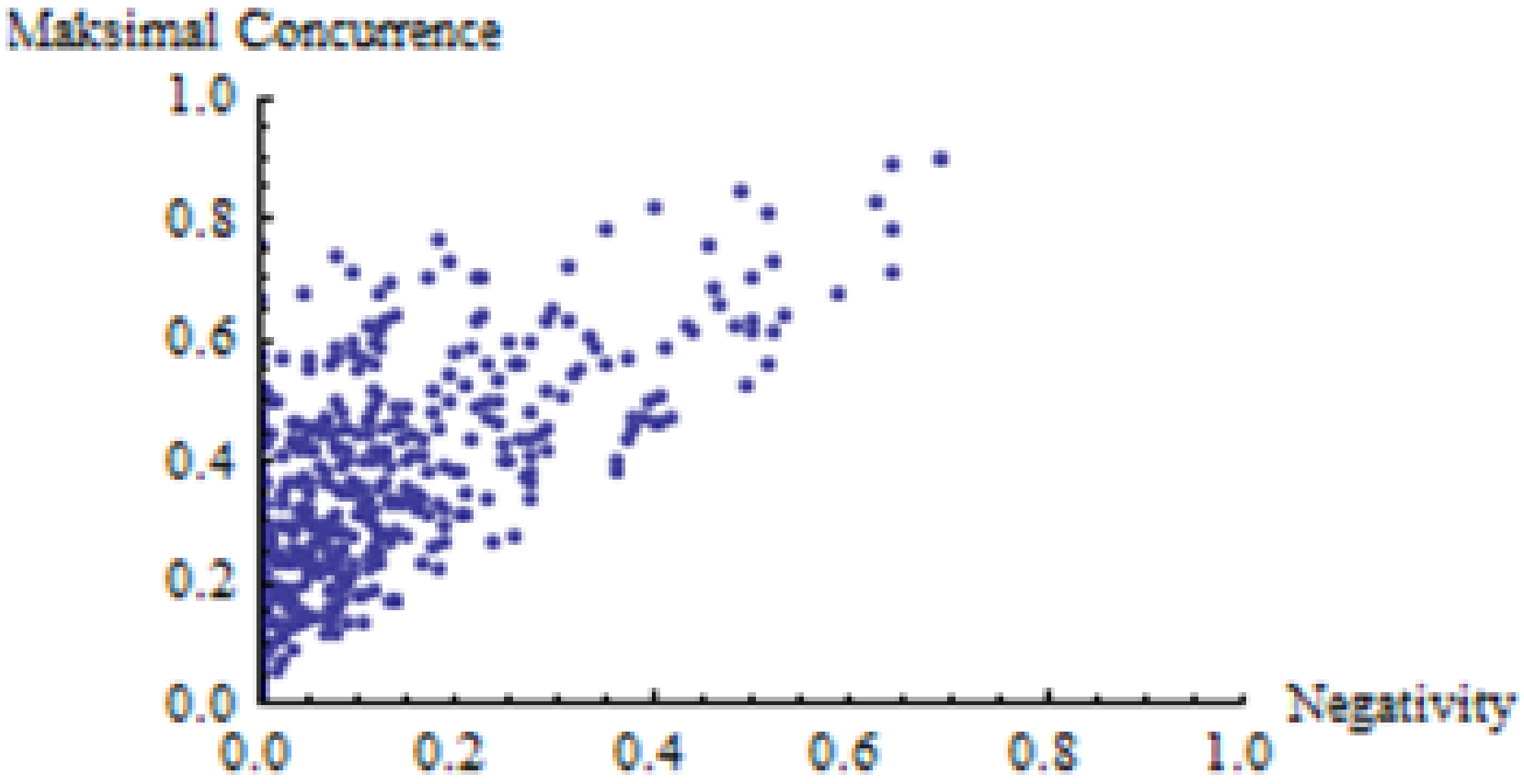

Şekil 6.14-İki Kübit 1000 adet rastgele türetilmiş sistem durumu için Negatiflik ve Maksimal Eş Zamanlılık sonuçlarının karşılaştırılması

| **Sınıf** | **Maksimal Eş Zamanlılık – Negatiflik Karşılaştırması** |
|---|---|
| 1 | $C_{max}(S1) > 0, N(S1) = 0$ |
| 2 | $C_{max}(S2) > N(S2)$ |

Tablo 6.6- Şekil 6.14'teki karşılaştırma sonucunda elde edilen sınıflar

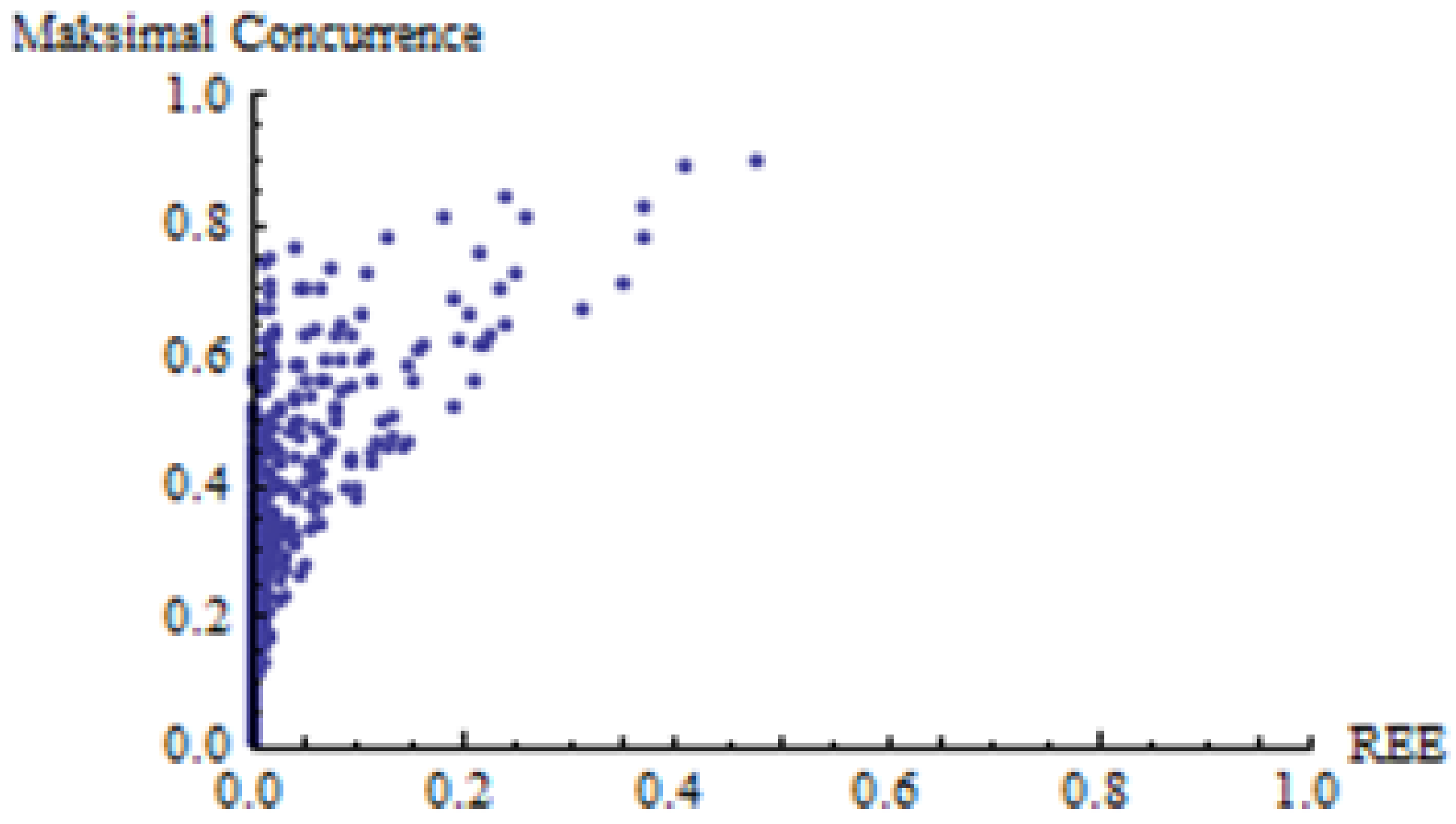

Şekil 6.15-İki Kübit 1000 adet rastgele türetilmiş sistem durumu için Dolanıklığın Göreceli Entropisi (REE) ve Maksimal Eş Zamanlılık sonuçlarının karşılaştırılması

| Sınıf | Maksimal Eş Zamanlılık – REE Karşılaştırması |
|---|---|
| 1 | $C_{max}$ *(S1)* > 0, *REE(S1)* = 0 |
| 2 | $C_{max}$ *(S2)* > *REE(S2)* |

Tablo 6.7 – Şekil 6.15'teki karşılaştırma sonucunda elde edilen sınıflar

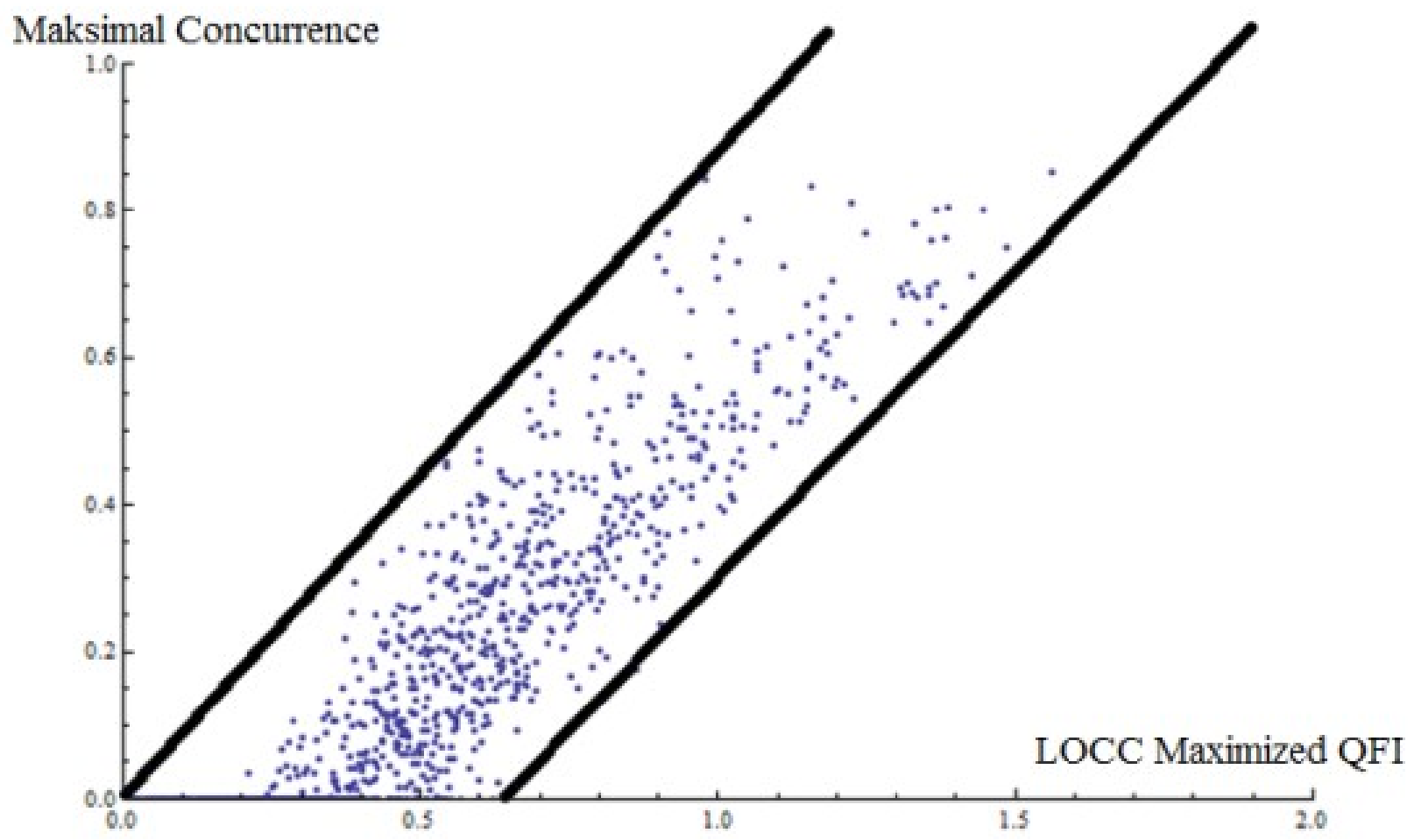

Şekil 6.16-1000 rastgele iki kübit sistem durumu için LOCC Maksimize KFB – Maksimal Eş Zamanlılık karşılaştırılması

Şekil 35'te LOCC maksimize edilmiş KFB ve Maksimal Eş Zamanlılık değerlerinin karşılaştırılması yapılmıştır. Elde edilen grafiğe göre bu 1000 rastgale sistem durumu için değerlerin grafikte $y = x$ ve $y = 0.65 + x$ doğruları arasında dağıldığını görmekteyiz. Bu sonuca göre karşılaştırılması yapılan değerlerin birbirlerine benzerliği aşikardır.

LOCC Maksimizasyonu ile elde ettiğimiz değerlerin KFB değerlerini dolanıklığın ölçümü açısından çok daha anlamlı hale getirdiğini bu sonuçlardan gözlemleyebilmekteyiz.

Bu bölümde 1000 adet iki kübit ve 1000 adet kübit-kütrit sistem durumları için Negatiflik ve Dolanıklığı Göreceli Entropisi değerleri hesaplanmış ve sonuçlar karşılaştırmalı olarak gösterilmiştir[61].

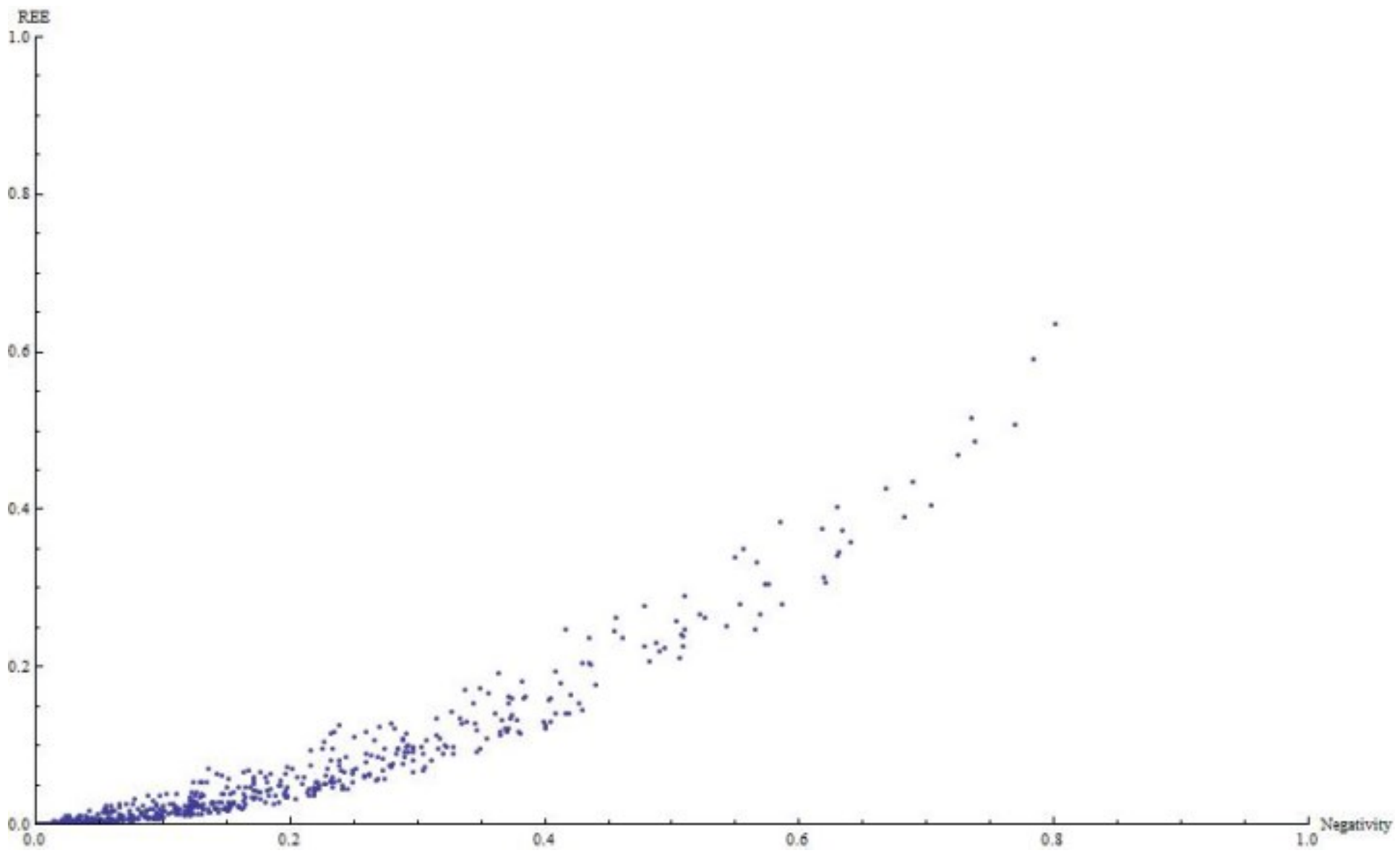

Şekil 6.17- 1000 adet iki kübitlik rastgele sistem durumu için Negatiflik ve REE sonuçlarının karşılaştırılması [61]

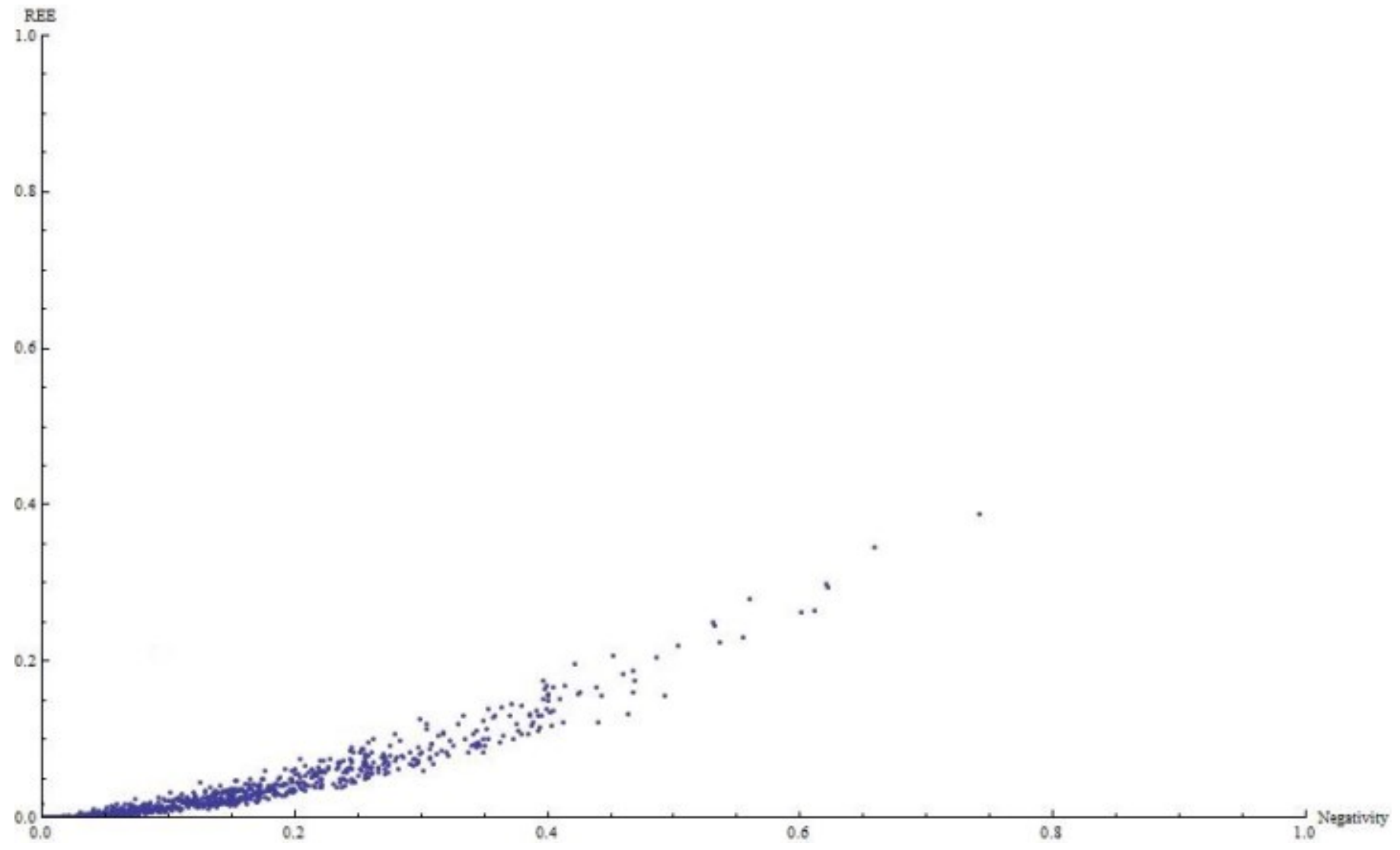

Şekil 6.18- 1000 adet kübit-kütrit rastgele sistem durumu için Negatiflik ve REE sonuçlarının karşılaştırılması [61]

| **Sınıf** | **Negatiflik – Dolanıklığı Göreceli Entropisi Karşılaştırılması** |
|---|---|
| 1 | *N(S1) > REE(S1)* |

Tablo 6.8- Kübit-Kütrit Sistem Durumları İçin Negatiflik – Dolanıklığı Göreceli Entropisi Karşılaştırılması

Yapılan karşılaştırma sonucunda reel yoğunluk matrislerine sahip iki kübit ve kübit-kütrit sistem durumlarının genel olarak Negatiflik değerlerinin REE değerlerinden daha büyük olduğu tespit edilmiştir.

Her iki durum için de neredeyse birebir aynı grafiğin çıkmış olması da sistem durum sıralaması için oldukça ilginç bir sonuçtur. Bu durumda sıralama problemi için yalnızca bir sınıf olduğu tespit edilmiştir. Bu sonuç, yalnızca [35,36] çerçevesinde değil, aynı zamanda kuantum bilgi teorisi ve kuantum haberleşme açısından oldukça önemlidir.

Gelecekteki çalışmalarda iki-kütrit veya daha büyük iki taraflı dolanık sistemler incelenirse sınırda dolanık (bound entangled) sistem durumlarının da tespit edilebileceği düşünülerek çok daha farklı karşılaştırma grafikleri olacağı düşünülmektedir.

İki kütrit sistemlerde Negatiflik ölçütü için kapalı formül mevcut olmasına rağmen yine REE ölçütü için kapalı bir formül olmadığı gibi bir kestirim metodu da mevcut değildir. Açık bir araştırma olarak bu konu isteyen araştırmacılar tarafından çalışılabilir.

## VI.5. Elde Edilen Sonuçların Genel Analizi

Kuantum Fisher Bilgisi tek başına bir dolanıklık ölçütü değildir ancak özellikle faz hassasiyeti gerektiren durumlar için bize çok önemli öngörüler elde etmemiz konusunda fikir vermektedir. Bu konudaki çalışmalarda açık alan olarak mevcut araştırmalarda Fisher Bilgisi'nin optimize edilmeden kullanıldığı tespit edilmiştir. Bu Tez çalışması kapsamında Lokal Operasyon Klasik Kanal kullanarak bir optimizasyon/maksimizasyon prosedürü KFB üzerinde uygulanmış ve elde edilen sonuçlardan yola çıkarak dolanıklık ölçütleri ile karşılaştırılmalı analizi yapılmıştır. Yine *Sistem Durum Sıralaması* problemi çerçevesinde sıralama ilişkileri üzerinde detaylı olarak durulmuştur.

Buradan elde edilen en temel sonuç KFB değerlerinin maksimize edilmesi durumunda özellikle REE değerleri için oldukça anlamlı sıralama ilişkilerinin elde edilebildiği gerçeğidir. Bu sonuca gore *LOCC'ye gore maksimize edilmiş KFB* değerlerinin dolanıklık açısından anlamlı olduğu ve sistem durum sıralaması problem için daha kullanışlı olduğudur. Buradan elde edilen sıralama sonuçlarının özellikle çoklu dolanık sistemlerin incelenmesi durumunda çok daha ilginç sonuçlar elde edilebileceği tahmin edilmektedir. Özellikle gelecek çalışmalara iki kütrit sistemler incelenerek başlanabileceğini düşünmekteyiz. Son bölümde de bahsedildiği gibi iki kütrit sistemlerde Negatiflik ölçütü için kapalı formül mevcut olmasına rağmen yine REE ölçütü için kapalı bir formül olmadığı gibi bir kestirim metodu da mevcut değildir. Açık bir araştırma olarak bu konu isteyen araştırmacılar tarafından çalışılabilir.

# VII. SONUÇ

Kuantum Bilgisayarların üretilmesi ile beraber o mimaride çözülmesi öngörülen benzer ve daha karmaşık büyük veri problemlerinin çözümünde bu bilginin kuantum sistemler için geliştirilmiş türü olan kuantum Fisher bilgisinin aktif olarak kullanılacağı öngörülmektedir.

Kuantum kriptografi, haberleşme, bilgisayarı gibi temel kuantum teknolojilerindeki birçok işte, GHZ, W gibi çok taraflı kuantum dolanık (multi-partite entangled) sistemlere ihtiyaç vardır. Faz hassasiyeti gerektiren işlerde, klasik sistemlerden daha iyi faz hassasiyeti verebilmesi için, bir kuantum sistemin sahip olması gereken niteliklerin başında, çok taraflı dolanık bir sistem olması gerekmektedir, ancak bu nitelik tek başına yeterli değildir. Sistemin Fisher bilgisinin hesaplanması ve belli bir seviyeyi tutturduğu da kanıtlanmalıdır.

Kuantum sistemlerde kullanılan hesaplama yöntemlerinin temeli dolanıklık kavramına dayanmaktadır. Kuantum dolanıklığı ölçmenin birçok yöntemi bulunmaktadır ve bunlara dolanıklık monotonları veya daha özel bir ifadesi ile dolanıklık ölçütleri denmektedir. Tez kapsamında dolanıklık ölçütleri ayrı bir bölümde detaylı olarak incelenmiştir. Bu kapsamda ağırlıklı olarak analiz edilen dolanıklık ölçütleri Eş Zamanlılık (Concurrence), Negatiflik (Negativity) ve Dolanıklığın Göreceli Entropisi (Relative Entropi of Entanglement) ölçütleridir. Bu ölçütler hem iki seviyeli (two kübit) ve kübit-kütrit kuantum sistemler açısından incelenmiş ve bu değerler hesaplanarak *Sistem Durum Sıralaması* problemi çerçevesinde irdelenmiştir.

Kuantum Fisher Bilgisi tek başına bir dolanıklık ölçütü değildir ancak özellikle faz hassasiyeti gerektiren durumlar için bize çok önemli yansımalar sağlamaktadır. Bu konudaki çalışmalarda açık alan olarak mevcut araştırmalarda Fisher Bilgisi'nin optimize edilmeden kullanıldığı tespit edilmiştir. Bu Tez çalışması kapsamında Lokal Operasyon Klasik Kanal kullanarak bir optimizasyon prosedürünü KFB üzerinde uygulanmış ve elde edilen sonuçlardan yola çıkarak dolanıklık ölçütleri ile karşılaştırılmalı analizi yapılmıştır. Yine *Sistem Durum Sıralaması* problemi çerçevesinde sıralama ilişkileri üzerinde detaylı olarak durulmuştur.

Buradan elde edilen en temel sonuç KFB değerlerinin maksimize edilmesi durumunda özellikle REE değerleri için oldukça anlamlı sıralama ilişkilerinin elde edilebildiği gerçeğidir. Bu sonuca gore *LOCC'ye gore maksimize edilmiş KFB* değerlerinin dolanıklık açısından anlamlı olduğu ve sistem durum sıralaması problem için daha kullanışlı olduğudur. Buradan elde edilen sıralama sonuçlarının özellikle çoklu dolanık sistemlerin incelenmesi durumunda çok daha ilginç sonuçlar elde edilebileceği tahmin edilmektedir. Özellikle gelecek çalışmalara iki kütrit sistemler incelenerek başlanabileceğini düşünmekteyiz. Son bölümde de bahsedildiği gibi iki kütrit sistemlerde Negatiflik ölçütü için kapalı formül mevcut olmasına rağmen yine REE ölçütü için kapalı bir formül olmadığı gibi bir kestirim metodu da mevcut değildir. Açık bir araştırma olarak bu konu isteyen araştırmacılar tarafından çalışılabilir.

Tez kapsamında yapılan çalışmalarda elde ettiğimiz bulguların Kuantum Bilgi Teorisi'nin çalışma alanlarına giren aşağıdaki alanlarda genişletilerek kullanılabileceği ve araştırmacılara yeni bir perspektif kazandırabileceğini düşünmekteyiz:

- Olasılık ve İstatistik tabanlı hesaplama uygulamaları
- Başarıya Kadar Tekrar Et (Repeat Until Success) Tabanlı Haberleşme ve Kodlama algoritmaları
- Çizge Kuramı tabanlı problemlerin çözümleri
- Büyük Veri ve Optimizasyon problemleri

# KAYNAKLAR

[1] M.A. Nielsen and I.L. Chuang, Quantum Computation and Quantum Information, Cambridge, UK: Cambridge Univ. Press, 2000.

[2] M. B. Plenio, “Logarithmic Negativity: A Full Entanglement Monotone That is not Convex”, Phys. Rev. Lett., vol. 95, no. 090503, 2005.

[3] K.Zyczkowski, P. Horodecki, A. Sanpera, and M. Lewenstein, “Volume of the set of separable states”, Phys. Rev. A, vol. 58, no. 883, 1998.

[4] J.Eisert and M.B.Plenio, “A Comparison of Entanglement Measures”, J. Mod. Opt., vol. 46, pp. 145-154, 1999.

[5] A. Zeilinger, M. A. Horne, H. Weinfurter, and M. Zukowski, “Three-Particle Entanglements from Two Entangled Pairs”, Phys. Rev. Lett., vol. 78, no. 3031, 1997.

[6] T. Tashima, S. K. Ozdemir, T. Yamamoto, M. Koashi, and N. Imoto, “Elementary optical gate for expanding an entanglement web”, Phys. Rev. A, vol. 77, no. 030302, 2008.

[7] T. Tashima, S. K. Ozdemir, T. Yamamoto, M. Koashi, and N. Imoto, “Local expansion of photonic W state using a polarization-dependent beamsplitter”, New J. Phys. A, vol. 11, no. 023024, 2009.

[8] T. Tashima, T. Wakatsuki, S. K. Ozdemir, T. Yamamoto, M. Koashi, and N. Imoto, “Local Transformation of Two Einstein-Podolsky-Rosen Photon Pairs into a Three-Photon W State”, Phys. Rev. Lett., vol. 102, no. 130502, 2009.

[9] S. Bugu, C. Yesilyurt and F. Ozaydin, “Enhancing the W-state quantum-network-fusion process with a single Fredkin gate”, Phys. Rev. A, vol. 87, no. 032331, 2013.

[10] C. Yesilyurt, S. Bugu and F. Ozaydin, “An Optical Gate for Simultaneous Fusion of Four Photonic W or Bell States”, Quant. Info. Proc., vol. 12, no. 2965, 2013.

[11] F.Ozaydin et al., “Fusing multiple W states simultaneously with a Fredkin gate”, Phys. Rev. A, vol. 89, no. 042311, 2014.

[12] Z. Ji et. al, “Parameter Estimation of Quantum Channels”, IEEE Trans. Info. Theory., vol. 54, no. 5172, 2008.

[13] B. M. Escher, M. Filho, and L. Davidovich, “General framework for estimating the ultimate precision limit in noisy quantum-enhanced metrology”, Nat. Phys., vol. 7, no. 406, 2011.

[14] X. Yi., G. Huang and J. Wang, “Quantum Fisher Information of a 3-Qubit State”, Int. J. Theor. Phys. vol. 51, p. 3458, 2012.

[15] N. Spagnalo et al., “Quantum interferometry with three-dimensional geometry”, Nature Sci. Rep., vol. 2, no. 862, 2010.

[16] Z. Liu, “Spin Squeezing in Superposition of Four-Kübit Symmetric State and W States”, Int. J. Theor. Phys., vol. 52, p. 820, 2013.

[17] F. Ozaydin, A. A. Altintas, S. Bugu and C. Yesilyurt, “Quantum Fisher Information of N Particles in the Superposition of W and GHZ States”, Int. J. Theor. Phys., vol. 52, p. 2977, 2013.

[18] F. Ozaydin, A. A. Altintas, S. Bugu, C. Yesilyurt and M.Arik, “Quantum Fisher Information of Several Qubits in the Superposition of A GHZ and two W States with Arbitrary Relative Phase” Int. J. Theor. Phys., vol. 53, p. 3259, 2014.

[19] P. Gibilisco, D. Imparato, and T. Isola, “Uncertainty principle and Quantum Fisher Information”, J. Math. Phys., vol. 48, no. 072109, 2007.

[20] A. Andai, “Uncertainty principle with Quantum Fisher Information”, J. Math. Phys., vol. 49, no. 012106, 2008.

[21] J. Ma, Y. Huang, X. Wang and C. P. Sun, “Quantum Fisher information of the Greenberger-Horne-Zeilinger state in decoherence channels”, Phys. Rev. A, vol. 84, no. 022302, 2011.

[22] Toth et al., “Spin squeezing and entanglement”, Phys. Rev. A, vol. 79, no. 042334, 2009.

[23] F.Ozaydin, A.A.Altintas, S.Bugu, C.Yesilyurt, “Behavior of Quantum Fisher Information of Bell Pairs under Decoherence Channels”, Acta Phys. Pol. A, vol. 125(2), p. 606, 2014.

[24] S.Z.Ang, G.I.Harris, W.P.Bowen and M.Tsang, “Optomechanical parameter estimation”, New J. Phys., vol. 15, no. 103028, 2013.

[25] K.Iwasawa et al., “Quantum-Limited Mirror-Motion Estimation”, Phys. Rev. Lett., vol. 111, no. 163602, 2013.

[26] M.Tsang, “Quantum metrology with open dynamical systems”, New J. Phys., no. 15, vol. 073005, 2013.

[27] M.Tsang and N.Ranjith, “Fundamental quantum limits to waveform detection”, Phys. Rev. A, vol. 86, no. 042115, 2012.

[28] M.Tsang, “Ziv-Zakai Error Bounds for Quantum Parameter Estimation”, Phys. Rev. Lett., vol. 108, no. 230401, 2012.

[29] M.Tsang. H.M.Wiseman, C.M.Caves, “Fundamental Quantum Limit to Waveform Estimation”, IEEE Conference on Lasers and Electro-Optics (CLEO) 2011.

[30] M.Tsang, J.H.Shapiro, S.Lloyd, Quantum Optical Temporal Phase Estimation by Homodyne Phase-Locked Loops, IEEE Conference on Lasers and Electro-Optics (CLEO) 2009.

[31] B. R. Frieden and R. A. Gatenby, Exploratory Data Analysis Using Fisher Information, Arizona, USA:Springer, 2007.

[32] L. Pezze, A. Smerzi, “Entanglement, Nonlinear Dynamics, and the Heisenberg Limit”, Phys. Rev. Lett., vol. 102, no. 100401, 2009.

[33] V. Giovannetti, S. Lloyd and L. Maccone, “Quantum-Enhanced Measurements: Beating the Standard Quantum Limit”, Science, vol. 306, no. 1330, 2004.

[34] P. Hyllus, O. Gühne and A. Smerzi, “Not all pure entangled states are useful for sub-shot-noise interferometry”, Phys. Rev. A, vol. 82, no. 012337, 2010.

[35] A. Miranowicz and A. Grudka, “A comparative study of relative entropy of entanglement, concurrence and negativity”, J. Opt. B, vol. 6, no. 542, 2004.

[36] A. Miranowicz and A. Grudka, “Ordering two-qubit states with concurrence and negativity”, Phys. Rev. A, vol. 70, no. 032326, 2004.

[37] B. Horst, K. Bartkiewics and A. Miranowicz, “Two-qubit mixed states more entangled than pure states: Comparison of the relative entropy of entanglement for a given nonlocality”, Phys. Rev. A, vol. 87, no. 042108, 2013.

[38] C. W. Helstrom, Quantum Detection and Estimation Theory, New York: Academic Press, 1976.

[39] A. S. Holevo, Probabilistic and Statistical Aspects of Quantum Theory, Amsterdam, The Netherlands: North-Holland, 1982.

[40] H. N. Xiong, J. Ma, W. F. Liu and X. Wang, “Quantum Fisher Information for Superpositions of Spin States”, Quant. Inf. Comp., vol.10 no.5&6, 2010.

[41] Cai et al., “Entanglement-Based Machine Learning on a Quantum Computer”, Phys. Rev. Lett., vol. 114, no. 110504 (2015).

[42] K. Takata and Y. Yamamoto, “Data search by a coherent Ising machine based on an injection-locked laser network with gradual pumping or coupling”, Phys. Rev. A, vol. 89, no. 032319, 2014.

[43] A. W. Harrow, A. Hassidim, and S. Lloyd. “Quantum algorithm for solving linear systems of equations.”, Phys. Rev. Lett., vol. 15(103) no: 150502, 2009.

[44] K. Temme, T. J. Osborne, K. G. Vollbrecht, D. Poulin, and F.Verstraete. “Quantum Metropolis sampling.” Nature, vol. 471(7336) pp. 87–90, 2011.

[45] M. B. Plenio and S. Virmani, “An introduction to entanglement measures”, Quant. Inf. Proc. vol. 7, pp. 1-51, 2007.

[46] C. H. Bennett et al., “Mixed-state entanglement and quantum error correction”, Phys. Rev. A, vol. 54, no. 3824, 1996.

[47] C. H. Bennett et al., “Purification of Noisy Entanglement and Faithful Teleportation via Noisy Channels”, Phys. Rev. Lett., vol. 76, no. 722, 1996.

[48] W. K. Wootters, “Entanglement of Formation of an Arbitrary State of Two Qubits”, Phys. Rev. Lett., vol. 80, no. 2245, 1998.

[49] Y. Zinchenko, S. Friedland and G. Gour, “Numerical estimation of the relative entropy of entanglement”, Phys. Rev. A, vol. 82, no. 052336, 2010.

[50] G. Vidal and R. F. Werner, “Computable measure of entanglement”, Phys. Rev. A, vol. 65, no. 032314, 2002.

[51] V. Vedral, M. B. Plenio, M. A. Rippin and P. L. Knight, “Quantifying Entanglement”, Phys.Rev. Lett., vol. 78, no. 2275, 1997.

[52] V. Vedral and M. B. Plenio, “Entanglement measures and purification procedures”, Phys. Rev. A, vol. 57, no. 1619, 1998.

[53] R. Jozsa, D. S. Abrams, J. P. Dowling, and C. P. Williams, “Quantum Clock Synchronization Based on Shared Prior Entanglement”, Phys. Rev. Lett., vol. 85, no. 2010, 2000.

[54] J. J. Bollinger, W. M. Itano, D. J. Wineland, and D. J. Heinzen, “Optimal frequency measurements with maximally correlated states”, Phys. Rev. A, vol. 54, no.R 4649 (1996).

[55] F.Ozaydin et al., Quantum Fisher Information of Bipartitions of W States, Acta Phys. Pol. A, to appear.

[56] A. Miranowicz, “Violation of Bell inequality and entanglement of decaying Werner states.” Phys. Lett. A, no. 327, pp. 272–283, 2004.

[57] V. Erol, S. Bugu, F. Ozaydin and A. A. Altintas, “An analysis of concurrence entanglement measure and quantum fisher information of quantum communication networks of two-qubits” in Proceedings of IEEE 22nd Signal Processing and Communications Applications Conference SIU 2014 (IEEE, 2014), pp. 317-320.

[58] V. Erol, F. Ozaydin and A. A. Altintas, “Analysis of Entanglement Measures and LOCC Maximized Quantum Fisher Information of General Two Qubit Systems”, Nature Sci. Rep., vol. 4, no. 5422, 2014.

[59] J. Rehacek and Z. Hradil, "Quantification of Entanglement by Means of Convergent Iterations", Phys. Rev. Lett., vol. 90, no. 127904, 2003.

[60] P. B. Slater, "Ratios of maximal concurrence-parameterized separability functions, and generalized Peres-Horodecki conditions", J. Phys. A: Math. Theor., vol. 42, no. 465305, 2009.

[61] V. Erol, F. Ozaydin and A. A. Altintas, "Analysis of Negativity and Relative Entropy of Entanglement Measures for Qubit-Qutrit Quantum Communication Systems" in Proceedings of IEEE 23nd Signal Processing and Communications Applications Conference SIU 2015, to appear.

[62] A. Peres, "Separability Criterion for Density Matrices", Phys. Rev. Lett., vol. 77, no. 1413, 1996.

[63] M. Horodecki, P. Horodecki, and R. Horodecki, "Separability of Mixed States: Necessary and Sufficient Conditions", Phys. Lett. A, vol 223, p. 1, 1996.

[64] F. Ozaydin, "Phase damping destroys quantum Fisher information of W states", Phys. Let. A, vol. 378, pp. 3161-3164, 2014.

# EKLER

# DOKTORA SÜRESİNCE BUGÜNE KADAR YAPILAN YAYIN LİSTESİ

1- **Volkan Erol**, Fatih Özaydın and Azmi Ali Altıntaş, Analysis of Entanglement Measures and LOCC Maximized Quantum Fisher Information of General Two Qubit Systems, *Nature Sci. Rep.* **4**, 5422; DOI:10.1038/srep05422 (2014). (***SCI/E Indexed Journal*** **2013 IF:5.078**)

2- Fatih Özaydın, Azmi Ali Altıntaş, Can Yeşilyurt, Sinan Buğu, **Volkan Erol**, Quantum Fisher Information of Bipartitions of W States, *Acta Physica Polonica A* (***SCI/E Indexed Journal***) *(Mayıs 2015'te basılacak).*

*3-* **Volkan Erol**, A Comparative Study of Concurrence and Negativity of General Three-Level Quantum Systems of Two Particles, *AIP Conf. Proc.* **1653**, *020037 (2015).* ***(Conference SCI/E Indexed Proceeding)***

*4-* **Volkan Erol**, Sinan Buğu, Fatih Özaydın, Azmi Ali Altıntaş, İki Kübitlik Kuantum Haberleşme Ağlarının Eş Zamanlılık Dolanıklık Ölçütü ile Kuantum Fisher Bilgisinin Analizi, *IEEE 22. Sinyal İşleme ve İletişim Uygulamaları Kurultayı (SIU 2014), 23-25 Nisan 2014, Trabzon* ***(IEEE Xplore Indexed Proceeding)***

*5-* **Volkan Erol**, Fatih Özaydın, Azmi Ali Altıntaş, Kübit-Kütrit Kuantum Haberleşme Sistemleri İçin Negatiflik ve Dolanıklığın Göreceli Entropisi Ölçütlerinin Analizi, *IEEE 23. Sinyal İşleme ve İletişim Uygulamaları Kurultayı (SIU 2015), 16-19 Mayıs 2015, Malatya (kabul edildi).* ***(IEEE Xplore Indexed Proceeding)***

*6-* Sinan Buğu, **Volkan Erol**, Can Yeşilyurt, Azmi Ali Altıntaş and Fatih Özaydın, Strategy with recycling for the enhanced setup for creating large-scale W state networks, *International Workshop on Quantum Communication Networks (QCN 2014), 9-10 Ocak 2014, Leeds, İngiltere*

7- **Volkan Erol**, Aslı Uyar Özkaya, Elektrokardiyografi (EKG) Sinyallerindeki Aritmilerin Sınıflandırılması, *Akademik Bilişim 2014, 5-7 Şubat 2014, Mersin*

*8-* **Volkan Erol**, Baran Sakallıoğlu, Bekir Tevfik Akgün, Sosyal Oyunlar, *Akademik Bilişim 2014, 5-7 Şubat 2014, Mersin*

*9-* Baran Sakallıoğlu, **Volkan Erol**, Bekir Tevfik Akgün, Oyun Nedir ve Oyun Türlerinin Tanımlanmasında Sosyal Oyunların Yeri, *Akademik Bilişim 2014, 5-7 Şubat 2014, Mersin*

*10-* Can Yeşilyurt and **Volkan Erol**, Integrating Gesture Recognition Techniques to Digital and Outdoor Marketing Scenarios, *International Conference on Computational Techniques and Mobile Computing (ICCTMC'2012) 14-15 Aralık 2012 Singapur*